\documentclass[12pt,a4paper]{JHEP3}
\usepackage{amssymb}
\usepackage{amsbsy}

\newcommand{\del}{\partial}
\newcommand{\p}{\partial}
\def\half{{1\over 2}}
\def\a{\alpha}
\def\b{\beta}
\def\g{\gamma}
\def\d{\delta}
\def\e{\epsilon}
\def\m{\mu}
\def\vp{\varphi}

\def\l{\lambda}

\def\s{\sigma}
\def\p{\partial}
\def\pd{{\dot +}}

\def\t{{\theta}}

\def\nn{{\nonumber}}
\def\ZZ{{\mathbb{Z}}}
\def\RR{{\mathbb{R}}}
\def\N{{\cal N}}
\def\M{{\cal M}}

\def\tf{\tilde\phi}
\def\tk{\tilde\kappa}
\def\ad{{\dot \a}}

\def\bd{\overline d}

\newcommand{\be}{\begin{equation}}
\newcommand{\ee}{\end{equation}}
\newcommand{\bea}{\begin{eqnarray}}
\newcommand{\eea}{\end{eqnarray}}

\newcommand{\pp}{{\mathchoice
          {
              \kern 1pt%
              \raise 1pt
              \vbox{\hrule width5pt height0.4pt depth0pt
                    \kern -2pt
                    \hbox{\kern 2.3pt
                          \vrule width0.4pt height6pt depth0pt
                          }
                    \kern -2pt
                    \hrule width5pt height0.4pt depth0pt}%
                    \kern 1pt
           }
            {
              \kern 1pt%
              \raise 1pt
              \vbox{\hrule width4.3pt height0.4pt depth0pt
                    \kern -1.8pt
                    \hbox{\kern 1.95pt
                          \vrule width0.4pt height5.4pt depth0pt
                          }
                    \kern -1.8pt
                    \hrule width4.3pt height0.4pt depth0pt}%
                    \kern 1pt
            }
            {
              \kern 0.5pt%
              \raise 1pt
              \vbox{\hrule width4.0pt height0.3pt depth0pt
                    \kern -1.9pt  
                    \hbox{\kern 1.85pt
                          \vrule width0.3pt height5.7pt depth0pt
                          }
                    \kern -1.9pt
                    \hrule width4.0pt height0.3pt depth0pt}%
                    \kern 0.5pt
            }
            {
              \kern 0.5pt%
              \raise 1pt
              \vbox{\hrule width3.6pt height0.3pt depth0pt
                    \kern -1.5pt
                    \hbox{\kern 1.65pt
                          \vrule width0.3pt height4.5pt depth0pt
                          }
                    \kern -1.5pt
                    \hrule width3.6pt height0.3pt depth0pt}%
                    \kern 0.5pt
            }
        }}

\newcommand{\mm}{{\mathchoice
                       {
                             \kern 1pt
               \raise 1pt    \vbox{\hrule width5pt height0.4pt depth0pt
                                  \kern 2pt
                                  \hrule width5pt height0.4pt depth0pt}
                             \kern 1pt}
                       {
                            \kern 1pt
               \raise 1pt \vbox{\hrule width4.3pt height0.4pt depth0pt
                                  \kern 1.8pt
                                  \hrule width4.3pt height0.4pt depth0pt}
                             \kern 1pt}
                       {
                            \kern 0.5pt
               \raise 1pt
                            \vbox{\hrule width4.0pt height0.3pt depth0pt
                                  \kern 1.9pt
                                  \hrule width4.0pt height0.3pt depth0pt}
                            \kern 1pt}
                       {
                           \kern 0.5pt
             \raise 1pt  \vbox{\hrule width3.6pt height0.3pt depth0pt
                                  \kern 1.5pt
                                  \hrule width3.6pt height0.3pt depth0pt}
                           \kern 0.5pt}
                       }}

\def\ad{{\kern0.5pt
                   \alpha \kern-5.05pt
\raise5.8pt\hbox{$\textstyle.$}\kern 0.5pt}}

\def\bd{{\kern0.5pt
                   \beta \kern-5.05pt \raise5.8pt\hbox{$\textstyle.$}\kern 0.5pt}}

\def\qd{{\kern0.5pt
                   q \kern-5.05pt \raise5.8pt\hbox{$\textstyle.$}\kern 0.5pt}}
\def\Dot#1{{\kern0.5pt
     {#1} \kern-5.05pt \raise5.8pt\hbox{$\textstyle.$}\kern 0.5pt}}

\title{Non-Critical Pure Spinor Superstrings}

\author{
Ido Adam$^1$, Pietro Antonio Grassi$^{2\ 3\ 4}$,  Luca
Mazzucato$^1$,

Yaron
Oz$^1$ and Shimon Yankielowicz$^1$,\\
$^1$Raymond and Beverly Sackler Faculty of Exact Sciences \\
School of Physics and Astronomy \\
Tel-Aviv University, Ramat-Aviv 69978, Israel \\
$^2$ Centro Studi e Ricerche E. Fermi,
Compendio Viminale, I-00184, Roma, Italy \\
$^3$ CERN, Theory Unit, CH-1211 Geneve, 23, Switzerland \\
$^4$ DISTA, Universit\`a del Piemonte Orientale,
Via Bellini 25/G, 15100, Alessandria, Italy and INFN, Torino, Italy \\
E-mails: \email{adamido@post.tau.ac.il},
\email{pietro.grassi@cern.ch}, \email{mazzul@post.tau.ac.il},
\email{yaronoz@post.tau.ac.il},
 \email{shimonya@post.tau.ac.il}}
\abstract{We construct non-critical pure spinor superstrings in
two, four and six dimensions. We find explicitly the map between
the RNS variables and the pure spinor ones in the linear dilaton
background. The RNS variables map onto a patch of the pure spinor
space and the holomorphic top form on the pure spinor space is an
essential ingredient of the mapping. A basic feature of the map is
the requirement of doubling the superspace, which we analyze in
detail. We study the structure of the non-critical pure spinor
space, which is different from the ten-dimensional one, and its
quantum anomalies. We compute the pure spinor lowest lying BRST
cohomology and find an agreement with the RNS spectra. The
analysis is generalized to curved backgrounds and we construct as
an example the non-critical pure spinor type IIA superstring on
$AdS_4$ with RR 4-form flux.}

\preprint{TAUP-2828/06 \\ hep-th/0605118} 

\begin{document}


\section{Introduction and summary}

The critical dimension for the superstrings in flat space-time is
$d=10$. In dimensions $d<10$, the Liouville mode is dynamical and
needs to be quantized as well. These superstrings are sometimes
called non-critical. The Liouville mode can be interpreted as a
dynamically generated dimension. Thus, if we start with
superstring theory in  $d<10$ space-time dimensions, we have
effectively $d+1$ space-time dimensions. The total conformal
anomaly vanishes  for the non-critical superstrings due to the
Liouville background charge. However, while this is a necessary
condition for the consistency of non-critical superstrings, it is
not a sufficient one. Much work has been done on the analysis of
non-critical strings in two and less dimensions. Consistent
superstring theories in linear dilaton backgrounds with even
dimensions have been constructed in \cite{Kutasov:1990ua}, and
studied in the RNS formalism.

There are various  motivations to study non-critical  strings.
First, non-critical  superstrings can provide alternative to
superstring compactifications. Second, the study of non-critical
superstrings in the context of the gauge/string correspondence may
provide dual descriptions of new gauge theories, and in particular
QCD \cite{Polyakov:1998ju,Polyakov:2004br}.

A complication in the study of  non-critical superstrings in
curved spaces is that, unlike the critical case, there is no
consistent approximation where  supergravity provides a valid
effective description. The reason being that the $d$-dimensional
supergravity low-energy effective action contains a cosmological
constant type term of the form
$$
S \sim \int d^d x \sqrt{G}e^{-2 \Phi}\left({d-10 \over
l_s^2}\right) \ ,
$$
which vanishes only for $d=10$. This implies that the low energy
approximation $E\ll l_s^{-1}$ is not valid when $d \neq 10$, and
the higher order curvature terms of the form $\left(l_s^2{\cal
R}\right)^n$ cannot be discarded. A manifestation of this is that
solutions of the  $d$-dimensional non-critical supergravity
equations have typically curvatures of the order of the string
scale $l_s^2{\cal R}\sim O(1)$ when  $d \neq 10$.

The second complication is that interesting target space curved
geometries include RR field fluxes. As in the critical superstrings
case, the RNS formulation is inadequate for the quantization of
superstrings in such backgrounds. In \cite{Grassi:2005kc} a covariant
description of non-critical superstrings in even dimensions has been
constructed using the hybrid type variables. The approach taken was to
construct a covariant description of non-critical superstrings on the
linear dilaton background and use the supersymmetric variables to
construct the non-critical superstrings $\sigma$-model action in
general curved target space backgrounds. The goal of this work is to
develop the pure spinor quantization procedure for non-critical
superstrings.

The pure spinor formalism for critical superstrings
\cite{Berkovits:2000fe} is based on integer worldsheet spin
variables. They are the target superspace coordinates
$(x^m,\theta^{\alpha})$, where $x^m, m=0,...,9$ are commuting
coordinates and $\theta^{\alpha}, \alpha=1,...,16$ are
anti-commuting coordinates, $p_{\alpha}$ the conjugate momenta to
$\theta^{\alpha}$, the bosonic spinor ghosts $\lambda^{\alpha}$,
which satisfy the (complex) pure spinor constraint
$$
\lambda^{\alpha}\gamma_{\alpha\beta}^m\lambda^{\beta} = 0 \ ,
$$
and their conjugate momenta $ w_\a$. By construction, the pure
spinor formalism is manifestly space-time supersymmetric and
provides a simple coupling to the NS-NS and RR fields.

Unlike the GS formalism, there is no $\kappa$-symmetry to gauge
fix. Instead, an important ingredient of the pure spinor formalism
is the nilpotent BRST operator
$$
Q = \int dz \lambda^{\alpha}d_{\a} \ ,
$$
where $d_{\a}$ acts as  the supersymmetric derivative in ten
dimensions. Physical states are elements in the BRST cohomology
modulo the pure spinor constraints. In addition, the invariance of
the action under the BRST symmetry can provide strong constraints
on possible quantum corrections. This has been used, for instance,
to prove that $AdS_5\times S^5$ is a consistent background for
type II superstrings \cite{Berkovits:2004xu}.

In this paper we will use a pure spinor formalism to describe
non-critical superstrings. The strategy of constructing the pure
spinor description of the non-critical superstrings is to first
map the bosonic and fermionic linear dilaton RNS variables to pure
spinor variables. A generic feature  of the map in all dimensions
is that it takes the RNS variables to the pure spinor variables on
a patch of the pure spinor space. The pure spinor space of
non-critical superstrings will be different from the pure spinor
space of critical superstrings.

By performing the map of the RNS variables to the pure spinor
variables, we will see that a basic feature is the requirement of
doubling the superspace. This can be done by enlarging the linear
dilaton superspace structure to include superspace coordinates
(with their conjugate momenta), which are not BRST invariant on
the RNS side. Similarly we will have a doubling of the
superderivates, with only half of them being physical, though all
being conserved. We will see that the current algebra of the
doubled superderivatives is not a supersymmetry algebra and only a
nonanomalous subsector closes on the spacetime supersymmetry.
Indeed, working in a doubled superspace with the pure spinor
variables will require an appropriate projection to this physical
subsector. However, the doubled superspace will allow to study
pure spinor superstrings in backgrounds with double the
supersymmetries of the linear dilaton background.

Let us illustrate the above discussion by briefly presenting the
pure spinor structure that we will find for two-dimensional
superstrings in linear dilaton background
 $$
\mathbb{R}_\vp\times U(1)_x \ .
 $$
The background has two bosonic dimensions, the noncompact
Liouville  direction $\vp$ and the $x$ direction compactified on a
circle with radius $R=2/Q$, where $Q=2$ is the Liouville
background charge. The doubled superspace (in the holomorphic
sector) has two fermionic coordinates $(\theta^+,\theta^\pd)$ with
their conjugate momenta $(p_+,p_\pd)$. The $(\theta^+, p_+)$ pair
are BRST invariant physical quantities, while $(\theta^\pd,p_\pd)$
are BRST non-invariant, but required by the map from the RNS
variables to the covariant variables. Similarly, the corresponding
two superderivatives are $d_+$ and $d_\pd$, where  the former is
BRST invariant, and the latter is not. The OPE between the
physical and unphysical $d$'s does not close a superalgebra, but
rather has a double pole
$$d_+(z)d_\pd(0)\sim-{Q\over z^2} + ...$$
The space-time zero-dimensional supersymmetry is realized by the
physical $d_+$, which is in fact nilpotent in this case.

The crucial ingredient in this covariant formalism is provided by
the pure spinor variables $(\lambda^{\alpha}, w_\a)$, which form a
curved beta-gamma system on the pure spinor space. This has
important consequences that will be discussed in detail in the
paper. First there is a coupling of the worldsheet curvature
$r^{(2)}$ to the holomorphic top form $\Omega$ on the pure spinor
space
$$S \sim  \int d^2z
r^{(2)}\log \Omega(\lambda)\ , $$ which modifies the stress
tensor, as well as the saturation rules for correlators. Second,
there are global obstructions to the definition of  the pure
spinor system on the worldsheet and on target space, associated
with the need for holomorphic transition functions relating
$(\lambda^{\alpha},w_{\alpha})$ on different patches of the pure
spinor space, which are compatible with their OPE. They are
reflected by quantum anomalies in the worldsheet and pure spinor
space holomorphic diffeomorphisms \cite{Witten:2005px}. The
critical superstring pure spinor space has a singularity at
$\lambda^{\alpha} = 0$. Blowing up the singularity results in an
anomalous theory. However, simply removing the origin leaves a
non-anomalous theory \cite{Nekrasov:2005wg}. The same holds for
the pure spinor spaces of non-critical superstrings. However, in
two and four dimensions, removing the origin results in
disconnected pure spinor spaces.

The pure spinor variables  in the two dimensional non-critical
string are $(\lambda^+, \lambda^\pd)$ satisfying the equation
$$\lambda^+
 \label{2dps}
\lambda^\pd = 0 \ . $$ This defines a complex dimension one pure
spinor space. Note for comparison that the pure spinor space of
critical superstrings in ten dimensions is a complex
eleven-dimensional space. The map from the RNS variables to the
pure spinor ones takes the RNS variables to the patch of the pure
spinor space defined by
  $(\lambda^+ \neq 0, \lambda^\pd =  0)$.

The superstring on a two dimensional linear dilaton background is
defined by the following stress tensor
 \bea
 T=&-\half\Pi^m\Pi_m-d_I\p\t^I+{Q\over2}\e_{IJ}\p\t^I\p\t^J+
{Q\over2}\p^2(\Pi^\varphi-i\Pi^x) \nn\\
 &+w_{I}\p\l^{I}-\half\p^2\log\Omega(\l) \ ,\nn
 \eea
 where $I=+,\pd$ and $m=x,\vp$.
This structure is very similar to that of pure spinor critical
superstrings and is convenient for generalization to curved
backgrounds. Note that the terms proportional to $Q=2$ are
features of the linear dilaton background.

In the BRST operator of the non-critical superstrings we will
include not only the physical superderivatives but also the
unphysical ones. Note in comparison that, in the critical
superstring in flat ten dimensions, all the superderivatives in
the BRST charge are physical. Here, the BRST operator
$$Q_B = \oint
\lambda^{I}d_{I}$$ where $I=+,\pd$, includes both the physical and
unphysical superderivatives. Note that, due to the double pole
in the $d$'s OPE,  the
 nilpotency of the BRST charge $Q_B$ requires, in addition
to the pure spinor condition $\l^+\l^{\dot+}=0$, that
$$\p\l^+\l^{\dot+}-\l^+\p\l^{\dot+}=0 \ .$$
However, this derivative condition is a consequence of the
algebraic one.

We will compute the cohomology and, since we are interested in the
space-time supersymmetry multiplets, we will consider only the
part of the cohomology that contains the variables that realize
the supersymmetry current algebra. The physical states are the
vertex operators at ghost number one and weight zero. We will find
two different kinds of such operators. The first is analogous to
the usual ten-dimensional vertex operator. In general this
contains the off-shell $d$-dimensional supergravity multiplet (for
the closed strings). The second type is peculiar to the linear
dilaton background. It contains a gauge multiplet, in which the
tachyon sits. This will reproduce the RNS computation of the short
supermultiplets.

In computing the cohomology of the BRST operator, we restrict to
the part of the vertex operators that contains the physical
fermionic coordinate $\t^+$. This indeed reproduces the RNS
computation of the short supermultiplets. The supergravity
multiplet does not exist in the case of two-dimensional
superstrings, while the supermultiplet in which the tachyon sits
has two bosonic and two fermionic degrees of freedom.

One essential feature of the pure spinor non-critical string is
that it can be generalized to describe other non-critical
backgrounds, as we are used to do in the ten-dimensional critical
superstrings. As an example, we will propose a sigma model for the
type IIA non-critical superstring on $AdS_4$ background with RR
four--form flux. This background is described by the supercoset
 $$
 {OSp(2|4)\over SO(1,3)\times SO(2)}\nn
 $$
and has eight real supercharges, which is the content of the
enlarged superspace in the four-dimensional non-critical string.
As we will see, the action for this supercoset will be suitable
for quantization and its BRST charge will be the usual pure spinor
BRST charge.

\vspace{0.2cm}

The paper is organized as follows. In section 2 we first give a
brief introduction to the pure spinor formalism for critical
ten-dimensional superstrings. We then
 construct a map from the RNS variables to
the pure spinor ones for critical ten-dimensional superstrings.
The map  makes the $\beta \gamma$-system structure of the pure
spinor variables explicit and we will gain an insight into the
global definition of the pure spinor space and the importance of
its holomorphic top form. In section 3 we discuss the RNS
non-critical superstrings in the linear dilaton background and
provide all the necessary ingredients for the rest of the paper.
In particular we summarize the space-time supersymmetry structure
and describe the spectrum of space-time supermultiplets.

In section 4 we pass to the construction of the pure spinor
two-dimensional non-critical superstring, whose main features have
been outlined above. We map the RNS variables to the pure spinor
ones, analyze the supersymmetry structure of the covariant
formulation and its pure spinor space. We construct the pure
spinor action and stress tensor, introduce the BRST operator and
its cohomology and consider global issues and anomalies of the
pure spinor formalism. In section 5 and 6 we repeat the same
construction and analysis for the pure spinor non-critical
superstrings in four and in six dimensions.

In section 7 we comment on the computation of amplitudes and
discuss the non-critical pure spinor measure. In section 8 we
propose the type IIA pure spinor formulation of non-critical
$AdS_4$ background with RR flux. Section 9 is devoted to a
discussion of the open problems and the future directions. In the
appendices we will put the details of various computations. In
appendix A we explain the various notations for the spinors we
used through the main text. In appendix B and C we collect the RNS
generators of the $N=2$ superconformal algebra and we show some
details of the RNS computation of the short multiplets. In
appendix D and E we give some details of the pure spinor
computations. In appendix F we present a curious deformation of
the pure spinor theory we came accross.


\section{The pure spinor formalism}

In this section we will briefly review the main ingredients of the
pure spinor formalism for critical superstrings in flat
ten-dimensional target space
\cite{Berkovits:2000fe,Berkovits:2002zk}. These structures will
appear with some modifications in the pure spinor non-critical
superstrings. We will consider for simplicity the open
superstring. The generalization to the closed string case is
straightforward. Consider the supermanifold
$(x^m,\theta^{\alpha})$, where $x^m, m=0,...,9$ are commuting
coordinates and $\theta^{\alpha}, \alpha=1,...,16$ are
anti-commuting coordinates. One introduces $p_{\alpha}$ as the
conjugate momenta to $\theta^{\alpha}$ with the OPE \be
p_{\alpha}(z) \theta^{\beta}(0) \sim \frac{\delta_{\a}^{\b}}{z} \
. \ee Next we add bosonic spinor ghosts $\lambda^{\alpha}$, which
satisfy the (complex) pure spinor constraint \be
\lambda^{\alpha}\gamma_{\alpha\beta}^m\lambda^{\beta} = 0 \ ,
\label{const} \ee and their conjugate momenta $ w_\a$.
The system $( w_\a,\lambda^{\alpha})$ is a $(\beta,\gamma)$ system
of weights $(1,0)$. $\gamma_{\alpha\beta}^m$ are the symmetric
$16\times 16$ Pauli matrices in ten dimensions. The pure spinor
set of constraints (\ref{const}) is reducible. It defines a
complex eleven-dimensional space ${\cal M}$, which is a cone over
${\cal Q} = \frac{SO(10)}{U(5)}$.

The pure spinor constraint implies that $ w_\a$ are defined up to
the gauge transformation \be \delta  w_\a = \Lambda^m (\gamma_m
\lambda)_{\alpha} \ . \ee Therefore, $ w_\a$ appears only in gauge
invariant combinations. These are the Lorentz algebra currents
$M_{mn}$, the ghost number current $J_{(w,\lambda)}$ which assigns
ghost number $1$ to $\lambda$ and ghost number $-1$ to $ w $ \be
M_{mn} = \frac{1}{2}w\gamma_{mn}\lambda,~~~~~J_{(w,\lambda)} =
 w_\a\lambda^{\alpha} \ , \ee and the pure spinor
stress-energy tensor $T_{(w,\lambda)}$, which we will now discuss.

The gauge fixed worldsheet action is $S=S_0+S_1$, where \be S_0 =
\int d^2z \left(\frac{1}{2}\p x^m\bar{\p}x_m +
p_{\a}\bar{\p}\theta^{\alpha} -   w_\a\bar{\p} \lambda^{\alpha}
\right) \ , \ee and \be \label{topcop} S_1 = \int d^2z
\left(\frac{1}{4}r^{(2)}\log \Omega(\lambda)\right) \ . \ee $S_1$
is a coupling of the worldsheet curvature $r^{(2)}$ to the
holomorphic top form $\Omega$ of the pure spinor space ${\cal M}$
 \be \Omega = \Omega(\lambda)d \l^1\wedge ...\wedge
d\l^{11} \ .
 \ee
It is not yet clear, however, what is the non gauge-fixed form of
the action $S$. \footnote{There have been various attempts to
derive the pure spinor formalism from first principles
\cite{Matone:2002ft,Aisaka:2005vn,Gaona:2005yw}.} Note that the
$(w_\a,\lambda^{\alpha})$ action is holomorphic and does not
depend on their complex conjugates.

The stress tensor of the $(w,\lambda)$ system reads
 \be
T_{(w,\lambda)} =   w_\a\partial \lambda^{\alpha} -
\frac{1}{2}\p^2\log \Omega(\l) \ , \label{Ttop}
 \ee
and we will discuss the significance of the last term below. The
system $( w_\a,\lambda^{\alpha})$ is interacting due to the pure
spinor constraint. It has the central charge $c_{( w ,\lambda)} =
22$, which is twice the complex dimension of the pure spinor space
\be T_{( w ,\lambda)}(z)T_{( w ,\lambda)}(0) \sim
\frac{dim_{\mathbb{C}}({\cal M})}{z^4} + ... \ . \ee

The ghost number anomaly reads \be J_{( w ,\lambda)}(z)T_{( w
,\lambda)}(0) \sim -\frac{8}{z^3} + ... = \frac{c_1({\cal
Q})}{z^3}+ ...\ , \ee where ${c_1({\cal Q})}$ is the first Chern
class of the pure spinor cone base ${\cal Q}$.

The physical states are defined as the ghost number one cohomology
of the nilpotent BRST operator \be Q = \oint dz\,
\lambda^{\alpha}d_{\a} \ , \label{10ld}\ee where \be d_{\a} =
p_{\a} - \frac{1}{2}\gamma^m_{\a\b}\theta^{\b}\p x_m
-\frac{1}{8}\gamma^m_{\a\b}\gamma_{m\g\delta}\theta^{\b}
\theta^{\g}\p\theta^{\delta} \ . \ee This BRST operator is an
essential ingredient of the formalism but it is not clear how to
derive its form by a gauge fixing procedure.

The $d_{\a}$ are the supersymmetric Green-Schwarz constraints.
They are holomorphic and satisfy the OPE
 \be d_{\a}(z) d_{\b}(0)
\sim - \frac{\gamma^m_{\a\b}\Pi_m(0)}{z} \ , \ee and \be d_{\a}(z)
\Pi^m (0) \sim \frac{\gamma^m_{\a\b}\p \theta^{\b}(0)}{z} \ , \ee
where \be \Pi_m = \p x_m + \frac{1}{2} \theta\gamma_m \p \theta \
, \ee is the supersymmetric momentum. $d_{\a}$ acts on function on
superspace $F(x^m, \theta^{\a})$ as \be d_{\a}(z)
F(x^m(0),\theta^{\b}(0)) \sim \frac{D_{\a}
F(x^m(0),\theta^{\b}(0))}{z} \ , \ee where \be D_{\a} =
\frac{\p}{\p \theta^{\a}} +
\frac{1}{2}\gamma^m_{\a\b}\theta^{\b}\p_m \ , \ee is the
supersymmetric derivative in ten dimensions.

Massless states are described by the ghost number one weight zero
vertex operators \be {\cal V}^{(1)} =
\lambda^{\alpha}A_{\alpha}(x,\theta) \ . \ee The conditions
$Q{\cal V}^{(1)}=0$ and the gauge invariance $\d {\cal V}^{(1)} =
Q \Omega^{(0)}$ imply by explicit computation that $A_{\alpha}$ is
a super Maxwell spinor superfield in ten dimensions
\be
A_\alpha (x, \theta) = \frac{1}{2} (\gamma^m \theta)_\alpha a_m (x) +
\frac{i}{12} (\theta \gamma^{mnp} \theta) (\gamma_{mnp})_{\alpha
\beta} \psi^\beta (x) + O(\theta^3)
\ee
where $a_m(x)$ is the gauge field and $\psi^{\g}(x)$ is the
gluino. It is related to the gauge field $A_m$ by \be A_m =
\g_{m}^{\a\b}D_{\a}A_{\b} \ , \ee and $A_m(x,\theta) = a_m(x) +
O(\theta) $. Only in ten dimensions do these conditions give an
on-shell vector multiplet. In lower dimensions they describe an
off-shell vector multiplet.

The integrated ghost number zero vertex operator for the massless
states reads \be {\cal V} = \int dz \left(\p\theta^{\a}A_{\a} +
\Pi^m A_{m} + d_{\a}W^{\a} + \frac{1}{2}M_{mn}F^{mn}\right) \ ,
\ee where $W^{\a}$ and $F^{mn}$ are the spinorial and bosonic
field strength, respectively. The analysis of the massive states
proceeds in a similar way. At the first massive level the ghost
number one weight one vertex operator has the expansion
\cite{Berkovits:2002qx} \be {\cal U}^{(1)} = \p\lambda^{\a}A_{\a}
+   \lambda^{\a}\p \theta^{\b}B_{\a\b} + ... \ . \ee In curved
spaces the superspace field equations are derived by the
requirement that $\l^{\a}d_{\a}$ is holomorphic and nilpotent
\cite{Berkovits:2001ue}.

The construction of the closed superstrings is straightforward.
One introduces the right moving superspace variables
 $(\bar{p}_{\hat{\alpha}}, \bar{\theta}^{\hat{\alpha}})$, the pure spinor
 system $(\bar{ w }_{\hat{\alpha}},\bar{\lambda}^{\hat{\alpha}})$ and
the nilpotent BRST operator \be \bar{Q} = \oint
d\bar{z}\,\bar{\lambda}^{\hat{\alpha}}\bar{d}_{\hat{\a}} \ . \ee
The analysis of the spectrum proceeds by combining the left and
right sectors. For instance, the integrated ghost number zero
vertex operator for the massless states reads \be {\cal U} = \int
d^2z \left(\p\theta^{\a}A_{\a\hat{\b}}
\bar{\p}\bar{\theta}^{\hat{\b}} + \p\theta^{\a} A_{\a
m}\bar{\Pi}^m
 + ... \right) \ .
\ee

The pure spinor system $(\lambda^{\alpha},w_{\alpha})$ defines a
non-linear $\sigma$-model due to the curved nature of the pure
spinor space (\ref{const}). There are global obstructions to
define  the pure spinor system on the worldsheet and on target
space \cite{Witten:2005px,Nekrasov:2005wg}. They are associated
with the need for holomorphic transition functions relating
$(\lambda^{\alpha},w_{\alpha})$ on different patches of the pure
spinor space, which are compatible with their OPE. They are
reflected by quantum anomalies in the worldsheet and target space
(pure spinor space) diffeomorphisms. The conditions for the
vanishing of these anomalies are the vanishing of the integral
characteristic classes \label{anomalies} \be
\frac{1}{2}c_1(\Sigma)c_1(\M) = 0,~~~~~~\frac{1}{2}p_1(\M)=0 \ ,
\ee $c_1(\Sigma)$ is the first Chern class of the worldsheet
Riemann surface, $c_1(\M)$ is the first Chern class of the pure
spinor space $\M$, and $p_1$ is the first Pontryagin class of the
pure spinor space. The vanishing of $c_1(\M)$ is needed for the
definition of superstring perturbation theory and it implies the
existence of the nowhere vanishing holomorphic top form
$\Omega(\l)$ on the pure spinor space $\M$, that appears in the
stress tensor (\ref{Ttop}).

The pure spinor space (\ref{const}) has a singularity at
$\lambda^{\alpha} = 0$. Blowing up the singularity results in an
anomalous theory. However, simply removing the origin leaves a
non-anomalous theory. This means that one should consider the pure
spinor variables as twistor-like variables. Indeed this is a
natural intrepretation of the pure spinor variables considering
them from the twistor string point of view.

Finally, although we will not discuss the computation of loop
amplitudes, it is worth mentioning that unlike the RNS
superstrings, all the variables that we use in the pure spinor
superstring are of integer worldsheet spin and there is no need to
sum over spin structures.


\subsection{The ten-dimensional map}
\label{10mapsection}

In this section we will construct a map from the RNS variables to
the pure spinor ones. We will make use of a parameterization of
the pure spinor components that would make the $\beta
\gamma$-system structure of the pure spinor variables explicit. In
this way we will gain a new insight into the global definition of
the pure spinor space and the importance of its holomorphic top
form. The pure spinor stress tensor we will obtain by the map will
contain the contribution of the holomorphic top form on the pure
spinor space. Indeed this term is necessary for a consistent
definition of the pure spinor $\beta\gamma$-system
\cite{Nekrasov:2005wg,Witten:2005px}. Note that in
\cite{Berkovits:2001us} a similar map from the RNS variables to
the pure spinor ones has been constructed, but with no
consideration of the  $\beta \gamma$-system structure  and the
holomorphic top form.

In the following we will consider the holomorphic sector. It is to
be complemented by the anti-holomorphic sector for obtaining the
Type II superstring. The holomorphic supercharges in the
$-\frac{1}{2}$ picture of the RNS superstring are given by the
spin fields
\begin{equation}
  q_{\boldsymbol{s}} = e^{-\phi / 2 + \sum_{I=1}^5 s_I H^I} \ ,
\end{equation}
where the $H^I$s are the bosons obtained from the bosonization of
the RNS worldsheet matter fermions and the $s_I$'s take the values
$\pm\frac{1}{2}$. These supercharges decompose into two Weyl
representations.

In order to proceed with the map, one must first solve the pure
spinor constraint $\lambda \gamma^m \lambda = 0$, going to one
patch of this manifold. In each patch a different component of the
pure spinor is non-zero. The field redefinition we will use maps
the RNS description into one patch of the pure spinor manifold.
For concreteness we will work on one of the patches which is most
conveniently described by the $SU(5) \times U(1)$ decomposition of
the pure spinor $\l^\a=(\l^+,\l^a,\l_{ab})$. The component of the
pure spinor assumed to be non-zero is $\lambda^+$ corresponding to
the representation $1_\frac{5}{2}$ of this decomposition. In this
patch one can solve for the $5_{-\frac{3}{2}}$ components
$\lambda^a$ in terms of $\lambda^+$ and the components in the
$10_\frac{1}{2}$ representation $\lambda_{ab}$.

On this patch the supercharge $q_+$ which is the singlet of
$SU(5)$ is raised to the $+\frac{1}{2}$ picture:
\begin{equation}
  q_+ = b \eta e^{3 \phi / 2 + i \sum_a H^a / 2} + i \sum_a \partial
  (x^a + i x^{a+5}) e^{\phi / 2 + i \sum_b H^b / 2 - i H^a} \ ,
\end{equation}
while the supercharges $q_a$, corresponding to the pure spinor
components $\lambda^a$ we solved for, remain in the $-\frac{1}{2}$
picture. Together they form a part of the original ten-dimensional
supersymmetry algebra.  One then defines the fermionic momenta
\begin{equation}
  p_+ = b \eta e^{3 \phi / 2 + i \sum H^a / 2} \ , \quad p_a = q_a
\end{equation}
and their conjugate coordinates $\theta^+$ and $\theta^a$. Note
that the OPE's of the fermionic momenta among themselves  are all
non-singular.

The heart of the map is the introduction of two new fields $\tilde
\phi$ and $\tilde \kappa$ using
\begin{equation}
\label{field10}
  \eta = e^{\tilde \phi + \tilde \kappa} p_+ \ , \qquad b = \quad e^{(\tilde \phi
  - \tilde \kappa) / 2} p_+ \ ,
\end{equation}
yielding
\begin{eqnarray}
  \tilde \phi & = & - {3i\over4} \sum_a H^a -  \kappa - {9\over4}
   \phi + \half \chi \ ,\\
  \tilde \kappa & = & {i\over4}\sum_a H^a - \kappa - {3\over4}
  \phi - \half\chi  \ ,
\end{eqnarray}
whose OPE's are
\begin{equation}
  \tilde \phi (z) \tilde \phi (0) \sim - \log z \ , \quad \tilde
  \kappa (z) \tilde \kappa (0) \sim \log z \ .
\end{equation}
The reason why we choose the particular field redefinition
(\ref{field10}), explained in \cite{Berkovits:2001us}, is that the
pure spinor formalism is, loosely speaking, equivalent to the RNS
formalism when we take into account all the different pictures at
the same time, which is achieved by working in the large Hilbert
space, that is including the zero modes of the ghost $\xi$. But
the usual cohomology of the RNS BRST charge $Q_{RNS}$ in the small
Hilbert space is equivalent to the cohomology of
$Q_{RNS}+\oint\eta$ in the large Hilbert space. With the
redefinition (\ref{field10}), we are then mapping the $\oint\eta$
term of this extended BRST charge directly to the part $\oint
\l^+d_+$ of the Berkovits BRST operator (\ref{10ld}).

By substituting the map into the RNS energy-momentum tensor one
obtains
\begin{eqnarray}
  T & = & T_\mathrm{m} + T_\mathrm{gh} = - \frac{1}{2} \sum_m
  (\partial x^m)^2 - p_+ \partial \theta^+ - \sum_a p_a \theta^a -
  \nonumber\\
  && {} - \frac{1}{2} (\partial \tilde \phi)^2 + \frac{1}{2} (\partial
  \tilde \kappa)^2 + \partial^2 \tilde \phi + \partial^2 \tilde \kappa \ .
  \label{eq:critical-energy-momentum}
\end{eqnarray}
This can be verified to have still a vanishing central charge
\begin{displaymath}
  c = (10)_x + (-12)_{p \theta} + (2)_{\tilde \phi \tilde \kappa} = 0 \ .
\end{displaymath}
The pure spinors are reconstructed by the ordinary bosonization of
a $\beta \gamma$-system \cite{Friedan:1985ge}
\begin{equation}
  \lambda^+ = e^{\tilde \phi + \tilde \kappa} \ , \quad
  w_+ = \partial \tilde \kappa e^{-\tilde \phi - \tilde \kappa} \ ,
\end{equation}
whose OPE is
\begin{equation}
  w_+ (z) \lambda^+ (0) \sim \frac{1}{z} \ .
\end{equation}
But the naive stress tensor one would expect for this  $\beta
\gamma$-system
\begin{displaymath}
  w_+ \partial \lambda^+ = - \frac{1}{2} (\partial \tilde \phi)^2 +
  \frac{1}{2} (\partial \tilde \kappa)^2 - \frac{1}{2} \partial^2 \tilde
  \phi - \frac{1}{2} \partial^2 \tilde \kappa \ ,
\end{displaymath}
does not coincide with the one we got from the map
(\ref{eq:critical-energy-momentum}). This shows that the pure
spinor stress tensor is not simply $w_+\p\l^+$ but actually
\begin{equation}
  T_\lambda = w_+ \partial \lambda^+ - \frac{1}{2} \partial^2 \log
  \Omega(\l) \ ,
\end{equation}
where $\Omega$ is the coefficient of a top form defined on the
pure spinor space \cite{Nekrasov:2005wg}. By comparison we can
read off the top form itself
\begin{equation}\label{topp}
  \Omega = e^{-3 (\tilde \phi + \tilde \kappa)} = (\lambda^+)^{-3} \ .
\end{equation}

At this point, we can map the RNS saturation rule for amplitudes
on the sphere
 \be
 \langle c\p c\p^2 c e^{-2\phi}\rangle=1,
 \ee
to the pure spinor variables, obtaining
 \be
 \langle (\l^+)^3(\t^a)^5\rangle=1,
 \ee
which is the prescription for the saturation of the zero modes in
the Berkovits formalism \cite{Berkovits:2000fe}. Note that the
third power of the pure spinor is consistent with the expression
of the holomorphic top form (\ref{topp}) we just reconstructed.

The next step in performing the map is that we have to
covariantize this superstring by adding the missing coordinates
and momenta. So, following \cite{Berkovits:2001us}, we add a BRST
quartet consisting of ten $(1, 0)$ $b c$-systems $(p_{ab},
\theta^{ab})$ and ten $(1, 0)$ $\beta \gamma$-systems $(w_{ab},
\lambda^{ab})$. They have opposite central charges, so the total
central charge remains unchanged. In this way we recover the full
pure spinor stress tensor
 \be
 T=-\half\p x^m\p
 x_m-d_\a\p\t^\a+w_\a\p\l^\a-\half\p^2\log\Omega(\l).
 \ee
The BRST charge has to be modified to assure that these extra
degrees of freedom are not included in physical states. Since we
will be concerned with non-critical supertrings in this paper, we
leave for a future work the study of how the BRST cohomology of
the RNS superstring is mapped to the pure spinor cohomology.

Let us mention that the study of the global properties of the pure
spinor space $\M$ is crucial in order to obtain the correct
cohomology of the superstring. The curved space $\M$ can be
covered by sixteen patches, one for each component of the pure
spinor $\l^\a$ that we can take to be nonzero. On a single patch,
the pure spinor action reduces to the sum of eleven free
$\beta\gamma$--systems. If we want to recover the spectrum of the
superstring from the cohomology of the free $\beta\gamma$--system,
we need to add to the Berkovits operator (\ref{10ld}) the Cech
operator $\d_{Cech}$ on the pure spinor space $\M$, such that the
total BRST charge of the pure spinor $\beta\gamma$--system is
actually
 \be
 Q_{BRST}=\oint \l^\a d_\a-\d_{Cech},
 \ee
which computes the Cech cohomology on $\M$ with values in the BRST
cohomology of (\ref{10ld}). This will be discussed in a separate
work.


\section{RNS non-critical superstrings}
\label{rnssection}

In this section we will consider the RNS description of
superstrings propagating in the $d+2$ dimensional background
\cite{Kutasov:1990ua,Kutasov:1991pv}
 \be
\RR^{1,d-1}\times \RR_\vp \times U(1)_x\, ,
 \label{backgr}\ee
with flat metric in the string frame and a linear dilaton
 $$
 \Phi={Q\over2}\vp \ .
 $$
The effective string coupling $g_s=e^\Phi$ varies as we move along
the $\vp$ direction and when considering scattering processes one
needs to properly regularize the region in which the coupling
diverges. We will only consider the weak coupling region
$\vp=-\infty$, where perturbative string computations are valid
and we can safely analyze the string spectrum.

In the following we will focus on the holomorphic sector of the
closed superstring. The $d+2$ dimensional RNS superstring is
described in the superconformal gauge by $2n+1$ superfields
$X^\mu$, with $\mu=1,\dots,d=2n$, and $X$ and by a Liouville
superfield $\Phi_l$. In components we have $X^\mu=(x^{\mu},
\psi^{\mu})$, $X=(x,\psi_x)$ and $\Phi_l=(\varphi, \psi_l)$, where
the $\psi$'s are Majorana-Weyl fermions.

The $d=2n$ coordinates $x^\mu$ parameterize the even dimensional
flat Minkowski part of the space, while the coordinate $x$ is
compactified on a circle of radius $R=2/Q$, whose precise value is
dictated by the requirement of space-time supersymmetry, as we
will see below. The coordinate $\vp$ parameterizes the linear
dilaton direction with a background charge $Q$. As usual, we need
to add the superdiffeomorphisms ghosts $(\beta, \gamma)$ and
$(b,c)$. The central charge of the system is
 $$
 c=(3/2)(2n+1)_{\{X^\mu,X\}}+(3/2+3Q^2)_{\{\Phi_l\}}+(11)_{\{\b\gamma\}}-(26)_{\{bc\}}
 $$
and the requirement that it vanishes fixes the slope of the
dilaton to $Q(n) = \sqrt{4-n}$. For $n=4$, the background charge
vanishes and we have eight flat coordinates plus $\vp$ and $x$,
getting back to the flat ten-dimensional critical superstring.
When $n\neq 4$ we have non-critical superstrings.

\subsection{Space-time supersymmetry}

In even dimensions $d=2n$, at the particular value of the radius
$R=2/Q$ the worldsheet theory has a global $N=2$ superconformal
symmetry. Before showing its generators, let us define $\Psi =
\psi_l + i \psi_x$ and $\Psi^{I} = \psi^I + i \psi^{I+n}$ (with $I
= 1, \dots, n$) and bosonize them in the usual way by introducing
the bosonic fields $H$, $H^I$ and setting
\begin{eqnarray}
  \Psi \Psi^\dagger & = & 2 i \p H \ , \quad \Psi^I \Psi^{I\,\dagger} = 2 i \p H^I
  \ ,
\end{eqnarray}
where $\dagger$ denotes Hermitian conjugation in field space and
the $H$'s have canonical OPE's $H^I(z) H^J(w) \sim - \delta^{I J}
\log(z-w)$. In this way we can define the spin fields
$\Sigma^{\pm} = e^{{\pm } {i \over 2} H}$ in the $(x,\vp)$
direction and the spin fields $\Sigma^{a} = e^{\pm {i \over
2}H^{1}  \dots \pm {i \over 2} H^{n}}$, where the index $a$ runs
over the independent spinor representation of $SO(2n)$. We list
below the matter part of the $\N=2$ superconformal generators,
their ghost part is collected in appendix A.

The matter stress tensor is
 \bea T_{m} &=&-\half
\sum^{2n}_{\mu=1} (\p x^{\mu})^{2} - \half\sum_{I=1}^n(\p H^I)^2
-\half(\p x)^2  - \half (\p \vp )^{2} + {Q\over 2} \p^{2}
\vp - \half(\p H)^2 ,\nn\\
 \eea
the two supercurrents are
\begin{eqnarray}
  G^+ & = & i \sum_{I=1}^n e^{-i H^I} \p (x^I + i x^{I+n}) + i e^{-i
  H} \p (\varphi + i x + i Q H) \ ,\label{gplus} \\
  G^- & = & i \sum_{I=1}^n e^{i H^I} \p (x^I - i x^{I+n}) + i e^{i H}
  \p (\varphi - i x - i Q H) \ ,\label{gminus}
\end{eqnarray}
and the $U(1)$ current is
\begin{equation}
  J  = - i \sum_{I=1}^n \p H^I - i \p H + i Q \p x \ .
\end{equation}

The worldsheet $\N=2$ superconformal symmetry gives rise as usual
to space-time supersymmetry. Since it is present only at $R=2/Q$,
we call it the supersymmetric radius.\footnote{The stability of
the linear dilaton background away from the supersymmetric radius
has been recently discussed in \cite{Itzhaki:2005zr}.} In other
words, the radius changing operator is not an $N=2$ primary. For
the $(2n+2)$-dimensional superstrings we can construct $2^{n+2}$
candidates for space-time supercurrents in the $-\half$ picture
 \be
 q \sim e^{-{\phi \over 2} + {i \over
2}\left(\pm H \pm H^1 \pm ...\pm H^n \pm Q x \right)} \ ,
 \ee
with the usual bosonization of the superghosts $\beta=\p\xi
e^{-\phi}$ and $\gamma=e^\phi\eta$.

Only $2^n$ of them are mutually local and BRST invariant.
Combining the left and right sectors, we can realize a space-time
supersymmetry algebra with $2^{n+1}$ real supercharges that close
on the $SO(d)$ translation along the flat $\RR^{d}$ part of the
space-time ($d=2n$). In the case of even $n$, the supercharges are
in different $SO(d)$ spinor representations, while if $n$ is odd
they are in the same spinor representation. The circle on which
$x$ is compactified is related to the R-symmetry: the momentum
along the circle corresponds to the R-charge and is measured by
the affine current
 \be J_R =  {2 i \over Q(n)} \p x
\,. \label{R}
 \ee

In the RNS formalism, space-time supersymmetry only closes up to
picture changes. We will need to consider supercharges in the
$+\half$ picture as well, so we will make use of the picture
raising operator $Z_+$
\begin{equation}
Z_+ = \{Q_B, \xi\} = 2 \del \phi b \eta e^{2 \phi} + e^\phi (G^+ +
G^-) + 2 b \del \eta e^{2 \phi} + \del b \eta e^{2 \phi} + c \del
\xi \ .
\end{equation}
We will return later to the supersymmetry algebra in the various
dimensions and show explicitely its current algebra case by case.

\subsection{Spectrum}

In this section, we will collect some useful results about the
spectrum in various dimensions \cite{Murthy:2003es}, that we will
compare to the pure spinor covariant cohomology computation.

Consider first various  general features of the superstring, which
are valid in all non-critical dimensions on the backgrounds
(\ref{backgr}). The RNS computation of the spectrum follows an
indirect path, since an explicit BRST analysis has been done only
in the $d=0$ case. It can be done in three steps:

1) Identify the physical space-time supercharges and GSO project
\noindent the vertex operators.

2) Impose the on-shell condition $\Delta=1$ and the Dirac
equation.

3) Impose the level matching conditions on the operators with the
same momentum in the noncompact Liouville direction $\vp$.

4) Require that the vertex operators are all mutually local with
respect to each other.

The bosonic part of the lowest level spectrum is what, in a
familiar ten dimensional language, we would have called the
graviton, the dilaton and the $B$-field plus the appropriate odd
dimensional RR field strengths\footnote{In $d+2$ dimension, if
there is a $F^{(d+2/2)}_+$ form it is also self dual.} and a new
character, the tachyon, which in the non-critical case is non
tachyonic and survives the GSO projection. However, because the
theory is compactified on a small circle, the analysis of the
spectrum from the $d+2$ dimensional point of view is misleading
and we should think of them as on-shell  string modes in $d+1$
dimensions with certain winding and momentum around the circle. In
this picture, all the $(d+1)$--dimensional modes are massive and
the lowest lying state is always the tachyon (in $d+1=5$
dimensions the tachyon is massless). On the other hand, the
spectrum must arrange itself into representations of the
space-time symmetries, namely the super--Poincar\'{e} group acting
on the flat $\RR^{1,d-1}$ part of the space-time. We have to fit
the $(d+1)$--dimensional modes into supersymmetry multiplets of
$d$ dimensions. There are two crucial features here:

 {\it i)} From the $(d+1)$--dimensional point of view the string
modes are on-shell: the mass in their dispersion relation is
fixed, because we are reducing on the compact direction
$x$.\footnote{The dispersion relation comes from the condition
$\Delta=1$ for on-shell vertex operators.} When we further reduce
these modes along the Liouville direction down to $d$ dimensions,
however, they arrange themselves into {\it off-shell} $d$
dimensional supermultiplets. Because the momentum in the
noncompact Liouville direction $\vp$ is continuous, in fact, the
mass that appears in the dispersion relation for the
$d$--dimensional momentum $k_\mu$ is continuous, above a certain
mass gap.

 {\it ii)} The different winding and momentum modes of the same
parent $(d+2)$--dimensional RR field strength fit into {\it
different} $d$--dimensional supermultiplets.

There are two kinds of vertex operators in the theory: the
normalizable modes are particles propagating in the bulk of the
linear dilaton and they corresponds to states in the holographic
dual theory (in the AdS/CFT sense); the non--normalizable vertex
operators instead have a wavefunction exponentially supported in
the weak coupling region and they correspond to operators in the
dual theory. We will be interested in computing the deformations
of the worldsheet lagrangian, corresponding to variations of the
string background, so we will focus just on the non--normalizable
operators. In particular, the ones which are chiral primary fields
of the $N=2$ worldsheet superalgebra can be added to the action
without breaking the $N=2$ superconformal symmetry itself. This
means that they correspond in the dual theory to observables that
preserve space-time supersymmetry and they necessarily fit into
off-shell short representations of the space-time supersymmetry.

Let us now summarize the symmetries and the spectrum of the
non-critical superstring (to be concrete we will only consider
type IIB) on the background $\mathbb{R}^{d-1,1} \times
\mathbb{R}_\varphi \times U(1)_x$ in the various dimensions. We
always find two kinds of {\it off-shell} supermultiplets, a gauge
multiplet, to which the tachyon belongs, and a supergravity
multiplet. At the end, we will discuss how the holographic picture
of the background is realized on this spectrum.

\vskip 0.1cm
\medskip {\it d=0}
\vskip 0.1cm

This is the so called two-dimensional non-critical superstring. In
this case there is no Lorentz symmetry, but only $\N=2$
supersymmetry in zero dimensions with a $U(1)$ R-symmetry. There
is no supergravity sector in this case, the graviton in fact is
just a discrete state. We only have the gauge multiplet,
containing the RR scalar potential $C$, the tachyon $T$ and two
real fermions,\footnote{From the two dimensional point of view the
tachyon only gives rise to winding modes and usually it is not
included in the propagating spectrum of the two dimensional
non-critical string. But here we are classifying the spectrum
according to zero dimensional supersymmetry, so we include it in
the supermultiplet. Moreover, the RR potential is a two
dimensional chiral boson and the two fermions are chiral.} for a
total of $2\oplus2$ degrees of freedom.

\vskip 0.1cm
\medskip {\it d=2}
\vskip 0.1cm

We have $SO(1,1)$ Lorentz symmetry acting on the flat $\RR^{1,1}$
and in the type IIB case we find $\N=(4,0)$ supersymmetry with a
$U(1)\times \ZZ_2$ R-symmetry. We still have two supermultiplets.
The gauge supermultiplet, containing the tachyon, has $4\oplus4$
off-shell degrees of freedom. Then we have the two-dimensional
supergravity multiplet, containing $8\oplus8$ degrees of freedom.

\vskip 0.1cm
\medskip {\it d=4}
\vskip 0.1cm

In this case we have $\N=2$ super--Poincar\'{e} $SO(1,3)$ symmetry
acting on the flat $\RR^{1,3}$ part of the space-time. In addition
we have also a $U(1)$ R-symmetry. The spectrum contains an $\N=2$
off-shell gauge multiplet, which contains the tachyon and has
$8\oplus8$ degrees of freedom. Then we have the $\N=2$
supergravity multiplet with
$32\oplus32$ states.\\

We collect the various spectra in the following table:
 \be\begin{array}{cccc}
 \label{rehearsal}
                &\hspace{0.5cm} d=0       &\hspace{0.5cm}  d=2 &\hspace{0.5cm}  d=4 \\
 \textrm{gauge:}  & \hspace{0.5cm} 2\oplus2&\hspace{0.5cm}  4\oplus4 &\hspace{0.5cm}  8\oplus8  \\
 \textrm{supergravity:} & \hspace{0.5cm}  -       &\hspace{0.5cm}  8\oplus8 &\hspace{0.5cm}  32\oplus32
\end{array}
 \ee

Consider the holographic interpretation in the $d=4$ case
\cite{Giveon:1999zm}. The Liouville direction $\vp$ is the
holographic direction. At the weak coupling end of the space
$\vp=-\infty$ a four dimensional non-gravitational theory lives,
which is non local and is called Little String Theory. Its low
energy limit is a particular $\N=2$ supersymmetric gauge theory,
the $SU(2)$ Seiberg--Witten theory at the singular point in the
moduli space of vacua where a monopole becomes massless. This
system can also be realized as type IIB string theory on the
conifold at vanishing string coupling, in which case we know that
this is the dual gauge theory.

At the monopole point, the gauge theory is abelian and its field
content consists of an $\N=2$ gauge multiplet and a massless
hypermultiplet. The latter contains the massless monopole (which
can be realized as a D3-brane wrapping the vanishing 3-cycles of
the conifold). In the dual perturbative string spectrum we can
only see the off-shell vector  multiplet, which corresponds
precisely to the supermultiplet with $8\oplus8$ degrees of freedom
to which the tachyon belongs.

A similar picture exist in the lower dimensions, corresponding to
type IIB superstring on a higher dimensional conifolds at
vanishing string coupling. These give rise to lower dimensional
Little String Theories. The vector multiplet in their low energy
spectrum corresponds to the gauge multiplet (in which the
``non-tachyonic" tachyon sits) of the superstring in the linear
dilaton background.


\section{Two-dimensional superstrings}
\label{2dsection}

In this section we will construct the pure spinor superstring in
the linear dilaton background
 $$
\mathbb{R}_\vp\times U(1)_x \ .
 $$
This theory has two bosonic dimensions, the dilaton direction
$\vp$ and the $x$ direction. The latter is compactified at the
supersymmetric radius $R=2/Q$, where $Q=2$ is the Liouville
background charge. Since there are no transverse directions
($d=0$), this two dimensional case is the easiest case for
explaining the construction of the corresponding pure spinor
theory.

The strategy for the construction of the pure spinor non-critical
superstring is that we first map the RNS worldsheet variables to a
patch of the pure spinor space and then we reconstruct the
covariant formulation. As we will see, an important feature of the model
is that it is naturally embedded in a larger superspace, that we
will eventually reduce to the physical one. This will be a feature
that allows to generalize the non-critical pure spinor action to
other backgrounds with a larger amount of supersymmetries.

We first recall some facts about the RNS computation of the
spectrum and supersymmetry algebra
\cite{Itoh:1991qb,Seiberg:2005bx,Ita:2005ne}. Then we will
introduce the map to the pure spinor variables and reconstruct the
full covariant theory. We will compute the cohomology and show
that it agrees to the RNS spectrum. Eventually, we will present
some speculations about the possible generalization to other
non-critical two dimensional backgrounds.

\subsection{The RNS superstring}

The supersymmetry structure of the model is that of a
zero-dimensional space-time. Consider the holomorphic part. The
addition of the antiholomorphic sector in order to get the closed
superstring is straightforward.

The model has one real supercharge, and we can choose the BRST
invariant supercurrent $q_+(z)$ in the $-\frac{1}{2}$ picture
\begin{equation}
  q_+ = e^{-\frac{1}{2} \phi + \frac{i}{2} H - i x} \ .
\end{equation}
The corresponding supercharge $Q$ is given by
\begin{equation}
  Q_+ = \oint e^{-\frac{1}{2} \phi + \frac{i}{2} H - i x} \ ,
\end{equation}
and is nilpotent $Q_+^2=0$, as we expect from the fact that we
have no transverse space and hence zero dimensional
supersymmetry.\footnote{This theory is realized as type II
superstring on a Calabi--Yau fivefold in the limit of zero string
coupling.} We have another physical supercurrent
$q_-=e^{-\half\phi-{i\over2}H+ix}$ which is nonlocal with respect
to $q_+$. The choice of the $q_+,\bar q_+$ gives type IIB, while
the pairing $q_+,\bar q_-$ gives type IIA. We will only consider
the former case.

A basic feature of the map from the RNS to the pure spinor
variables is that it requires doubling the superspace. In
\cite{Grassi:2005kc} it has been noted the existence of another
supercurrent $q_\pd (z)$ of the form
\begin{equation}
q_\pd = e^{-\phi / 2 + i H / 2 + i x} \ . \label{tq}
\end{equation}
The supercurrents $q_+$ and  $q_\pd$ are mutually local and the
latter is a conserved current as well, so the correponding charge
is conserved. However, while $q_+$ is BRST invariant, $q_\pd$ is
not. Indeed, the model has only one physical real supersymmetry.
The supercharges satisfy \be \{Q_+, Q_\pd \} = \oint e^{- \phi + i
H} \ . \label{qq} \ee

However, (\ref{qq}) is not a supersymmetry algebra. Recalling that
in the RNS formalism supersymmetry only closes up to picture
changing,  we can picture raise the physical supercurrent $q_+$ to
the $\frac{1}{2}$ picture
\begin{equation}
  q_+^{(+1/2)} = b \eta e^{3 \phi / 2 + i H / 2 - i x} + i \del
  (\varphi + i x + i Q H) e^{\phi / 2 -i H / 2 - i x}  \ ,
\end{equation}
and compute the OPE between the physical  supercurrent
$q_+^{(+1/2)}$ and the unphysical one in the $-\frac{1}{2}$
picture
\begin{equation}
\label{alge2}
  q^{(+1/2)}_+ (z) q^{(-1/2)}_\pd (0) \sim \frac{Q}{z^2} +
  \frac{i}{z} \del (\varphi + i x + i Q H) (0) \ ,
\end{equation}
which is indeed not a supersymmetry current algebra, due to the
presence of the double pole. This anomalous term is proportional
to the Liouville background charge $Q=2$.

As we will see, working in a doubled superspace in the pure spinor
variables will require an appropriate projection to the physical
superspace. However, the doubled superspace will allow to study
pure spinor superstrings in two-dimensional backgrounds with four
supercharges, that is twice as many supersymmetries as in the
linear dilaton background.

\subsection{Multiplet spectra: RNS analysis}
\label{multitwo}

In the following we will construct the multiplet spectra of the
RNS superstring on the $\mathbb{R}_\varphi \times U(1)_x$
background. We first consider the holomorphic sector as a building
block for the closed superstring multiplet.

As we anticipated in Section~\ref{rnssection}, we will consider
non--normalizable vertex operators only and we normalize their
Liouville dependent part as \be V\sim e^{\beta\varphi}=
e^{{Q\over2}\varphi}e^{-E\varphi} \equiv g_S(\varphi)e^{-E\varphi}
\ , \ee such that the corresponding wavefunction $\Psi(E)\sim
e^{-E\vp}$ is localized in the weak coupling region $g_s(\varphi)
\rightarrow 0$. These operators satisfy the Seiberg bound
\cite{Seiberg:1990eb}
 \be \textrm{Re}\,\beta
\leq \frac{Q}{2},
 \ee
There are also operators with complex $\beta=Q/2+ik_\vp$ whose
imaginary part is the momentum of the wavefunction of a particle
moving in the $\varphi$ direction. They correspond to propagating
particles that can be scattered.

\subsubsection{Holomorphic sector}

We begin with the NS sector.

\vskip 0.3cm {\it NS sector} \vskip 0.3cm

The tachyon vertex operator in the $-1$ picture is
\begin{equation}
  T = e^{-\phi + i p x + \beta \varphi} \ .
\end{equation}
The condition for $T$ to have weight $\Delta(T)=1$ is
 \be p^2 -
\beta (\beta - Q) = 1 \ ,
 \ee
where $Q=2$. The requirement for mutual locality of $T$ with the
supercurrent $q_+$ (GSO projection) reads $p Q \in 2 \mathbb{Z}
+1$. Note that this condition also implies that $T$ is mutualy
local with $q_\pd$. The lowest lying states have $p=\pm{1\over
Q}$. The operator
 \be
 T_\pm=e^{-\phi+{1\over Q}(\varphi\pm ix)},
 \ee
is a worldsheet (anti)chiral primary (annihilated by $G^\pm$) of
$\Delta_{matter}(T_\pm) = \pm\frac{p}{2} = \frac{1}{2}$, with
space-time R-charge $R=\pm\frac{1}{2}$. Both are non-normalizable
operators as their Liouville momentum satisfies $\beta <
\frac{Q}{2}$. Note that the OPE of $T^+$ and $T^-$ has a branch
cut, so they are not both mutually local.

The NS sector may also contain a vector of the form
\begin{equation}
  V_\pm = e^{-\phi + i \epsilon H + i p x + \beta \varphi} \ ,\quad
  \epsilon = \pm 1 \ .
\end{equation}
The weight requirement is $p^2 = \beta (\beta - Q)$ and the mutual
locality condition  with the supercharge requires $p Q \in 2
\mathbb{Z}$. In order to be a primary of the worldsheet $N=1$ SCA
(i.e., having no double poles with the supercurrent $G_m$) we need
that $\beta = \epsilon p + Q$, which together with the weight one
requirement imply that $p=0$ and $\beta=Q$. This violates the
Seiberg bound and the operator is normalizable, so it does not fit
into space-time short supermultiplets.

\vskip 0.3cm {\it Ramond sector} \vskip 0.3cm

The Ramond sector operator in the $-\frac{1}{2}$ picture reads
\begin{equation}
  R_\pm = e^{-\phi / 2+ i \epsilon H / 2 + i p x + \beta x} \ .
\end{equation}
The weight one condition is $p^2 - \beta (\beta - Q) = 1$, and the
mutual locality with the supercharge implies that $2 p Q \in 4
\mathbb{Z} + \epsilon -1$. The Dirac condition
 \be \label{diracco}\oint \gamma
(G^++G^-)R=0 \ .
 \ee
is $\beta = \epsilon p + \frac{Q}{2}$, which satisfies the weight
one condition identically.

Consider the lowest space-time R-charge states. At $\epsilon=+1$,
the vertex operator with $p=0$ saturates the bound
$\beta=\frac{Q}{2}$. When $p=-1$ we get
\begin{equation}
  R_+ = e^{-\phi / 2 + i H /2 - i x} \ ,
\end{equation}
which is the supercharge $q_+$.

At $\epsilon=-1$  and $p=\frac{1}{2}$ we find the supersymmetric
partner $R_-$ of the tachyon $T_-$
\begin{equation}
  R_- = e^{-\phi / 2 - i H / 2 + (\varphi + i x) / Q} \ .
\end{equation}
$T_-$  and $R_-$ are of course mutually local.

\subsubsection{Closed superstring}

We consider the Type IIB theory with the supercharges $Q_+$ and
$\bar Q_+$ from  the holomorphic and anti-holomorphic sectors,
respectively. In order to construct the $\N=2$ $d=0$
supersymmetric multiplet we use the results of the holomorphic
sector from above.

The bosons of the closed string multiplet are the NS--NS operator
$T^- \bar T^-$ and the R--R operator $R_- \bar R_-$. The fermions
are the R--NS and NS--R operators $R_- \bar T_-$ and $T_- \bar
R_-$. They are arranged in the supermultiplet
\begin{displaymath}
  \begin{array}{ccccc}
    && T_- \bar R_- && \\
    & \nearrow_{Q_+} && \searrow_{\bar Q_+} & \\
    R_- \bar R_- & & & &  T_- \bar T_- \\
    & \searrow_{\bar Q_+} & & \nearrow_{Q_+} & \\
    & & R_- \bar T_- & &
  \end{array}
\end{displaymath}
for a total of $2\oplus2$ degrees of freedom.

\subsection{Pure spinor variables}
\label{2dpure}

The RNS superstring has four bosonic fields: $x$, $\varphi$,
$\beta$ and $\gamma$, and four fermionic fields: $\psi_x$,
$\psi_l$, $b$ and $c$. In the following we will map them to four
bosonic and four fermionic fields, which will be the pure spinor
superstring variables.

The fermionic variables are the dimension zero fermionic
coordinates of the doubled superspace $(\theta^+,\theta^\pd)$ and
their dimension one conjugate momenta $(p_+,p_\pd)$
\begin{eqnarray}
p_+ & =  b \eta e^{\frac{3}{2} \phi + \frac{i}{2} H - i x} \
,\qquad
\theta^+ & =  c \xi e^{-\frac{3}{2} \phi - \frac{i}{2} H + i x} \ ,\nn\\
p_\pd & =  e^{-\frac{1}{2} \phi + \frac{i}{2} H + i x} \ ,\qquad
\theta^\pd & =  e^{\frac{1}{2} \phi - \frac{i}{2} H - i x} \ .
\end{eqnarray}
Note that $(\theta^+,p_+)$ are BRST invariant while
$(\theta^\pd,p_\pd)$ are not BRST invariant.

These fermionic variables have the free field OPE's
\begin{equation}
p_+(z) \theta^+(0) \sim \frac{1}{z} \ , \quad p_\pd(z)
\theta^\pd(0) \sim \frac{1}{z} \ ,
\end{equation}
and all the other OPE's vanish.

Consider next the bosonic variables. We construct a map analogous
to the one we used for the ten dimensional critical superstring in
Section~\ref{10mapsection}
\begin{equation}
  \eta = e^{\tilde \phi + \tilde \kappa} p_+\ , \quad
  b = e^{(\tilde \phi - \tilde \kappa) / 2} p_+ \ ,
\end{equation}
$b$ and $\eta$ are the RNS fields and we introduced two new
variables $\tilde\phi$ and $\tilde\kappa$, which we will relate to
the pure spinor variables.

Their  OPE's are
\begin{eqnarray}
\tilde \phi(z) \tilde \phi (0) & \sim & - \log z \ ,\nn\\ \tilde
\kappa (z) \tilde \kappa (0) & \sim & \log z \ ,\label{tildeope}
\end{eqnarray}
and $\tilde \phi(z) \tilde \kappa (0) \sim  0$. It can be easily
verified that the OPE's of $\tilde \phi$ and $\tilde \kappa $ with
either the fermionic momenta or the fermionic coordinates are all
non-singular.

We can express  $\tilde\phi$ and $\tilde\kappa$ via the RNS
variables as
\begin{eqnarray}
  \tilde \phi & = & - \frac{9}{4} \phi - \frac{3}{4} i H + \frac{3}{2}
  i x - \kappa + \frac{1}{2} \chi \ ,\\
  \tilde \kappa & = & \frac{3}{4} \phi + \frac{i}{4} H - \frac{i}{2} x
  + \kappa + \frac{1}{2} \chi \ .
\end{eqnarray}

However, the OPE's of $x$ with the fermionic momenta and fermionic
coordinates as well as with $\tilde \phi$ and $\tilde \kappa$ are
singular. In order to fix this, we shift the $x$ coordinate and
define
\begin{equation}
x' = x + i \phi - H \ ,
\end{equation}
whose OPE is
\begin{equation}
x'(z) x'(0) \sim -\log z \ .
\end{equation}

Using these new variables the RNS energy-momentum tensor
$T=T_\mathrm{m}+T_\mathrm{gh}$ is mapped to
\begin{eqnarray}
  T & = & - \frac{1}{2} (\del \varphi)^2 - \frac{1}{2} (\del x')^2 +
  \del^2 (\varphi - i x') - p_+ \del \theta^+ - p_\pd \del
 \theta^\pd - \nonumber\\
  && {} - \frac{1}{2} (\del \tilde \phi)^2 + \frac{1}{2} (\del \tilde
  \kappa)^2 + \del^2 (\tilde \phi + \tilde \kappa) \ .
  \label{eq:2dRNSmapped-energy-momentum}
\end{eqnarray}

The pure spinor variable $\lambda^+$ and its conjugate momentum
$w_+$ are recovered in terms of $\tilde \phi$ and $\tilde \kappa$
as an ordinary beta gamma system
\begin{equation}
\lambda^+ = e^{\tilde \phi + \tilde \kappa} \ , \quad w_+= \del
\tilde \kappa e^{- \tilde \phi - \tilde \kappa} \ .\label{purespi}
\end{equation}
Their OPE reads
\begin{equation}
\quad w_+(z)\lambda^+(0) \sim \frac{1}{z} \ . \label{para}
\end{equation}

The pure spinor variable  $\lambda^+$ parameterizes the patch
  $(\lambda^+ \neq 0, \lambda^\pd =  0)$
of the complex dimension one pure spinor space
 \be \lambda^+
 \label{2dps}
\lambda^\pd = 0 \ .
 \ee
It will be a generic feature in all dimensions that the map takes
the RNS variables to the pure spinor variables on a patch of the
pure spinor space.

The total central charge of the theory still vanishes
\begin{displaymath}
c = (1 - 12)_{\{x'\}} + (1 + 12)_{\{\varphi\}} + (-2)_{\{p_+
\theta^+\}} + (-2)_{\{p_\pd \theta^\pd\}} + (2)_{\{\lambda^+w_+\}}
= 0.
\end{displaymath}

The pure spinor variables form a curved beta-gamma system. The
pure spinor energy-momentum tensor, that we get from $T_{\tf,\tk}$
on the patch is not simply $T_{(\lambda^+,w_+)} = w_+ \del
\lambda^+$. Rather, it has an additional term as in the critical
superstring case
 \be
 \label{twostress}
T_{(\lambda^+,w_+)} = w_+ \del \lambda^+ - \frac{1}{2} \del^2 \log
\Omega \ ,
 \ee
where $\Omega = e^{ - 3 (\tilde \phi + \tilde \kappa)} =
(\lambda^+)^{-3}$. This arises from the top form on pure spinor
space, which reads on the patch
 $(\lambda^+ \neq 0, \lambda^\pd =  0)$
 \be
 \label{omegatwo} \Omega(\lambda^+) = \frac{d \lambda^+}{(\lambda^+)^3} \ .
  \ee

Mapping the RNS saturation rule on the sphere
 to the pure spinor variables one gets
a requirement for an insertion of $(\lambda^+)^3$, which is
consistent with the measure corresponding to the top form we
obtained. This ghost number three insertion, required for a
nonvanishing amplitude, will be the same in all different
non-critical dimension and actually coincides with the ten
dimensional saturation rule, we derived in
Section~\ref{10mapsection}.

We can write the pure spinor stress tensor in a covariant way by
\begin{eqnarray}
  T & = & - \frac{1}{2} (\del \varphi)^2 - \frac{1}{2} (\del x)^2 +
  \del^2 (\varphi - i x) - p_{I}\del \theta^{I}
  + \nonumber\\
  && {} + w_{I}\del \lambda^{I} - \frac{1}{2} \del^2 \log \Omega(\lambda) \
  ,\label{purestress2}
\end{eqnarray}
where $I = +, \pd$, and we renamed $x'$ as $x$ for the simplicity
of notation.

The matter part of this stress tensor can be derived from the pure
spinor action
 \be
S = {1\over2\pi\a'}\int d^2z \left(\frac{1}{2}\p x \bar\p x +
\frac{1}{2}\p \varphi \bar\p \varphi + p_I\bar\p\t^I \right)
-{Q\over2}\int d^2z r^{(2)}(\vp-ix) \ ,\label{flat2d}
 \ee
where the last term is the Fradkin--Tseytlin term that couples the
space-time linear dilaton to the worldsheet curvature. For the
consistency of the FT term we have to compactify the $x$ direction
on a circle of radius $R=2/Q$, which in fact is the supersymmetric
radius we already know from RNS analysis. Note that the
hermiticity property of the action and the stress tensor implies
that the hermiticity properties of the variables is not the naive
one \cite{Berkovits:1996bf}.

\subsubsection{Supersymmetry structure}

An important ingredient in the construction of the pure spinor
non-critical superstring is the supersymmetry algebra and the
superspace structure. Let us construct the pure spinor
superstring. Recall that the map from the RNS variables to the
pure spinor variables imposed on us the introduction of an
additional fermionic coordinate and its conjugate momentum
$(\theta^\pd, p_\pd)$, which are not physical. The RNS OPE between
the supercharges in (\ref{alge2}) is mapped on the pure spinor
side to \be
 q_+(z)q_\pd(0)\sim{Q\over z^2}+{\p(\vp-ix)(0)\over z} \ ,
\ee where we denoted $x'$ by $x$ for simplicity of notation. The
corresponding algebra of the superderivatives has the opposite
sign as usual \be\label{dope}
 d_+(z)d_\pd(0)\sim-{Q\over z^2}-{\p(\vp-ix)(0)\over z} \ .
\ee

We introduce GS-like  constraints that reproduce this algebra
 \bea
 d_+=&p_+-\half\t^\pd\p(\varphi-ix)+\half Q\p\t^\pd,\nn\\
  d_\pd=&p_\pd -\half\t^+\p(\varphi-ix)-\half Q\p\t^+,
 \label{di2}
 \eea
where the crucial difference with respect to the flat background
is the last term, proportional to the background charge $Q$. This
term is responsible for the double pole in the algebra, which
signals the breaking  of the two-dimensional flat supersymmetry
algebra to the physical zero-dimensional supersymmetry algebra of
the linear dilaton background. We introduce the compact notation
 \be
 \label{dzero}
 d_I=p_I-\half\tau_{IJ}\t^J\p(\vp-ix)+{Q\over2}\e_{IJ}\p\t^J,
 \ee
where $I=+,\dot+$. The two dimensional matrices
$\tau_{IJ}=\s^1_{IJ}$ (the Pauli matrix) and $\e_{IJ}$, the usual
antisymmetric tensor normalized as $\e_{+\dot+}=1$, act in the
following on the index $I$, by which we denote the physical
supercoordinates, $I=+$, and the unphysical ones, $I=\dot+$.

\subsubsection{Cohomology}
\label{2Dcohom}

An essential ingredient of the pure spinor formulation is the BRST
operator, that in the two dimensional background is
 \be Q_B = \oint
\lambda^{I}d_{I} \ , \label{QB}
 \ee
where $I=+,\pd$. In the critical superstring in flat ten
dimensions, all the superderivatives in the BRST charge
$Q=\oint\l^\a d_\a$ are physical. The first issue that needs to be
dealt with in the linear dilaton background (and essentially in
any case with reduced supersymmetry) is whether to include in
(\ref{QB}) only the physical superderivatives or also the
unphysical ones. The strategy we follow is to
 include always in the BRST operator  all the
$d$'s, in our case both $d_+$ and $d_\pd$,  and compute the
cohomology thereof. In a second step, since we are interested in
the space-time supersymmetry multiplets, we will consider only the
part of the cohomology that contains the variables that realize
the supersymmetry current algebra, in this case $\t^+$. This will
reproduce the RNS computation of the short supermultiplets. Recall
that the full current algebra of the $d$'s is anomalous, i.e. it
has a double pole in the OPE between $d_+$ and $d_\pd$: the non
anomalous subalgebra is in fact the space-time supersymmetry
algebra, in our case the zero dimensional supersymmetry $d_+^2=0$.

The two dimensional pure spinor constraint (\ref{2dps}) is
sufficient to prove the nilpotency of the BRST charge $Q_B$.
Looking at the algebra (\ref{dope}) we see that $Q_B^2=0$,
provided the following conditions are satisfied
 \be\label{2dnilpo}
\l^+\l^{\dot+}=0,\qquad \p\l^+\l^{\dot+}-\l^+\p\l^{\dot+}=0.
 \ee
The first condition is the pure spinor constraint itself. The
derivative condition is a consequence of the first condition.
There are various ways to see this \cite{Grassi:2005jz}. The
simplest is to expand the condition $\l^+\l^{\dot+}=0$ in modes
 \be\label{mode2d}
 \l^+_0\l^{\dot+}_0=0,\qquad
 \l^+_0\l^{\dot+}_1+\l^+_1\l^{\dot+}_0=0, \,\,\,\,\,\,\ldots
 \ee
Then for a solution of the zero mode $\l^+_0\ne 0$, the second
equation implies $\l^+_0\l^{\dot+}_1=0$ and so on. Hence, the
derivatives separately vanish, $\l^+\p\l^{\dot+}=0$. The same
argument on the other solution $\l^{\dot+}_0\ne0$ implies that
$\p\l^+\l^{\dot+}=0$. In particular, we see that the two
derivative terms in (\ref{2dnilpo}) vanish separately and the BRST
charge is nilpotent.\footnote{Another way to prove that the
derivatives vanish separately is to use the Ward identities coming
from the Lorentz current and the ghost current OPE's
\cite{Grassi:2005jz}.}

To simplify the computation of the cohomology it is convenient to
introduce the new notation $Z=\vp+ix$ and $\bar Z=\vp-ix$ with OPE
$$
Z(z)\bar Z(0)\sim -2\log z \ ,
 $$
so that the GS-like contraints read
 $$
 d_+=p_+-\t^\pd \p \bar Z+Q\p\t^\pd,\qquad d_\pd=p_\pd \ ,
 $$
whose OPE is still (\ref{dope}). The physical states are the
vertex operators at ghost number one and weight zero, we find two
different kinds of operators. The first is analogous to the usual
ten dimensional vertex operator. In general this contains the
off-shell $d$ dimensional supergravity multiplet, even if in the
particular $d=0$ there are no such multiplets. The second type is
new and peculiar to the linear dilaton background. It contains
what we called the ``gauge multiplet" in the rehearsal of the
spectrum (\ref{rehearsal}), including the tachyon, which in fact
is the peculiarity of the linear dilaton background also in the
RNS.

Usually in the pure spinor formalism, massless states are given by
the zero weight ghost number one vertex operator. Since the only
worldsheet fields with weight zero are the zero modes of the
fields $\varphi$, $x$ and $\theta^I$, this vertex operator has to
be expressed by these alone with a single power of $\lambda^I$ so
that it be of ghost number one.  Hence the vertex operator is of
the form
\begin{equation}
\label{2dvertex}
  \mathcal{V}^{(1)} = \lambda^I A_I(\bar Z, \theta^J) \ ,
\end{equation}
where
\begin{equation}
  A_I(\bar Z, \theta^J) = B_I(\bar Z) + \theta^J C_{IJ}(\bar Z) + \theta^I \theta^J D_{IJK}(\bar Z) \ .
\end{equation}
Next we have to require that it be BRST closed when the pure
spinor constraint $\lambda^+ \lambda^{\dot +}=0$ is imposed but
not solved. This leaves us with just
\begin{eqnarray}
  A_+ & = & B_+ + \theta^{\dot +} C_{+ \dot +} - 2 i \theta^+
  \theta^{\dot +} \partial_Z B_+ \ ,\\
  A_{\dot +} & = & B_{\dot +} + \theta^+ C_{\dot + +} \ .
\end{eqnarray}
These superfields still have some gauge freedom given by zero
weight ghost number one $Q_B$-exact terms. Parameterizing the
general weight zero ghost number zero gauge transformation
superfield as
\begin{equation}
  \Omega = \eta + \theta^+ \xi_+ + \theta^{\dot +} \xi_{\dot +} + 2
  \theta^+ \theta^{\dot +} \Lambda_{+ \dot +} \ ,
\end{equation}
so that
\begin{eqnarray}
  \delta \mathcal{V}^{(1)} & = & Q_B \Omega^{(0)} = \lambda^+ \left( \xi_+ + 2
  \theta^{\dot +} \Lambda_{+ \dot +} + 2 i \theta^{\dot +} \partial_Z
  \eta - 2 i \theta^+ \theta^{\dot +} \partial_Z \xi_+ \right) +
  \nonumber\\
  && {} + \lambda^{\dot +} \left( \xi_{\dot +} - 2 \theta^+ \Lambda_{+
  \dot +} \right) \ ,
\end{eqnarray}
the gauge transformations of the components of the vertex operator
are given by
\begin{eqnarray}
  B_+ & \to & B_+ + \xi_+ \ , \quad
  C_{+ \dot +} \to C_{+ \dot +} + 2 \Lambda_{+ \dot +} + 2 i
  \partial_Z \eta \ ,\nonumber\\
  B_{\dot +} & \to & B_{\dot +} + \xi_{\dot +} \ , \quad
  C_{\dot+  +} \to C_{\dot+ \dot +} - 2 \Lambda_{+ \dot +} \ ,
\end{eqnarray}
so that the entire vertex operator is pure gauge and this sector
of the cohomology is trivial.

The second type of vertex operator is peculiar to the linear dilaton
background. Due to the term $\del^2 \bar Z $ appearing in the stress
tensor (\ref{purestress2}), we see that the operator $\exp{(-{Z\over
Q})}$ has weight $-1$. We can obtain again a weight zero ghost number
one operator by the following procedure.  Consider the weight one
ghost number one operator, which in the ten dimensional background
corresponds to the first massive level \cite{Berkovits:2002qx}
\be\label{massi2} \mathcal{U}^{(1)} = \del \lambda^I A_I + \lambda^I
\del \theta^J B_{IJ} + \lambda^I d_J C_I^J + \lambda^I \Pi^{\bar Z}
H_I + \lambda^+ J^+_+ F_{++}^+ + \lambda^{\dot +} J_{\dot +}^{\dot +}
F_{\dot + \dot +}^{\dot +} \ , \ee where $A_I,B_{IJ},C_{I},H_I^J,F_I$,
for $I,J=+,\pd$, are generic superfields constructed with the $\t^I$
coordinates and $J=w_I\l^I$ is the pure spinor $U(1)$ current. However
now we restrict the wavefunctions in these superfields to be
$e^{-Z/Q}$, so the total weight of the operator (\ref{massi2}) is
zero.  The gauge invariance of this vertex operator is \be \d {\cal
U}^{(1)}=Q_{B}\Lambda^{(0)}, \ee where $\Lambda^{(0)}$ is a weight
one, ghost number zero vertex operator constructed out of the \be
 \label{ground}
\Lambda^{(0)}= \p\t^I \Omega_I+\p(\vp-ix)\Gamma+p_I
\Lambda^I+J\Phi \ ,
 \ee
and again $\Omega_I, \Gamma,\Lambda,\Phi$ are generic superfields,
whose wavefunctions are chosen to be $e^{-Z/Q}$, such that the
total weight of $\Lambda^{(0)}$ vanishes. We postpone the details
about the cohomology computation to the appendix. The result is
that the only operator that survives in the cohomology is
 \be
 \label{tacchy}
 {\cal U}^{(1)}=(\l^\pd \p \t^+  )B_{\pd+},\qquad
 D_{\dot+} B_{\pd+}=0.
 \ee
The chiral superfield $B_{\pd+}=T+\t^+ R$ contains $1\oplus1$
states, $T$ being a real boson and $R$ a Majorana-Weyl fermion.
Since in the pure spinor formalism we do not have to worry about
GSO projections, the closed string spectrum is just given by the
left--right producet of the open string one
 $$
 \textrm{closed}=\textrm{open}\otimes\overline{\textrm{open}}\ ,
 $$
Therefore we find $2\oplus2$ states, reproducing the RNS result
for the ``gauge" supermultiplet.

\subsubsection{Curved non-critical backgrounds}
\label{curved2}

In this section we suggest a generalization of the linear dilaton
action to a generic two dimensional non-critical background with
at most four real supercharges.

Even if the linear dilaton background has only zero dimensional
supersymmetry, we introduce the momenta that will be useful when
generalizing the model to backgrounds with extra supersymmetry
 \be
 \Pi^\vp=\p\vp+\half\tau_{IJ}\t^I\p\t^J,\quad
 \Pi^x=\p x-{i\over 2}\tau_{IJ}\t^I\p\t^J \ .
 \ee
We have the  following OPE's \bea
 d_I(z)\Pi^\vp(0) \sim {\tau_{IJ}\p\t^J(0)\over z},\qquad
 d_I(z)\Pi^x(0) \sim -i{\tau_{IJ}\p\t^J(0)\over
 z} .
\eea The stress tensor (\ref{purestress2}) can be cast in the
following form
 \bea
 T=&-\half\Pi^m\Pi^n\eta_{mn}+{Q\over2}\e_{IJ}\p\t^I\p\t^J-d_I\p\t^I+
{Q\over2}\p^2(\Pi^\phi-i\Pi^x) \nn\\
 &+w_{I}\p\l^{I}-\half\p^2\log\Omega(\l),\label{stresscu}
 \eea
 where $I=+,\pd$ and $m=x,\vp$ and we note the presence of the extra terms proportional to
$Q=2$, which is a feature  of the linear dilaton background. We
would like to generalize the two dimensional pure spinor action
(\ref{flat2d}) to a generic curved two dimensional non-critical
background.\footnote{In the non-critical string it is not clear
whether the concept of a background makes sense. Since the
curvature is of the order of the string length, the classical
supergravity approximation is not valid in general.} Consider the
matter part of the action (\ref{flat2d}). By using the following
identity
 \be
 \half \p x^m\bar\p x_m+p_I\bar\p\t^I=\half
 \Pi^m\bar\Pi_m+{Q\over2}\e_{IJ}\p\t^I\bar\p\t^J-{1\over4}\tau_{IJ}\t^I\left(\bar\p Z\p\t^J-\p
 Z\bar\p\t^J\right)+
 d_I\bar\p\t^I,
 \ee
which in ten dimensions is usually referred to as Siegel's trick
\cite{Siegel:1985xj}, we can covariantize the matter part of the
type II action in linear dilaton background (\ref{flat2d}) in the
following way
 \bea
 S = &{1\over2\pi\a'}\int d^2z \left(\frac{1}{2}G_{MN}(Y)\p Y^M\bar\p Y^N + E_{M}^I(Y)
  d_I\bar\p Y^M +E_{M}^{\hat I}(Y)
  \bar d_{\hat I}\p Y^M \right)\nn\\
&-\int d^2z r^{(2)}\Phi(Y) \ ,\label{curve2d}
 \eea
where we introduced the curved supercoordinates
$Y^M=(x^m;\t^I,\bar\t^{\hat I})$. Note that $m$ is a curved two
dimensional vector index, while $I$ is a curved two dimensional
spinor index. The $E_M^A$ are the zweibein superfields. We are
following the notations of \cite{Berkovits:2001ue}, in which the
critical pure spinor action was studied in a generic ten
dimensional background. In the linear dilaton case, the background
superfields take the following values, the only surprise being in
the metric:

{\it i)} The zweibeins $E_M^A$ are the two dimensional flat ones.

{\it ii)} The dilaton superfield is linear
$\Phi={Q\over2}(\vp-ix)$ and the higher components of the
superfield vanish.\footnote{The top form coupling to the
worldsheet curvature (\ref{topcop}) might be considered a
component of the dilaton along the pure spinor space as well.}

{\it iii)} The metric $G_{MN}$ is constant. However in addition to
the usual terms we have in flat background, in the linear dilaton
background we also have a flat spinorial part
 \be
 G_{IJ}=Q \e_{IJ},
 \ee
which is proportional to the background charge $Q=2$ and is
responsible for the contribution $\e_{IJ}\p\t^I\p\t^J$ to the
stress tensor (\ref{stresscu}).
 We regard this as a specific feature of the linear dilaton
superspace structure in the pure spinor formalism. This explicitly
breaks the original $SO(2)$ Lorentz invariance of the action,
preserving the $U(1)_x$ R symmetry.

It is suggestive to think of (\ref{curve2d}) as the matter part of
the non-critical pure spinor action in a generic curved two
dimensional background. It would be interesting to develop further
this suggestion, in particular to work out the coupling of the
pure spinor action to the curved background, in analogy to the
critical case \cite{Berkovits:2001ue}.

\subsubsection{Anomalies}

The pure spinor space
$$
\lambda^+ \lambda^\pd = 0,
$$
is a one-complex dimensional cone ${\cal M}$ with a conical
singularity at $\lambda^{I}=0, I = +, \pd$. The pure spinor
$\lambda^{I}$ is a map from the worldsheet Riemann surface
$\Sigma$ to the pure space ${\cal M}$. A way to eliminate the
singularity is by deforming the equation to \be \lambda^+
\lambda^{\dot+}= \mu \ , \label{cone} \ee where $\mu$ is a complex
deformation parameter. The resulting (compact) space is the
2-sphere $CP^1$. However, since $c_1(CP^1) = 2$, we cannot define
the pure spinor system globally on any Riemann surface $\Sigma$
except on the 2-torus, which is unacceptable if we wish to have a
complete definition of the superstring perturbation series.

Another way to eliminate the singularity is to remove the singular
point. We get a disconnected space, which is the disjoint union of
$C^{*}$. In this way, the anomalies are avoided. As we discussed
before, a similar phenomenon occurs is the critical superstring.
The difference is that while in the critical superstring case the
removal of the origin still gives a connected space, here the
space has two disconnected components. Note that the RNS
non-critical superstring mapped to one patch of the space.


\section{Four-dimensional superstrings}

In this section we will construct the pure spinor superstring in
the four-dimensional linear dilaton background
 $$
\mathbb{R}^{1,1}\times \mathbb{R}_\vp\times U(1)_x \ .
 $$
The four-dimensional superstring has $d+1=3$ noncompact directions
$(x^1,x^2,\varphi)$ and the compact $U(1)_x$ direction $x$ with
radius $R=2/Q$, where $Q=\sqrt{3}$ is the Liouville background
charge. The strategy will be similar to the two dimensional case.

\subsection{Multiplet spectra: RNS analysis}

In the following we will compute the spectrum of the RNS
superstring in the four-dimensional linear dilaton background. We
look for the short multiplets of the space-time  supersymmetry in
the $\RR^{1,1}$ directions.  We follow closely the analysis of
section 4.2, but postpone the details of the computation to
appendix \ref{spectra4d}. Here we briefly collect the results
regarding the lowest lying operators, in particular the primaries
of the worldsheet $N=2$ superconformal algebra at zero momentum in
the transverse direction.

When $d=2n$ with odd $n$, the space-time supercharges are in the
same $SO(1,d-1)$ spinor representation. In this case the two
physical supercharges are
 \bea
 \label{susy4}
 Q_{+1}=&\oint e^{-\half\phi+{i\over2}(H+H_1)- {i\over2}Q x},\nn\\
 Q_{+2}=&\oint e^{-\half\phi-{i\over2}(H-H_1) +{i\over2}Q x} \ .
 \eea
They have the same $SO(1,1)_L$ Lorentz charge and $\mp 1$ R-charge
(\ref{R}). Their OPE is
 $$
 Q_{+1}(z)Q_{+2}(0)\sim{1\over z}e^{-\phi+iH_1}(0).
 $$
The other set of physical supercharges, which are nonlocal with
respect to the ones above, is
 \bea
 \label{susy4b}
 Q_{-1}=&\oint e^{-\half\phi+{i\over2}(H-H_1)- {i\over2}Q x},\nn\\
 Q_{-2}=&\oint e^{-\half\phi-{i\over2}(H+H_1) +{i\over2}Q x} \ .
 \eea
They have opposite $SO(1,1)$ chirality with respect to
(\ref{susy4}). In the type IIB superstring we GSO project both
holomorphic and antiholomorphic sector with the supercharges
(\ref{susy4}), in the type IIA we project the antiholomorphic
sector with (\ref{susy4b}) instead.

\subsubsection{Holomorphic sector}

\vskip 0.3cm {\it NS sector} \vskip 0.3cm

The tachyon is non-tachyonic but is massive. Its lowest lying
modes are
 \be
 \label{lowtac4}
 T_\pm=e^{-\phi+{1\over Q}(\varphi\pm ix)},
 \ee
and it is a worldsheet  (anti)chiral primary
$\Delta_{matter}(T_\pm) = \pm\frac{q}{2} = \frac{1}{2}$
annihilated by $G^\pm$, with space-time R-charge
$R=\pm\frac{2}{3}$. $T_+$ and $T_-$ are not mutually local.
However, we are interested in the mutual locality only when
matching holomorphic and antiholomorphic sectors, so we will
discuss locality only below.

The other NS operators are analogous to the ``vectors" in the ten
dimensional superstring. Their lowest lying states with
$p=\beta=0$ are
 \be
 J^\mu=e^{-\phi\pm H_1} \ ,
 \ee
where $\mu$ is an $SO(1,1)$ Lorentz vector index. They are
worldsheet $N=2$ primaries (they have only single poles with
$G^\pm$) and are not charged under $U(1)_x$.

\vskip 0.3cm {\it R sector} \vskip 0.3cm

The lowest lying components of the R vertex operators at zero
momentum $k_\mu=0$ in the transverse $\RR^{1,1}$ directions are
 \bea
 \label{lowR4}
R_{++}=&e^{-\half\phi+{i\over2}(H+H_1)-i{Q\over2}x},\qquad
&R_{--}=e^{-\half\phi-{i\over2}(H+H_1)+{i\over2Q}x+{1\over Q}\vp}\nn \ ,\\
R_{+-}=&e^{-\half\phi+{i\over2}(H-H_1)-{i\over 2Q}x+{1\over
Q}\vp},\qquad &R_{-+}=e^{-\half\phi-{i\over2}(H-H_1)+i{Q\over2}x}
\ .\nn\\
 \eea
In the following table we write the R charge and supersymmetry
transformations of these vertex operators
 $$
 \begin{array}{cccc}
  & U(1)_R & \d_{Q_{+1}} & \d_{Q_{+2}} \\
 R_{++}       & -1      & 0  & J^\mu \\
 R_{-+}       & 1      & J^\mu  & 0 \\
 R_{--}       & +1/3   & T_-  & 0\\
 R_{+-}       & -1/3      & 0 & T_+
 \end{array}
 $$
In this way we can identify which supersymmetry multiplet they
fall into. The last two columns list the transformations of each R
vertex operator, obtained by applying the supercharges in
(\ref{susy4}).

\subsubsection{Closed superstring}

We match left and right vertex operators in IIB for concreteness,
the antiholomorphic sector being a copy of the holomorphic one we
just described. In type IIB we have two-dimensional $\N=(4,0)$
spacelike SUSY in the flat noncompact directions. Because of the
requirement of mutual locality of the vertex operators, the RNS
spectrum is not just the left right product of the holomorphic
sector.

\medskip {\it NS--NS sector}
\vskip 0.3cm

We have two lowest lying closed string tachyons
 \be
 \label{taclow4}
 T_\pm\bar T_{\pm}=e^{-\phi-\bar\phi\pm{i\over Q}(x+\bar x)+{1\over
 Q}(\vp+\bar\vp)},
 \ee
with R-charges $\pm \frac{4}{3}$. The other NS--NS operators which
are primary fields of the $N=2$ are
 \be G^{\mu\nu}=e^{-\phi-\bar\phi\pm iH_1\pm i\bar H_1} \ ,
\label{gr}
 \ee
which are neutral under $U(1)_x$ .

\vskip 0.3cm
\medskip {\it R--R sector}
\vskip 0.3cm

We match the left and right R states given in (\ref{lowR4}). By
imposing the mutual locality condition we find that the lightest
surviving $R-R$ fields are the six operators
 \be
 \label{RR4}
R_{-+}\bar R_{-+},\, R_{-+}\bar R_{++},\, R_{++}\bar R_{-+},\,
R_{++}\bar R_{++},\,R_{--}\bar R_{--},\,R_{+-}\bar
 R_{+-}.
 \ee
They are worldsheet $N=2$ primaries as well.

\vskip 0.3cm
\medskip {\it R--NS and NS--R sectors}
\vskip 0.3cm

The fermions need not be primaries, because they are not
supersymmetric deformations of the dual space-time lagrangian and
only some linear combination of fermions is a definite component
of a space-time short multiplet. In the following table we list
the physical fermions that belong to short multiplets
 \be
 \label{fer4}
 \begin{array}{ccccc}
      \textrm{NS-R}: &T_-\bar R_{--},& T_+\bar R_{+-},&J^\mu\bar R_{++}, &J^\mu\bar R_{-+}\\
&&&&\\
\textrm{R-NS}: & R_{--}\bar T_-,& R_{+-}\bar T_+, &R_{++}\bar
J^\mu, &R_{-+}\bar J^\mu
\end{array}
 \ee
We identify which multiplet each fermion sits in by looking at its
supersymmetry variation, when hit by the supercharges.\\

{\it Summary}: The four NS-NS operators $G^{\mu\nu}$, the first
four R-R in (\ref{RR4}) and the eight fermions in the last two
columns of (\ref{fer4}) sit in a $\N=(4,0)$ $d=2$ off-shell
supergravity multiplet, with a total of $16=8\oplus 8$ states. We
plot this multiplet according to its supersymmetry transformations
as follows
  $$
\begin{array}{ccccc}
 &          &  J^\mu\bar R_{++},J^\mu\bar R_{-+}     &       &\\
 &\nearrow_Q  &       &\searrow_{\bar Q}&\\
R_{-+}\bar R_{-+}, R_{-+}\bar R_{++}&          &       &       &G^{\mu\nu}\\
 R_{++}\bar R_{-+},R_{++}\bar
 R_{++}&          &       &       &\\
 &\searrow_{\bar Q}  &    &\nearrow_Q    &\\
 &          & R_{++}\bar J^\mu,R_{-+}\bar J^\mu &&
 \end{array}
 $$
where $Q$ and $\bar Q$ represent the holomorphic and
antiholomorphic supercharges. We then find a gauge supermultiplet
with $4\oplus4$ degrees of freedom, obtained as a combination of
two chiral multiplets, whose top components are the two tachyons.
One multiplet is
 $$
 \begin{array}{ccccc}
 &          &  {T}_+\bar R_{+-}    &       &\\
 &\nearrow_Q  &       &\searrow_{\bar Q}&\\
R_{+-}\bar R_{+-}&          &       &       &T_+\bar T_+\\
 &\searrow_{\bar Q}  &    &\nearrow_Q    &\\
 &          & R_{+-}\bar
T_+&&
 \end{array}
 $$
and analogously for the other one $T_-\bar{T}_-,R_{--}\bar
T_-,T_-\bar R_{--},R_{--}\bar R_{--}$.

Collecting the results, we find the {\it physical spectrum} of
$12\oplus12$ operators in (\ref{rehearsal}).

\subsection{Pure spinor variables}
\label{map4d}

The RNS superstring in the four dimensional linear dilaton
background has two supercharges both in the right moving and in
the left moving sectors. Let us focus on the holomorphic sector
only. The closed superstring in this formalism will just be the
left right product of the two sectors, without the complications
of the mutual locality conditions we found in the RNS. We follow
the same strategy discussed in section \ref{2dpure} for the two
dimensional linear dilaton, but here we will introduce an
additional ingredient, the flat $\RR^{1,1}$ part of the
space-time.

The physical RNS supercharges $Q_{+1},Q_{+2}$ are given in
(\ref{susy4}). As noted in \cite{Grassi:2005kc}, there exist other
two additional supercharges
  \bea
 \label{fake4}
 Q_{\dot+1}=&\oint e^{-\half\phi-{i\over2}(H_1-H)+ {i\over2}Q x},\nn\\
 Q_{\dot+2}=&\oint e^{-\half\phi-{i\over2}(H_1+H) -{i\over2}Q x}
 \,
 \eea
which survive GSO projection, are mutually local with respect to
the physical ones (\ref{susy4}) and are conserved, i.e. they have
at most double poles with the stress tensor. However, they are not
in the BRST cohomology, as they do not correspond to any physical
space-time symmetry. They play a key role in the construction of
the covariant formalism.

As we will see, the RNS variables will be mapped into a patch of
the pure spinor space. The pure spinor degrees of freedom of the
four-dimensional superstring are an $SO(1,3)$ Dirac spinor $\l^A$,
$A=1,\ldots,4$ satisfying the conditions
 \be \l\Gamma^m\l=0 \ ,
\label{ps4}
 \ee where $\Gamma^m$ are the $4\times4$
four-dimensional Dirac matrices. To perform the map it is most
convenient to solve the pure spinor constraint by breaking
$SO(1,3)$ to $U(2)$ and decompose the Dirac spinor as a
$(\l,\l^a,\l_{ab})$, namely a singlet, a two component vector and
a one component antisymmetric irreducible representations, for the
details see the Appendix. The pure spinor conditions become
 \be
 \l \l^a=0,\quad \l_{ab}\l^a=0 \ .
\label{patch}
 \ee
They can be solved by going to the patch where $\l^a=0$, so we are
left with $(\l ,\l_{ab})$.\footnote{Another way to write the patch
is to parameterize the pure spinor degrees of freedom as the Weyl
and anti-Weyl spinors $(\lambda^{\alpha},\lambda^{\dot \alpha})$
with the pure spinor constraints
$\l^\a\s^m_{\a\dot\b}\l^{\dot\b}=0$. Since $\sigma^m$ is a
complete basis in the space of two-dimensional bispinors, the pure
spinor condition becomes simply $ \l^\a\l^{\dot\a}=0$. In fact, we
can identify the two parametrizations by noting that $\l^\a=\l^a$,
$\l^{\dot\a}=(\l ,\l_{ab})$, so the patch $\l^a=0$ reads
$\l^\a=0$. In the following we will map the RNS theory to this
patch.}

Let us consider the four RNS supercharges, both physical
(\ref{susy4}) and unphysical ones (\ref{fake4}). We take $Q_{+2}$
in the $+\half$ picture and all the others in the $-\half$ picture
and we recast them in the a $U(2)$ notation: $Q_{+1}\equiv Q_+$,
$(Q_{+2},Q_{\dot+1})\equiv Q_a$, $Q_{\dot+2}\equiv Q^{ab}$. We
define
\begin{eqnarray}
\label{ferti} \theta^+=c\xi
e^{-{3\over2}\phi-{i\over2}H_1-{i\over2}(H-Qx)},&
p_+=b\eta e^{{3\over2}\phi+{i\over2}H_1+{i\over2}(H-Qx)},\\
\theta^{a}=e^{{1\over2}\phi\mp{i\over2}(H_1-H)-{i\over2}Qx},&
p_a=e^{-{1\over2}\phi\pm{i\over2}(H_1-H)+{i\over2}Qx},\\
\theta_{ab}=e^{{1\over2}\phi+{i\over2}H_1+{i\over2}(H+Qx)},&
p^{ab}= e^{{-1\over2}\phi-{i\over2}H_1-{i\over2}(H+Qx)},
\end{eqnarray}
where the $\theta$'s are the conjugate variables to the $p$'s. The
OPE of the first three conjugate pairs correspond to free fields
\begin{equation}
p_+(z)\theta^+(w)  \sim
 {1\over (z-w)},~~~~~~
p_a(z)\theta^b(w)  \sim
 {\delta_a^b\over (z-w)},~~~~~~
\end{equation}
where $a,b=1,2$ and all the others vanishing. However, the OPE of
$p^{ab}$ and $\t_{ab}$ have poles with all the others, so we drop
these two components. We will recover them later as a BRST
quartet. \footnote{Note that while $(\theta^+,p_+)$ and
$(\theta^{1},p_{1})$ are BRST invariant, the remaining
$(\theta^2,p_2)$ are used in order to enlarge the superspace
structure but are not physical. We will recover the appropriate
physical superspace when discussing the pure spinor global
symmetries.} The way to map the bosonized RNS variables to the
pure spinors is to use the $U(2)$ singlet $p_+$ in the $+\half$
picture and set
\begin{equation}\label{etab}
\eta=p_+e^{\tf+\tk},\qquad b=p_+e^{{1\over2}(\tf-\tk)}.
\end{equation}
The bosons $\tilde\phi$ and $\tk$ satisfy the OPE's
(\ref{tildeope}). We redefine $x\to x'$ to have the correct free
field OPE's among our new variables
\begin{equation}\label{tx}
x'={1\over2}\left(x-Q(H_1+H-i\phi)\right).
\end{equation}
The total RNS stress tensor is mapped to
 \begin{eqnarray}
 \label{stremap}
T_m+T_{gh}=&-{1\over2}\sum_{\mu=1}^{2}(\partial
x^\mu)^2-{1\over2}(\partial x')^2
-{1\over2}(\partial\varphi)^2+{Q\over2}\partial^2(\varphi-ix')\nn\\
&-p_{+}\partial\theta^+- p_{a}\partial\theta^a-{1\over2}(\partial
\tf)^2+\partial^2\tf+{1\over2}(\partial \tk)^2+\partial^2\tk,
 \end{eqnarray}

The second step is to use the $\tf,\tk$ as the ordinary
bosonization of a $\beta,\gamma$ system of weight $(1,0)$ as in
(\ref{purespi}), representing now the $U(2)$ singlet components of
the pure spinor and its conjugate momentum as in (\ref{patch}).

If we take a closer look at the stress tensor for the bosonized
pure spinors $\tf,\tk$ we see that the same story as in two
dimensions is repeated. There is a mismatch between the stress
tensor we get from the map from the RNS and the naive stress
tensor $w_+\p\l^+$ one would expect for the free beta-gamma
system. This is the contribution coming from the coupling of the
top form on the pure spinor manifold $\Omega(\l)$ to the
worldsheet curvature (\ref{topcop}). We find again that the pure
spinor stress tensor we get from the map is (\ref{twostress}) and
the top form is (\ref{omegatwo}). Due to the coupling
(\ref{topcop}) of the top form to the action, we need three powers
of $\l^+$ in the saturation rule to get a nonvanishing amplitude.

The total central charge of the system still  vanishes
\begin{equation}
\nn c=(-2)_{p_+\theta^+}+
(-4)_{p_{a}\theta^{a}}+(2)_{x_1,x_2}+(1+12)_\varphi+(1-12)_{x'}+(2)_{w\l}=0
\ .
\end{equation}

In four dimensions there is a new feature with respect to the
two-dimensional case. We need to add a topological quartet with
central charge $c=0$ to reconstruct the target space structure on
the pure spinor side. We add a fermionic $bc$ system of weight one
$(p^{ab},\theta_{ab})$ and a bosonic $\beta\gamma$ system of
weight one with the same quantum numbers $(w^{ab},\l_{ab})$.

We suitably modify the currents of the twisted $N=2$ by adding a
term that depends on the quartet. In this way we can recast the
stress tensor (\ref{stremap}) in the following covariant way
 \bea
 T&=&-\half(\p x^\mu)^2-\half(\p
 x')^2-\half(\p\vp)^2+{Q\over2}\p^2(\vp-ix')-p_{Ii}\p\t^{Ii}\nn\\
 &&+w_{Ii}\p\l^{Ii}-\half\p^2\log\Omega(\l) \ .
\label{stress4}
 \eea
The index $I=+,\dot+$, where $I=+$ corresponds to the physical
superspace coordinates and conjugate momenta and $I=\dot+$
represents the variables of the enlarged superspace. The index $I$
keeps track of the $SO(1,1)$ Lorentz spinor chirality, namely
$I=+$ is $+\half$ Lorentz charge and $I=\dot+$ is $-\half$; the
index $i=1,2$ labels two different spinors with the same
chiralities. Note that we do not have any spinor index in the game
since all the spinors are in the Majorana Weyl representation.
Moreover, we reconstructed the full pure spinor $\l^{Ii}$. The
matter part of this stress tensor can be derived from the pure
spinor action
 \bea\label{fouraction}
S = {1\over2\pi\a'}\int d^2z \left(\half\p x^\mu\bar\p
x^\nu\eta_{\mu\nu}+\frac{1}{2}\p x \bar\p x + \frac{1}{2}\p
\varphi \bar\p \varphi + p_{Ii}\bar\p\t^{Ii} \right) \nonumber\\
-{Q\over2}\int d^2z r^{(2)}(\vp-ix) \ ,
 \eea
where the last term is the Fradkin--Tseytlin term that couples the
space-time linear dilaton $\Phi=-{Q\over2}(\varphi-ix)$ to the
worldsheet curvature (we denoted $x'$ by $x$). For the consistency
of the FT term we have to compactify the $x$ direction on a circle
of radius $R=2/Q$, which in fact is the supersymmetric radius we
already know from RNS analysis.

Mapping the RNS saturation rule on the sphere
 to the pure spinor variables one gets
a requirement for an insertion of $\lambda_+^3$, which is
consistent with the measure corresponding to the top form we
obtained. Finally, the ghost current on the patch\footnote{We
obtain this current by mapping the RNS ghost current, adding to it
the contribution of the quartet and adding the term
$-{3\over2}(\p\tf-\p\tk)$, which does not alter the anomalies nor
the ghost number but might be useful for the Lorentz properties.}
$J_{gh}=-\p\tf+w^{ab}\l_{ab}$ can be written coviariantly as
\begin{equation}
 J_{gh}=w_{Ii}\l^{Ii}.
\end{equation}


\subsubsection{Supersymmetry structure}

When constructing the map from the RNS to the pure spinors, we
doubled the superspace. Namely, we supplement the physical
supercharges $Q_{+1},Q_{+2}$ with two additional supercharges
$Q_{\dot+1},Q_{\dot+2}$ that are not BRST invariant, thus not
physical, although they are conserved. We proceed as in the two
dimensional case by looking at the algebra we get from the RNS.
Since the RNS supersymmetry algebra closes up to picture changing,
we take the supercharge $Q_{+2}$ in (\ref{susy4}) in the $+\half$
picture
 \bea
 q_{+2}=&b\eta e^{{3\over2}\phi+{i\over2}H_1+{i\over2}(H-Qx)}
 +\p(x_1+ix_2)e^{\half\phi-{i\over2}H_1+{i\over2}H-{i\over2}Qx}+\nn\\&+
\p(\varphi+ix+iQH)e^{\half\phi+{i\over2}H_1-{i\over2}H-{i\over2}Qx}
\ , \label{qplus4}
 \eea
 and the other three in the
$-\half$ picture. The OPE's between them are
 \bea
q_{+1}(z)q_{+2}(0)&\sim{1\over z}\p(x_1+ix_2)(0),\nn\\
q_{+2}(z)q_{\dot+1}(0)&\sim {Q\over z^2}+{1\over z}\p(\vp+i
x+iQH)(0)
 \eea
and $q_{+2}(z)q_{\dot+2}\sim\,reg$. If we map this algebra to the
pure spinor variables we find that the first equation remains
unchanged, while the second reads
 \bea
 \label{superc}
q_{+2}(z)q_{\dot+1}(0)&\sim {Q\over z^2}+{1\over z}\p(\vp-i x)(0)
\ ,
 \eea
where we denoted $x'$ by $x$ for simplicity of notation. As usual,
the algebra is realized on the superderivatives with opposite
signs.

With these variables we construct the following GS-like
constraints
 \be
 \label{ds4}
 d_{Ij}=p_{Ij}-\half\tau_{ij}\left(\d_{IJ}\t^{Ji}\p(x_1+ix_2)+\tau_{IJ}\t^{Ji}\p(\vp-ix)-Q\e_{IJ}
 \p\t^{Ji}\right),
 \ee
where $\tau$ is the $\s^1$ Pauli matrix and $\e_{IJ}$ is the
antisymmetric matrix. The GS-like constraints are similar to the
$d=0$ case (\ref{dzero}), except for the new term proportional to
$\d_{IJ}$, which realizes the $d=2$ space-time supersymmetry. The
notations are explained in (\ref{stress4}). These constraints
reproduce the algebra we mapped from the RNS superderivatives
(given by the supercurrent algebra (\ref{superc}) but with the
opposite signs)
 \be
 \label{ddope4}
d_{Ii}(z)d_{Jj}(0)\sim -\tau_{ij}{\e_{IJ}Q\over
z^2}-\tau_{ij}{\delta_{IJ}\over
z}\p(x_1+ix_2)(0)-\tau_{ij}{\tau_{IJ}\over z}\p(\vp-ix)(0).
 \ee
Let us briefly discuss this algebra. The second term tells us that
the physical $d_{+1}$, $d_{+2}$ close on the space-time $SO(1,1)$
holomorphic translation generator, exactly reproducing the RNS
supersymmetry. The two unphysical superderivatives $d_{\dot+1}$,
$d_{\dot+2}$ close on the same translation generator. However, the
two physical and unphysical superderivatives have a double pole in
their OPE, proportional to the Liouville background charge $Q$.
Therefore the non anomalous worldsheet current algebra realizes
$\N=(2,0)$ supersymmetry in two dimensions. In the closed type IIB
theory we get just the left right product of the two sectors,
realizing $\N=(4,0)$ two dimensional supersymmetry, as we expect
from the RNS analysis.

Let us recall the space-time symmetries. In each of the two
sectors, the expectation value of the dilaton and the
compactification of $x$ on a circle break the four dimensional
Lorentz symmetry with four supercharges to two real
supersymmetries and a bosonic $ SO(1,1) \times \left(U(1)\times
\ZZ_2\right)_R$. Under the bosonic $ SO(1,1) \times U(1)_x$
symmetry the $d$'s are charged according to
 \be\label{charges4}
\begin{array}{ccc}
     &  SO(1,1)  & U(1)_x \\
d_{+1}  & +\half &+1\\
d_{+2}  & +\half &-1\\
d_{\dot+1} & -\half &+1\\
d_{\dot+2}  & -\half &-1
\end{array}
 \ee
The physical worldsheet holomorphic current algebra realizes
space-time $\N=(2,0)$ supersymmetry in two dimensions
 \be
 d_{+1}(z) d_{+2}(0)\sim-{\p( x_1+ix_2)(0)\over z}.
 \ee
When acting on superfields depending on the field zero modes only,
the $d_{Ii}$ looks like a superderivative
 \be
 d_{Ii}(z)\Phi(\bar Z, x_1,x_2, \t^{Ii})(0)\sim-{1\over z}D_{Ii}\Phi(\bar Z, x_1,x_2, \t^{Ii})(0)
 \ee
where
 \be\label{superstardj}
 D_{Ii}=\partial_{\t^{Ii}}+\tau_{ij}(\d_{IJ}\p_{x_1-ix_2}+\tau_{IJ}\p_{Z})
 \ ,
 \ee
and the notations is the same as in the two dimensional case
$Z=\vp+ix$, $\bar Z=\vp-ix$.

\subsubsection{Cohomology}

The pure spinor BRST charge in the four-dimensional non-critical
string is
 \be
 Q_B=\oint\l^{Ii} d_{Ii} \ ,
 \ee
where the GS-like constraints $d_{Ii}$ in the linear dilaton
background are defined in (\ref{ds4}). As discussed in section
\ref{2Dcohom}, we included all the physical as well as the
unphysical $d$'s in the BRST charge. We will compute the
cohomology in two steps. First we compute the cohomology of $Q_B$
in the enlarged superspace containing all the $\t^{Ii}$, then we
restrict to the part of the cohomology that contains the variables
that realize the supersymmetry current algebra, namely the
$\t^{+i}$.

Let us discuss the nilpotency of the BRST charge $Q_B$. The
four-dimensional pure spinor constraint (\ref{ps4}) can be recast
according to the notation of the linear dilaton background in the
form\footnote{\label{diracdil} The constraint
$\l^A\Gamma^M_{AB}\l^B=0$ can be written in flat Weyl notation as
$\l^\a\l^\ad=0$ where $\l^A=(\l^\a,\l^\ad)$. Then we identify the
flat Weyl indices with our linear dilaton quantum numbers as
$Ii=+1\rightarrow \a=1$, $Ii=\dot+1\rightarrow \a=2$,
$Ii=+2\rightarrow \dot\a=\dot1$ and $Ii=\dot+2\rightarrow
\ad=\dot2$.}
 \be\label{pscond4}
 \tau_{ij}\l^{Ii}\l^{Jj}=0 \ ,
 \ee
for $I,J=+\dot+$ and $i,j=1,2$. Due to the OPE (\ref{ddope4}), the
nilpotency of the BRST charge requires the following conditions
 \be
 \tau_{ij}\d_{IJ}\l^{Ii}\l^{Jj}=0, \qquad \tau_{ij}\tau_{IJ}\l^{Ii}\l^{Jj}=0
\ ,
 \ee
which are implied by the pure spinor condition (\ref{pscond4}).
Due to the double pole in (\ref{ddope4}), however, we have an
additional derivative constraint
 \be
 \label{psder}
 \e_{IJ}\tau_{ij}\p\l^{Ii}\l^{Jj}=0 \ .
 \ee
One can show that the pure spinor condition (\ref{pscond4})
implies that both $\tau_{ij}\l^{+i}\p\l^{\dot+j}$ and
$\tau_{ij}\p\l^{+i}\l^{\dot+j}$ vanish separately. The way to
prove it is analogous to the $d=0$ case we discussed in
(\ref{mode2d}), by looking at the mode expansion of the pure
spinor constraint (\ref{pscond4}) or alternatively by analyzing
the OPE's involving the Lorentz generator and the ghost current
\cite{Grassi:2005jz,Wyllard:2005fh}.

Let us compute the cohomology now. As discussed in the
two-dimensional case, there are two different kinds of vertex
operators at ghost number one and weight zero. The first one is
 \be\label{ghostfir}
{\cal V}^{(1)}=\l^{Ii}A_{Ii}(\bar Z,x^\mu,\t^{Ii}) \ .
 \ee
Whereas in the previous case (\ref{2dvertex}) we found that this
operator was exact (there is no gravity in $d=0$), now we will
find an off-shell two dimensional supergravity multiplet. Imposing
that ${\cal V}$ is BRST closed on the pure spinor condition
(\ref{pscond4}), we get
 \be
 D_{(I1}A_{J1)}=0,\qquad D_{(I2}A_{J2)}=0 \ ,
 \ee
where $D_{Ii}$ is defined in (\ref{superstardj}). Due to the
algebra
 \be
 \{D_{I1},D_{J1}\}=0=\{D_{I2},D_{J2}\} \ ,
 \ee
we can solve these two equations by choosing $A_{I1}=D_{I1}B$ and
$A_{I2}=D_{I2}C$, for two generic superfields $B$ and $C$. Let us
take equivalently a linear combination $B=M+N$ and $C=M-N$, so
that $A_{I1}=D_{I1}(M+N)$ and $A_{I2}=D_{I2}(M-N)$. Now we require
that this vertex operator is not BRST exact, that is we mod out by
the following gauge invariance
 \be
 \d {\cal V}^{(1)}=Q_B \Omega^{(0)} \ ,
 \ee
where $\Omega^{(0)}(\bar Z,x^\mu,\t^{Ii})$ is a generic ghost
number zero and weight zero superfield. If we choose $\Omega=-M$,
we are left with
 \be
 \label{cohove}
 A_{I1}=D_{I1}N,\qquad A_{I2}=-D_{I2}N \ ,
 \ee
for a generic superfield $N$. It is easy to see that the degrees
of freedom encoded in ${\cal V}$ are $4\oplus4$, which is the
result of \cite{Grassi:2005sb,Grassi:2005jz}.\footnote{This is
done most quickly in the notations in the previous footnote, by
which $A_{I1}\equiv A_{\a}$ and $A_{I2}\equiv A_{\ad}$. If we
introduce the supercurvature $F_{AB}=D_{(A}A_{B)}$, for
$A,B=(\a,\ad)$ then (\ref{cohove}) is equivalent to the following
conditions on the curvature
$F_{\a\b}=F_{\ad\dot\b}=F_{\a\dot\b}=0$, which are the usual
superspace constraints defining the $\N=1$ off-shell vector
multiplet in four dimensions.} Since we are interested in the
$SO(1,1)$ supersymmetry multiplet, we have to eliminate the
$\t^{\dot+i}$ components in the vertex operator, keeping only the
$\t^{+i}$, the latter entering in the holomorphic space-time
$\N=(2,0)$ current algebra. The physical states in the holomorphic
sector consist finally of $2\oplus2$ degrees of freedom of an
off-shell $SO(1,1)$ vector supermultiplet.

We are interested in the closed string spectrum, which is the
tensor product of the holomorphic and antiholomophic sectors. The
cohomology computation of the closed string vertex operator
 \be
 {\cal V}^{(1,1)}=\l^{Ii}\bar\l^{Jj}A_{Ii,Jj}(\bar Z,x^\mu,
 \t^{Ii},\bar \t^{Kk}),
 \ee
where the $\bar \l$ and $\bar \t$ are the antiholomorphic
variables, gives a total of $8\oplus8$ degrees of freedom, that
fit into an $\N=(4,0)$ $d=2$ supergravity multiplet, reproducing
the RNS computation.

The second type of vertex operator is the generalization of the
two-dimensional one that we introduced in (\ref{massi2}). This
accounts for the gauge multiplet, which is the other character in
the cohomology of the linear dilaton superstring. In this case the
computation of the cohomology is more tedious, and we present the
details in appendix D. After taking into account the equations of
motion modulo the gauge symmetries and projecting to the physical
supercoordinates $\t^{+i}$, the vertex operator for this gauge
supermultiplet in the holomorphic sector is
 \be\label{tachyqua}
 {\cal U}_T^{(1)}=\l^{Ij}\p\t^{+i}D_{Ij}T_i(\t^{+i}) \ ,
 \ee
where the wavefunctions of the superfields $T_i$ are given by
$e^{-Z/Q}$, by which we get a weight zero vertex operator. To
compare this vertex operator to the RNS, we need to take into
account the structure of the pure spinor space. The
four-dimensional pure spinor space $\tau_{ij}\l^{Ii}\l^{Jj}=0$ is
the union of two disconnected patches and our vertex operator
(\ref{tachyqua}) is defined globally on the two patches exchanged
by $\l^{I1}\leftrightarrow\l^{I2}$. The RNS formalism is mapped
onto just one of the two patches, let us choose the $\l^{I2}\ne 0$
patch. In order to compare the RNS cohomology with the pure spinor
result, we will consider one of the two disconnected patches, so
the tachyon supermultiplet on the patch $\l^{I2}\ne 0$ has the
form
 \be
 {\cal U}^{(1)}=\l^{I2} \p\t^{1}D_{I2} T_1(\t^1,\t^{\dot1}) \ ,
 \ee
containing $2\oplus2$ degrees of freedom. In the closed string, we
have to take the product of the holomorphic and antiholomorphic
vertex operators. Since we are comparing the pure spinor
computation with the RNS cohomology at zero momentum, we need to
mod out by the center of the space-time symmetries $SO(1,1)\times
U(1)_x \times \ZZ_2$, which is just the $\ZZ_2$, so that we
recover the closed superstring $4\oplus4$ supermultiplet of
(\ref{rehearsal}).

Let us make a few comments. In the two-dimensional case, the gauge
supermultiplet (\ref{tacchy}) was the only physical multiplet
present in the second type vertex operator at ghost number one and
weight zero. This is in agreement with the known fact that in the
two-dimensional non-critical superstrings in the linear dilaton
background the tower of ``massive" string states is absent. In the
present four-dimensional case, however, this second type of vertex
operator contains other physical states in addition to the gauge
supermultiplet which we just described. These additional states
are some of the tower of higher states in the cohomology of the
superstring, which we expect to be present.\footnote{We would not
call these higher states massive, since, as we explained in
section 3, in the linear dilaton background the space-time
supermultiplets are off-shell, in the sense that we do not have
any dispersion relation.} The study of the higher states in the
superstring cohomology might be interesting by itself.

\subsubsection{Curved non-critical backgrounds}

We follow the discussion in section \ref{curved2} and suggest a
generalization of the non-critical pure spinors to generic four
dimensional curved backgrounds with at most eight real
supercharges.

Even if the linear dilaton has only flat two dimensional
supersymmetry, we introduce now the momenta $\Pi^m$, for
$m=1,2,\vp,x$, that will be useful when casting the theory in a
background with twice as many supersymmetries
 \bea
 \Pi^1=\p x^1+\half\d_{IJ}\tau_{ij}\t^{Ii}\p\t^{Jj},\quad
 \Pi^2=\p
 x^2+{i\over2}\d_{IJ}\tau_{ij}\t^{Ii}\p\t^{Jj},\nn\\
 \Pi^\vp=\p \vp+\half\tau_{IJ}\tau_{ij}\t^{Ii}\p\t^{Jj},\quad
 \Pi^x=\p
 x-{i\over2}\tau_{IJ}\tau_{ij}\t^{Ii}\p\t^{Jj}.
 \eea
They satisfy the following algebra
 \be
 \Pi^n(z)\Pi^m(0) \sim -{\eta^{mn}\over z^2},
 \ee
 \bea
d_{Ii}(z)\Pi^1(0) \sim {\d_{IJ}\tau_{ij}\over z}\p\t^{Jj}(0),
\qquad d_{Ii}(z)\Pi^2(0) \sim {i\d_{IJ}\tau_{ij}\over z}\p\t^{Jj}(0)\nn\\
  d_{Ii}(z)\Pi^\vp(0) \sim {\tau_{IJ}\tau_{ij}\over z}\p\t^{Jj}(0),
  \qquad d_{Ii}(z)\Pi^x(0) \sim {i\tau_{IJ}\tau_{ij}\over z}\p\t^{Jj}(0)
\label{jaco4}
 \eea

The stress tensor (\ref{stress4}) can be cast in the form
 \bea
 T=&-\half\Pi^m\Pi^n\eta_{mn}-d_{Ii}\p\t^{Ii}+{Q\over 2}\e_{IJ}\tau_{ij} \p\t^{Ii}\p\t^{Jj}+
{Q\over2}\p (\Pi^\phi-i\Pi^x) \nn\\
 &+w_{Ii}\p\l^{Ii}-\half\p^2\log\Omega(\l) \ .
 \eea
Note the presence of the extra terms proportional to $Q=\sqrt{3}$,
which is a feature  of the linear dilaton background. Similarly,
we can write the action in these variables, which correspond to
the supervielbeins of the general four-dimensional pure spinor
superstring backgrounds
 \bea
 S = &{1\over2\pi\a'}\int d^2z \left(\frac{1}{2}G_{MN}(Y)\p Y^M\bar\p Y^N + E_{M}^{A}(Y)
  d_{A}\bar\p Y^M +E_{M}^{A}(Y)
  \bar d_{A}\p Y^M \right)\nn\\
&-\int d^2z r^{(2)}\Phi(Y) \ ,\label{curve4d}
 \eea
where we introduced the curved supercoordinates
$Y^M=(x^m;\t^{A},\bar\t^{\hat A})$ and $m$ is a curved four
dimensional vector index, while $A=(\a,\ad)$ is a curved four
dimensional Dirac spinor index; the relation between $A$ and $Ii$
is explained in the appendix. The variables without bars are in the
holomorphic sector, the ones with bars are in the antiholomorphic
sector. The $E$'s are the vierbein superfields. The whole story is
a generalization of the two dimensional case. In the linear
dilaton case, the background superfields take the following
values. The vierbeins $E$'s are the four dimensional flat ones.
The dilaton superfield is linear $\Phi={Q\over2}(\vp-ix)$. The
metric $G_{MN}$ is constant, however in addition to the usual
terms appropriate for a flat background, we have a flat spinorial
part as well
 \be
 G_{\a\ad}=iQ \s^2_{\a\ad}\, \ ,\qquad G_{\hat \a\hat \ad}=iQ \s^2_{\hat \a\hat\ad}\,\ ,
 \ee
which is proportional to the background charge $Q=\sqrt{3}$ and is
responsible for the contribution
$Q\e_{IJ}\tau_{ij}\p\t^{Ii}\p\t^{Jj}$ to the stress tensor
(\ref{stresscu}). We regard this as a specific feature of the
linear dilaton superspace structure in the pure spinor formalism.
This explicitly breaks the parent $SO(1,3)$ Lorentz invariance of
the action down to $SO(2)$, preserving the $U(1)_x$ R symmetry.

It is suggestive to think of (\ref{curve4d}) as the matter part of
the non-critical pure spinor action in a generic curved four
dimensional background. We will discuss the non-critical $AdS_4$
example in section 8.

\subsubsection{Anomalies}

The pure spinor degrees of freedom in the Weyl notations are
$(\lambda^{\alpha},\lambda^{\dot \alpha})$ with the conditions
$$\lambda^{\alpha}\lambda^{{\dot \alpha}} = 0 \ .$$
These complexified equations define a two-complex dimensional
space with a conical singularity at $\lambda^{\alpha}=\lambda^{\dot
\alpha} = 0$.

As in the two-dimensional space, deforming this set of four
complex equations results in a space whose first Chern class is
nonvanishing, leading to an anomalous theory. Again, we can
eliminate the singularity by removing the singular point. We get a
disconnected space, which is the disjoint union of
 $C^{*}\times C^*$'s. In this way, the anomalies are avoided and
as in the two-dimensional case the resulting space is
disconnected.


\section{Six-dimensional superstrings}

In this section we discuss the six dimensional non-critical
superstring in the linear dilaton background
 $$
 \RR^{1,3}\times \RR_\vp\times U(1)_x \ .
 $$
We will closely follow the analysis in the previous Sections on
$d=0$ and $d=2$, so we will skip some details. The flat
$\RR^{1,3}$ coordinates are $x^\mu$, while $x$ is compactified on
a circle of radius $2/Q$, the Liouville background charge is
$Q=\sqrt{2}$ and we have $d=2n=4$.

\subsection{Multiplet spectra: RNS analysis}

The space-time supercharges have different chiralities of the
Lorentz group $SO(1,3)$. There are sixteen candidates for the
supercharges. They form two groups of eight mutually local
physical operators. One such group is given by (in the
 $-\frac{1}{2}$ picture)
 \begin{eqnarray}
   q_{+1} & =  e^{-\half\phi + i(H^1 - H^2 - H + Q x)/2} \ ,\qquad
   q_{+2} & =  e^{-\half\phi + i(-H^1 + H^2 - H + Q x)/2} \ , \nn\\
   q_{+\dot1} & =  e^{-\half\phi + i(H^1 + H^2 + H - Q x)/2} \
   ,\qquad
   q_{+\dot2} & =  e^{-\half\phi + i(-H^1 - H^2 + H - Q x)/2} \
   ,\nn\\ \label{susy6}
 \end{eqnarray}
Defining $\sigma_1 = \mathbf{1}$, $\sigma_2 = \tau_1$, $\sigma_3 =
i \tau_3$ and $\sigma_4 = \tau_2$, where the $\tau$s are the Pauli
matrices, we can rewrite the above OPE's in the concise form
\begin{equation}
\label{susya6}
  q_{+ \alpha}(z) q_{+ \dot \alpha} (0) \sim \frac{1}{\sqrt{2} z}
  \sigma_{i \alpha \dot \alpha} e^{-\phi} \psi^i (0) \ .
\end{equation}
The other set of supercharges $q_{-\a},q_{-\ad}$, which are
physical but nonlocal with respect to (\ref{susy6}) is listed in
the Appendix. We will choose the $q_+$ set in the holomorphic
sector. The choice of the $\bar q_+$ set in the antiholomorphic
sector as well defines type IIB superstring. The other choice of
the $\bar q_-$ set in the antiholomorphic sector defines the type
IIA superstring. In both cases we realize $\N=2$ space-time
supersymmetry in the flat $SO(1,3)$ directions. In the following
we will stick to the type IIB case.

Let us collect the short space-time supermultiplets of the type
IIB superstring. The details of the computations are listed in the
Appendix.

\subsubsection{Holomorphic sector}

The holomorphic sector is obtained by requiring mutual locality
with the supercharges $q_{+ \alpha}$ and $q_{+ \dot \alpha}$. This
will serve as a building block for the closed superstring states
we will consider in the next section.

The first NS state at zero transverse momentum is the tachyon
whose two lowest lying states are
\begin{equation}
 \label{tachy6}
  T^\pm = \varphi e^{-\phi + {1\over Q}(\varphi \pm i x)} \ ,
\end{equation}
which carry R-charges $\pm1$. The factor of $\vp$ in front of the
vertex operator comes from the requirement of non-normalizability
and is specific to $d+2=6$.

At the next level there are the NS vectors. The zero momentum
states are
\begin{equation}
 \label{vector6}
  J_\mu= e^{-\phi \pm i H^I} \ .
\end{equation}

We now turn to the Ramond sector. We use Polchinski's notation
$(\a,F)$ \cite{Polchinski:1998rr}, where $\alpha$ is the
space-time fermion index and $F$ is the worldsheet spinor index,
and denote the R operators by $R^F$. The zero momentum R states
with $F=1$ are
\begin{eqnarray}
  R^1_{+++} & =  e^{-\phi / 2 + i (H+H^1 + H^2) / 2 - i x / Q} \
  ,\quad
  R^1_{--+} & =  e^{-\phi / 2 + i (-H+H^1 - H^2) / 2 + i x / Q} \ ,\nn\\
  R^1_{-+-} & =  e^{-\phi / 2 - i (H+H^1 - H^2) / 2 - i x / Q} \
  ,\quad
  R^1_{+--} & =  e^{-\phi / 2 - i (-H+H^1 + H^2) / 2 + i x / Q} \
  ,\nn\\\label{ram1}
\end{eqnarray}
that are the four physical supercharges (\ref{susy6}).\footnote{In
terms of a full $SO(6)$ spinor representation, these R states have
an odd number of +'s and correspond to a $\underline{4}\in
SU(4)\simeq SO(6)$.} According to the supersymmetry algebra
(\ref{susya6}), they are mapped to the four NS states in
(\ref{vector6}), schematically $ \delta_{susy} R^1= J$. The $F=0$
states with zero momentum and zero R-charge have
$\beta=\frac{Q}{2}$ so we have to pick the non-normalizable vertex
operators:
\begin{eqnarray}
  R^0_{-++} & =  \varphi e^{-\phi / 2 + i (-H+H^1 + H^2 ) / 2 + \varphi
  / Q} \ ,\quad
  R^0_{+-+} & =  \varphi e^{-\phi / 2 + i (H+H^1 - H^2 ) / 2 + \varphi
  / Q} \ ,\nn\\
  R^0_{++-} & =  \varphi e^{-\phi / 2 - i (-H+H^1 - H^2) / 2 + \varphi
  / Q} \ , \quad
  R^0_{---} & =  \varphi e^{-\phi / 2 - i (H+H^1 + H^2 ) / 2 + \varphi
  / Q} \ .\nn\\\label{ram0}
\end{eqnarray}
These four operators\footnote{In terms of a full $SO(6)$ spinor
representation, these R states have an even number of +'s and
correspond to a $\underline{4}\in SU(4)\simeq SO(6)$.} are mapped
by the supercharges into the tachyon (\ref{tachy6}), schematically
$Q_\a R^0_{+\,\cdot\,\cdot}=T^+$ and $Q_\ad
R^0_{-\,\cdot\,\cdot}=T^-$.

\subsubsection{Closed superstring}

The closed superstring spectrum is obtained as the product of
holomorphic and antiholomorphic vertex operators subject to level
matching and mutual locality conditions. We want the Type IIB
spectrum here, so the GSO projection in the antiholomorphic sector
is the same as in the holomorphic one.

The allowed combinations are
\begin{displaymath}
  T^\pm \bar T^\pm \ , \ R^0 \bar T^\pm \ , \ T^\pm \bar R^0 \ , \ R^0
  \bar R^0 \ , \ G_{\mu\nu}\equiv J_\mu \bar J_\nu \ , \ J_\mu \bar R^1 \ , \
  R^1 \bar J_\nu \ , R^1 \bar R^1 \ .
\end{displaymath}

These degrees of freedom can be arranged in supermultiplets as
follows. The supergravity multiplet can be depicted as
\begin{displaymath}
  \begin{array}{ccccc}
    && R^1 \bar J_\mu && \\
    & \swarrow_{Q} && \nwarrow_{\bar Q} & \\
    G_{\mu\nu} &&&& R^1 \bar R^1 \\
    & \nwarrow_{\bar Q} && \swarrow_{Q} & \\
    && J_\mu\bar R^1 &&
  \end{array}
\end{displaymath}
It contains $32\oplus32$ states and is off-shell because for the
zero-momentum case the transversality condition does not impose
any restriction on the sign of $H^I$.

Then we have the supermultiplet in which the tachyon sits, which
is an off-shell $\N=2$ $SO(1,3)$ gauge multiplet. Since we are
working at zero momentum in the $SO(1,3)$ directions, we have to
mod out the vertex operators by the center of the group, namely we
identify $R^0_{++-}\sim R^0_{+-+}$ and $R^0_{+++}\sim R^0_{+--}$.
Then we can depict the multiplet as
\begin{displaymath}
  \begin{array}{ccccccccc}
    & & & & R^0_{+\,\cdot\,\cdot}\bar T^+ & & & & \\
    & & & \nearrow_{\bar Q_\a} & & \searrow_{Q_\a} & & \\
    & & R^0_{+\,\cdot\,\cdot}\bar R^0_{+\,\cdot\,\cdot} & & & & T^+\bar T^+ & & \\
    &  & & \searrow_{Q_\a} & & \nearrow_{\bar
    Q_\a} & & & \\
    & & & & T^+ \bar R^0_{+\,\cdot\,\cdot}& & & &  \\
  \end{array}
\end{displaymath}
The full supermultiplet consists of other three parts, which are
generated by acting with $Q_\ad$ and $\bar Q_\ad$ on
$R^0_{-\,\cdot\,\cdot}\bar R^0_{-\,\cdot\,\cdot}$ and
$R^0_{\pm\,\cdot\,\cdot}\bar R^0_{\mp\,\cdot\,\cdot}$. In total we
have $8\oplus8$ states, which represent an off-shell $\N=2$
$SO(1,3)$ supercurrent.

\subsection{Pure spinor variables}
\label{map6d}

 The RNS bosonic space-time coordinates are $x,\varphi,
x^\mu$. We proceed as above to map the RNS to a patch of the pure
spinor space. The first step is to realize, following
\cite{Grassi:2005kc} that there exist another set of four
supercharges, which we denote by $q_{\dot+\a}$ and $q_{\dot+\ad}$,
which are mutually local with respect to the ones in
(\ref{susy6}), are conserved, but not BRST invariant. Their
explicit form is given in the Appendix.

The six dimensional pure spinor consists of two complex Weyl
spinors $\l^A_i$, for $i=1,2$, in the $\underline{4}$ of
$SU(4)=SO(6)$. The pure spinor constraint is written using the
$4\times4$ off-diagonal antisymmetric Pauli matrices of the six
dimensional Dirac matrices as $\e^{ij}\l^A_i\s^m_{AB}\l^B_j=0$. In
terms of $U(3)$ representations, that we will conveniently use in
the map, each pure spinor splits into a singlet and a vector
$\l^A_i=(\l^+_i,\l^a_i)$, for $a=1,2,3$, and we can rewrite the
pure spinor constraint as
 \be
 \e^{ij}\l_i^+\l^a_j=0 \ ,\qquad \e_{abc}\e^{ij}\l^a_i\l^b_j=0 \
 . \label{eq:6dpure-spinor-constraint}
 \ee
We refer to the appendix for the relations between all the
different spinor representations. Working in the patch
$\l^+_1\neq0$ we can solve the pure spinor constraint as
$\l_2^a={\l^+_2\over\l^+_1}\l^a_1$ and see that a pure spinor in
six dimensions has five independent components
\cite{Grassi:2005sb}. We map the RNS variables to this patch of
the pure spinor space. However, since the map is quite similar to
the case $d=2$ in section \ref{map4d}, we present the details in
the Appendix. Here we only quote the results.

The RNS stress tensor, with the appropriate addition of a $c=0$
quartet, is mapped to the following pure spinor stress tensor
\begin{eqnarray}
  T' & =  -\frac{1}{2}\p x^\mu\p x^\nu\eta_{\mu\nu}- \frac{1}{2} (\del
  \varphi)^2 - \frac{1}{2} (\del x')^2 + \frac{Q}{2} \del^2 (\varphi -
  i x')  \nonumber\\
  & -  p_{I \a} \del \theta^{I\a} -  p_{I \ad} \del \theta^{I\ad}+ w_{I \a}
  \del \lambda^{I\a} + w_{I \ad}
  \del \lambda^{I\ad}- \frac{1}{2} \del^2 \log \Omega \
  ,\label{6dstress}
\end{eqnarray}
where $I=+,\dot+$ and $I=+$ stands for the physical
supercoordinates, while $I=\dot+$ denotes the unphysical ones. The
indices $(\a,\ad)$ refer to the Weyl and anti-Weyl spinor indices
of $SO(1,3)$ Lorentz symmetry. $\Omega$ is the holomorphic top
form on the pure spinor space, on the patch $\lambda^{+ \dot
1}\ne0$ it reads
\begin{equation}
  \Omega =  (\lambda^{+ \dot 1})^{-3}
  \ .
\end{equation}

The six dimensional pure spinor constraint, written according to
the space-time $SO(1,3)$ Lorentz symmetry, reads
 \bea
 \d_{IJ}\l^{I\a}\l^{J\ad}=0,\nn\\
 \e_{IJ}\e_{\a\b}\l^{I\a}\l^{J\b}=0,\label{pures6d}\\
 \e_{IJ}\e_{\ad\dot\b}\l^{I\ad}\l^{J\dot\b}=0\nn,
 \eea

Mapping the RNS saturation rule for the amplitudes on the sphere
 to the pure spinor variables one gets
a requirement for an insertion of $(\lambda^+)^3$, which is
consistent with the measure corresponding to the top form we
obtained.

\subsubsection{Supersymmetry algebra}

We consider just the holomorphic sector in the following. In the
RNS formalism we have eight conserved supercurrents, out of which
only four are physical. We want to find their algebra. As above we
take $q_{+\dot1}$ in the $+\half$ picture and all the rest in the
$-\half$ picture and compute their OPE's. If we take these OPE's
and map them to the pure spinor variables, we find the following
algebra
 \bea
 &q_{+\a}(z)q_{+\dot\a}(0)\sim{1\over z}\s^m_{\a\dot\a}\p x_m(0),&
 q_{\dot+\a}(z)q_{\dot+\dot\a}(0)\sim{1\over z}\s^m_{\a\dot\a}\p x_m(0),
 \nn\\
&q_{+\a}(z)q_{\dot+\b}(0)\sim{\e_{\a\b}Q\over z^2}+{\e_{\a\b}\over
z}\p(\vp-ix)(0),
 &q_{+\dot\a}(z)q_{\dot+\dot\beta}(0)\sim{\e_{\ad\dot\b}Q\over z^2}+{\e_{\ad\dot\b}\over
z}\p(\vp-ix)(0),\nn\\ \label{fake6a}
 \eea
and $q_{+\a}q_{\dot+\dot\a} \sim 0$, $q_{+\dot\a}q_{\dot+\a} \sim 0$. We see that
the same story follows as in the $d=0$ and $d=2$ cases. There are
two sets of supercharges, the physical ($+$) and the unphysical
($\dot+$) ones, which separately close on a flat four dimensional
supersymmetry algebra. However, the cross OPE's between the
physical and the unphysical sets have an anomalous double pole.
This means that the non anomalous current algebra realizes the
$SO(1,3)$ space-time supersymmetry.

We want to construct GS-like constraints that reproduce the
algebra (\ref{fake6a}) with opposite overall signs, as usual. They
are the direct generalization of the four dimensional ones in
(\ref{ds4})
 \bea
d_{I\a}=&p_{I\a}-\half\delta_{IJ}[\p
x^m-{1\over4}f^m(\t)](\s_m\t^J)_\a-\half\e_{\a\b}
[\tau_{IJ}\t^{J\b}\p(\vp-ix)-\e_{IJ}Q\p
\t^{J\b}],\nn\\
d_{I\ad}=&p_{I\ad}-\half\delta_{IJ}[\p
x^m-{1\over4}f^m(\t)](\t^J\s_m)_\ad-\half\e_{\ad\dot\b}
[\tau_{IJ}\t^{J\dot\b}\p(\vp-ix)-\e_{IJ}Q\p
\t^{J\dot\b}],\nn\\\label{ds6d}
 \eea
where we introduced the notation $
 f^m(\t)= \d_{IJ}\left(\p\t^I\s^m\t^J-\t^I\s^m\p\t^J\right)$,
to save space. We are using the two by two Pauli matrices
 $\g^m_{\a\ad}$ of the $SO(1,3)$ Lorentz group. The $d$'s realize
 the $q$'s algebra (\ref{fake6a}) we obtained from the RNS map, but with opposite signs
 \bea
d_{I\a}(z)d_{J\ad}(0)\sim-\e_{IJ}{\e_{\a\b}Q\over
z^2}-\d_{IJ}{\s^m_{\a\ad}\Pi_m(0)\over z}-\tau_{IJ}{\e_{\a\b}\over
z}\p(\vp-ix)(0),\label{fake6al}
 \eea
and all the others vanish.

Let us discuss the space-time symmetries. To identify the quantum
numbers of the superderivates, we refer to their RNS origin.
 \be
\begin{array}{cccc}
          &SO(1,3)&\times&U(1)_x \\
d_{+\a}       & (\half,0)& & +1 \\
d_{\dot+\dot\a} & (0,\half) & & +1 \\
d_{\dot+\a}   & (\half,0) & & -1 \\
d_{+\dot\a}& (0,\half) &  & -1 \\
\end{array}\label{dcharges}
 \ee
In the RNS description, even if all the $d$'s are conserved
currents, only half of them are actually physical. In the pure
spinor, due to the double pole in the OPE's (\ref{fake6al}), the
worldsheet current algebra only realizes $\N=1$ supersymmetry in
four dimensions in the holomorphic sector. The physical
supersymmetry is the one generated by
 \bea
d_{+\a}(z)d_{+\ad}(0)\sim-{\s^m_{\a\ad}\Pi_m(0)\over
z},\label{phys6al}
 \eea
 The closed superstring will realize $\N=2$ four
dimensional supersymmetry.

\subsubsection{Cohomology}

The pure spinor BRST charge in the six dimensional non-critical
superstring is
 \be
 Q_{B}=\oint (\l^{I\a} d_{I\a}+\l^{I\ad} d_{I\ad}).
 \ee
where the GS-like constraint in the linear dilaton background are
(\ref{ds6d}). The strategy to compute the cohomology will be the
same as in the previous cases. First we compute the cohomology in
the enlarged superspace, containing all the $\t^{I\a}$ and
$\t^{I\ad}$, then we will restrict to the physical ones $\t^{+}$,
that enter in the supersymmetry current algebra, and drop the
$\t^{\dot+}$.

Due to the OPE's (\ref{fake6al}), the nilpotency of the BRST
charge requires the following conditions on the pure spinors
 \be
 \d_{IJ}\l^{I\a}\l^{J\ad}=0,\nn
 \ee
 \be
 \e_{\a\b}\l^{+\a}\l^{\dot+\b}+\e_{\ad\dot\b}\l^{+\ad}\l^{\dot+\dot\b}=0, \label{ps6dil}
 \ee
 \be\label{derps}
 \e_{\a\b}(\p\l^{+\a}\l^{\dot+\b}-\l^{+\a}\p
 \l^{\dot+\b})+\e_{\ad\dot\b}(\p\l^{+\ad}
 \l^{\dot+\dot\b}-\l^{+\ad}\p\l^{\dot+\dot\b})=0,\nn
 \ee
The first two conditions are directly implied by the pure spinor
constraint (\ref{pures6d}). The last derivative condition is
implied by the pure spinor constraint as well. To show this, we
derive the Ward Identities as in
\cite{Berkovits:2002qx,Grassi:2005jz} and we prove that the
derivative constraints (\ref{derps}) are implied by the other
constraints.

It will be convenient to use an $SU(4)$ notation for the six
dimensional spinors, see the appendix for the details. We start by
noting that the product of two pure spinors $\l^{A}_{i}$, that
belong to $(\underline{4},\underline{2})$ representation of $SU(4)
\times SO(2)$, can be decomposed into representations
$(\underline{10},\underline{3}) \oplus
(\underline{10},\underline{1}) \oplus
(\underline{6},\underline{3}) \oplus
(\underline{6},\underline{1})$. Moreover, since the pure spinors
are commuting variables only the representations
$(\underline{10},\underline{3})\oplus
(\underline{6},\underline{1})$ survive. The vector representation
of $SO(6)$ is present only in $(\underline{6},\underline{1})$ and
it gives the correct pure spinor constraint. The latter can be
written as follows \be\label{new6dps} \l^{[A}_{i} \l^{B]}_{j}
=0\,, ~~~~~i,j=1,2\,, \ee since for
$(\underline{6},\underline{3})$ is automatically
satisfied.\footnote{In the four dimensional case ($d=2$ in our
notations), we have the pure spinor constraints $\l \Gamma^{m}
\l=0$ where $m=1,\dots,4$, but, due to the commuting nature of
$\l$'s, also the constraint $\l \Gamma^{m}\Gamma^{5}\l =0$ is
trivially satisfied. From the first constraint, it follows that
$\l\Gamma^{m} \p \l =0$. In addition from the Ward Identities one
can prove that $\l\g^{m}\g^{5}\p\l=0$ \cite{Grassi:2005jz}.} It
implies that $\l^{[A}_{i} \p \l^{B]}_{j} \e^{ij}=0$, and we will
use the Ward Identities to prove also $\l^{[A}_{(i} \p
\l^{B]}_{j)} =0$.

The pure spinor constraints (\ref{pures6d}) imply the gauge
invariance of the conjugated variables $\delta w_{A i} =
\Lambda_{AB} \e_{ij} \l^{B j}$ (where $\Lambda_{[AB]} \in
(\underline{6},\underline{0})$ is the gauge parameter) and it can
be shown that the only gauge invariant combinations are $J_{(ij)}
= {1\over 2}(w_{A i} \l^{A}_{j} + w_{A j} \l^{A}_{i})$ and
$J_{A}^{~B} = w_{A i} \l^{B i}$ (notice that they are in the
representations $(\underline{0},\underline{3})$,
$(\underline{15},\underline{0})\oplus
(\underline{1},\underline{0})$, where the trace of the second
operator is the ghost charge). Following
\cite{Berkovits:2002qx,Grassi:2005jz}, using the free OPE's for
the pure spinor and their conjugates, one finds \be\label{wiA}
:J_{(ik)} \l^{A}_{j} \l^{k B}: - {1\over 2} :J_{C}^{~B}\l^{A}_{j}
\l^{C}_{i}: = \ee
$$=
{\a'\over 2} \left( \l^{A}_{j} \p \l_{i}^{B} - \e_{ij} \l^{A}_{k}
\p \l^{k B} - \l_{i}^{A} \p \l^{B}_{j} + \l^{B}_{j} \p \l^{A}_{i}
\right)\,.
$$
Antisymmetrizing over the indices $A$ and $B$ and using
(\ref{new6dps}) (which implies also that $:J_{C}^{[B} \l^{A]}_{j}
\dots : =0$ under the normal ordering sign) one can conclude that
$\l^{[A}_{i} \p \l^{B]}_{j} =0$ for any $i,j$. In particular we
get $\l^{[A}_{(i} \p \l^{B]}_{j)} =0$ which is a stronger version
of the derivative constraints (\ref{derps}).

Let us now look at the cohomology. Following the strategy of the
previous $d=0$ and $d=2$ cases, there will be two kinds of vertex
operators contributing to the lowest lying cohomology at ghost
number one. The first kind is the usual weight zero operator
 \be\label{verte6}
 {\cal U}^{(1)}=\l^{I\a}A_{I\a}(\bar Z, x^\mu, \t)+\l^{I\ad}A_{I\ad}(\bar Z, x^\mu,
 \t).
 \ee
We are not presenting the details of the computation, which are
not very illuminating,\footnote{The cohomology computation for the
vertex operators (\ref{verte6}) is totally analogous to the flat
six dimensional case considered in
\cite{Grassi:2005jz,Wyllard:2005fh}.} but just state the results.
Imposing that
 \be
 Q_{B}{\cal U}^{(1)}=0,\qquad \d {\cal U}^{(1)}=Q_B\Omega^{(0)},
 \ee
gives $8\oplus8$ states, whose superfields depend both on the
physical $\t^+$ and the unphysical $\t^{\dot+}$. By considering
only the operators that belong to the physical supersymmetry
current algebra (\ref{phys6al}), namely the ones containing only
$\t^+$, we get $4\oplus4$ states. The closed string spectrum is
given by the product of the holomorphic and antiholomorphic vertex
operators and at the end of the day we get $32\oplus32$ states
which arrange in the $\N=2$ off-shell four--dimensional
supergravity multiplet, which reproduces the RNS computation.

The second kind of vertex operator is the one in which the
massless tachyon sits. This is the six-dimensional generalization
of the massive vertex operator that we used for two-dimensional
and four-dimensional superstrings. We leave the details of this
computations for a future analysis.

\subsubsection{Curved non-critical backgrounds}

We follow the discussion in section \ref{curved2} and suggest a
generalization of the non-critical pure spinors to generic six
dimensional curved backgrounds with at most sixteen real
supercharges. The computations are very similar, so we just
present the covariantized linear dilaton action for the matter
part
 \bea
 S = &{1\over2\pi\a'}\int d^2z \left(\frac{1}{2}G_{MN}(Y)\p Y^M\bar\p Y^N + E_{M}^{iA}(Y)
  d_{iA}\bar\p Y^M +E_{M}^{i\hat A}(Y)
  \bar d_{ i\hat A}\p Y^M \right)\nn\\
&-\int d^2z r^{(2)}\Phi(Y) \ ,\label{curve2d}
 \eea
where we introduced the curved six dimensional supercoordinates
$Y^M=(x^m;\t^{iA},\bar\t^{i\hat A})$. Note that $m$ is a curved
six dimensional vector index, while $A$ is a curved six
dimensional Weyl index and $i=1,2$ enumerates different Weyl
spinors. The $E$'s are the vielbein superfields. In the linear
dilaton case, the background superfields take the following
values: the supervielbeins are the six dimensional flat ones, the
dilaton superfield is linear $\Phi={Q\over2}(\vp-ix)$, the metric
$G_{MN}$ is constant, but in addition to the usual flat components
we have a new flat spinorial part, whose nonvanishing components
in four dimensional Weyl notations are
 \bea
 G_{I\a,J\b}=&-Q \e_{IJ}\e_{\a\b},\qquad G_{I\ad,J\dot\b}=&-Q
 \e_{IJ}\e_{\ad\dot\b},\nn\\
 G_{I\hat \a,J\hat \b}=&-Q \e_{IJ}\e_{\hat \a\hat \b},\qquad G_{I\hat{ \ad},J\hat{\dot\b}}=&-Q
 \e_{IJ}\e_{\hat {\ad}\hat{\dot\b}},
 \eea
which is proportional to the background charge $Q=\sqrt{2}$.
 We regard this as a specific feature of the linear dilaton
superspace structure in the pure spinor formalism. This explicitly
breaks the original $SO(6)$ Lorentz invariance of the action to
$SO(4)$, while preserving the $U(1)_x$ R symmetry.

It is suggestive to think of (\ref{curve2d}) as the matter part of
the non-critical pure spinor action in a generic curved six
dimensional background.

\subsubsection{Anomalies}

The pure spinor degrees of freedom are two Weyl spinors and anti
Weyl spinors $(\lambda^{I\alpha},\lambda^{I\dot \alpha})$, for
$I=+,\pd$ with the condition
 \bea
 \d_{IJ}\l^{I\a}\l^{J\ad}=0,\nn\\
 \e_{IJ}\e_{\a\b}\l^{I\a}\l^{J\b}=0,\\
 \e_{IJ}\e_{\ad\dot\b}\l^{I\ad}\l^{J\dot\b}=0\nn,
 \eea
These six equations define a five dimensional space with a conical
singularity at $\lambda^{I\alpha}= \lambda^{I\dot\a} =
0\,\,\forall I,\a,\ad$. Again, we expect that the correct way to
cure the singularity is by just removing it. We leave for the
future the proof that this procedure yields a pure spinor space
with vanishing first Chern class and first Pontryagin class. This
pure spinor space is different from the two and four-dimensional
cases we discussed above, in that the manifold we obtain after the
removal of the singularity at the origin is still connected. This
resembles the critical ten-dimensional pure spinor space
structure.

\section{The pure spinor measure}

In this section we will comment on the computation of tree level
string theory amplitudes in the non-critical pure spinor formalism.
In this formalism, a crucial issue in the evaluation of string
amplitudes is to construct a proper prescription for the integration
of the zero modes of the pure spinor variables. In ten dimensions,
this prescription was introduced by Berkovits in
\cite{Berkovits:2000fe} and further discussed in
\cite{Berkovits:2004px}. However, in the case of lower dimensional
pure spinor theories \cite{Grassi:2005sb,Wyllard:2005fh}, it is not
yet clear how to define this prescription for the zero mode
integration measure. In the following, we will present such a
prescription in the case of the non-critical type II superstrings.

Before proceeding, let us make some comments on the interpretation
of our integration measure. In the linear dilaton background, which
is the main subject of this paper, the string perturbation theory is
not well defined, due to the strong coupling regime. Therefore, the
linear dilaton prescription that we will discuss only makes sense
once we properly define the calculation in the strong coupling
regime, for instance by introducing the Liouville interaction, or by
replacing the $(x',\varphi)$ part of the space with a cigar CFT. On
the other hand, just as the flat ten-dimensional prescription is the
starting point for generalization to curved backgrounds, we expect
this lower dimensional prescription to be useful to study
non-critical superstrings on other curved backgrounds in which
string perturbation theory is well defined.

In the RNS formalism, the zero mode prescription for the ghosts in
a tree level amplitude is dictated by
$$\langle c\partial c
\partial^2c e^{-2\phi}e^{Q\vp}\rangle=1 \ .$$
We would like
 to obtain the analogous prescription in the pure spinor
variables for non-critical superstrings. The simplest way to do
that is to use the maps from the RNS variables to the pure spinor
variables in various dimensions. Doing that we obtain a generic
saturation rule $\langle\lambda^3\theta^{d/2}
e^{Q(\vp-ix)}\rangle$, where $d$ is the number of flat directions
in which the supersymmetry is realized, that we can list
explicitely
\begin{equation}
\label{pstree} \langle\lambda^3\theta e^{2(\vp-ix)}\rangle_{
d=0},\quad\langle\lambda^3\theta^2e^{\sqrt{3}(\vp-ix)}\rangle_{
d=2},\quad \langle\lambda^3\theta^3e^{\sqrt{2}(\vp-ix)}\rangle_{
d=4} \ ,
\end{equation}
where $x$ denotes $x'$ for simplicity of notation. The
$e^{Q(\vp-ix)}$ term is required to soak up the background charge
$Q$. Also we see that we need three ghost number one vertex
operators for a nonvanishing tree level amplitude.

Consider next a definition analgous to the pure spinor measure for
the critical superstrings \cite{Berkovits:2004px}. Let us recall
first the critical case. The ghost number anomaly reads
$$
J_{gh}(z)T(0)\sim{Q_{gh}\over z^3}+\ldots,
$$
with $Q=-8$. The generic pure spinor measure is $d^{11}\lambda$ as
the pure spinor space is eleven-dimensional. One writes this
measure as
$$d^{11}\lambda=[{\cal D}\lambda]
\lambda^3 \ , $$ where $[{\cal D}\lambda]$ is a Lorentz invariant
measure with ghost charge $-Q_{gh}=8$, and we are left with three
additional factors of $\lambda$.

Consider now the measure for the superspace variables $\theta^\a$.
In ten dimensions we have 16 supercharges, therefore the
integration measure is $d^{16}\theta$. We need  to insert as many
picture changing operators
$Y=C_\alpha\theta^\alpha\delta(C_\alpha\lambda^\alpha)$ as the
number of independent components of the pure spinor
$\lambda^\alpha$, where $C_\alpha$ are irrelevant constant spinors
\cite{Berkovits:2004px}. There are eleven picture changing
insertions, one for each of the eleven components of the pure
spinor, and since every PCO $Y$ carries a factor of $\theta$, we
are left with
$$d^{16}\theta Y^{11}\hspace{0.3cm}\sim \hspace{0.3cm}d^{16}\theta
\theta^{11} \delta^{11}(\lambda)\hspace{0.3cm}\sim\hspace{0.3cm}
d^5\theta \delta^{11}(\lambda) \ .$$ In this way we get that the
tree level pure spinor measure is defined by $\langle
\lambda^3\theta^5\rangle=1$ :
$$
d^{11}\lambda d^{16}\theta Y^{11}\hspace{0.3cm}\sim
\hspace{0.3cm}[{\cal D}\lambda]_{-Q_{gh}} \lambda^3 d^5
\theta\hspace{0.3cm}\Rightarrow\hspace{0.3cm} \langle
\lambda^3\theta^5\rangle=1.
$$

Consider next the non-critical superstrings. The number of pure
spinor degrees of freedom and the ghost number anomaly in the
various non-critical dimensions $d+2$ is
$$\begin{array}{cccc}
& d=0 \hspace{0.5cm}&d=2\hspace{0.5cm}& d=4 \\
 \textrm{dim}_\mathbb{C}({\cal M})\hspace{0.3cm}& 1\hspace{0.5cm}&2\hspace{0.5cm}&5\\
 Q_{gh}\hspace{0.5cm}& 2\hspace{0.5cm}&1\hspace{0.5cm}&-2
\end{array}
$$

Following the critical superstrings case, we propose that we can
write the pure spinor integration measure as
$$\begin{array}{cccc}
\hspace{0.5cm}& d=0 \hspace{0.5cm}&d=2\hspace{0.5cm}& d=4 \\
 \textrm{measure}\hspace{0.5cm}& d\lambda=[{\cal D}\lambda]_{-2}\lambda^3\hspace{0.5cm}
 &d^2\lambda=[{\cal D}\lambda]_{-1}\lambda^3\hspace{0.5cm}&
 d\lambda=[{\cal D}\lambda]_{2}\lambda^3
\end{array}
$$

Consider now the integration over the superspace coordinates
$\theta$. As we noted before, in the pure spinor superstring (as
well as in the hybrid formalism \cite{Grassi:2005kc}), the number
of fermionic coordinates is doubled with respect to the RNS
formalism. Thus, we get the following integration over the
superspace
$$\begin{array}{cccc}
\hspace{0.4cm}& d=0\hspace{0.4cm} &d=2\hspace{0.4cm}& d=4 \\
 \textrm{superspace}\hspace{0.4cm}& d^2\theta\hspace{0.4cm}&d^4\theta \hspace{0.4cm}&d^8\theta
\end{array}
$$
Now we have to insert in the amplitude as many picture changing
operators $Y=C\theta\delta(C\lambda)$ as the number of independent
components of the pure spinor, and since each PCO carries a factor
of $\theta$ we get
 \bea d=0:&\qquad d^2\theta Y\hspace{0.3cm}=\hspace{0.2cm}d^2\theta
\theta\delta(\lambda)\hspace{0.2cm}=\hspace{0.2cm}d\theta\delta(\lambda), \nn\\
d=2:&\qquad d^4\theta Y^2\hspace{0.3cm}=\hspace{0.2cm}d^4\theta
\theta^2\delta^2(\lambda)\hspace{0.2cm}=\hspace{0.2cm}d^2\theta\delta^2(\lambda),
\\
d=4:&\qquad d^8\theta Y^5\hspace{0.3cm}=\hspace{0.2cm}d^8\theta
\theta^5\delta^5(\lambda)\hspace{0.2cm}=\hspace{0.2cm}d^3\theta\delta^5(\lambda),\nn
 \eea
We see that in this way we reobtain the same prescription for the
pure spinors and $\theta$'s in tree level amplitudes as the one
obtained by the direct map (\ref{pstree}) from the RNS measure.


\section{Curved backgrounds}

As we noted in the introduction, due to the presence of
 a cosmological constant type term
which vanishes only for $d=10$, the low energy approximation $E\ll
l_s^{-1}$ is not valid for non-critical superstrings. Indeed, the
higher order curvature terms $\left(l_s^2{\cal R}\right)^n$ cannot
be discarded.

One may write an action for the lightest fields (which are always
massive), whose bosonic part takes the form \be S = {1 \over 2
k_d^2} \int d^d x \sqrt{G}\left(e^{-2 \Phi}\left(R + 4(\p \Phi)^2
+ {10 -d \over \alpha'} - {1 \over 2\cdot 3!}H^2\right) - {1 \over
2 \cdot n!} F_n^2 \right) \ , \ee where we have not included the
``non-tachyonic" tachyon field. Of course, solutions to the field
equations will have string scale curvature.

An interesting class of backgrounds of type IIA non-critical
superstrings are $AdS_d$ spaces with a constant dilaton $e^{2\Phi}
= {1 \over N_c^2}$ and a $d$-form RR field $F_d$ \be l_s^2F_d^2 =
2(10-d)d! N_c^2 \ . \ee The background has a string scale scalar
curvature $l_s^2{\cal R} = d-10$. While this backgrounds cannot be
studied via supergravity, it can be studied in our pure spinor
formalism. Let us sketch some of the details.

As we have seen in the previous sections, a basic feature of the
pure spinor formalism is a doubling of the superspace. This we did
by enlarging the linear dilaton superspace structure to include
BRST non-invariant superspace coordinates and their conjugate
momenta. In the linear dilaton background, working in a doubled
superspace in the pure spinor variables required an appropriate
projection to the physical superspace. However, the doubled
superspace allows us to study pure spinor superstrings in
backgrounds with twice as many supersymmetries as in the linear
dilaton background.

For instance, the non-critical superstring on $AdS_4$ with $N_c$
units of RR four form flux $F_4$ is obtained from the supercoset
 \be
 \frac{OSp(2|4)}{SO(1,3)\times SO(2)} \ ,
 \ee
and has eight real supercharges, which is precisely the content of
the doubled superspace in the non-critical superstring on the
four-dimensional linear dilaton background $\mathbb{R}^{1,1}\times
\mathbb{R}_\vp\times U(1)_x$. The supercharges are Majorana
spinors $Q_{I}^{\a}$, $I= 1,2$, $\a=1,...,4$. We can decompose the
Majorana index $\a$ into two Weyl indices $a, \dot a$ and we
identify the supercharges $Q_{1}^{a}, Q_{1}^{\dot a}$ with the
left moving sector charges and $Q_{2}^{a}, Q_{2}^{\dot a}$ with
those of the right-moving sector. The symmetry $SO(2)$ acts as the
R-symmetry on the space.

The   $OSp(2|4)$ left invariant 1-form is expanded  in the basis
of generators of the supergroup  (following Metsaev and Tseytlin
\cite{Metsaev:1998it}) as $L_\mu P^\mu + L_{\mu\nu} J^{\mu\nu} +
L_{IJ} \Lambda^{IJ} + L_\a^I Q_I^\a$. The pure spinors action then
consists of three terms
 $$
S_{ps}=S_{GS}+S_{\kappa}+S_{gh},
 $$
where the first term is the $\kappa$--symmetric GS action
\cite{Polyakov:2004br} \be\label{gsA} S_{GS} = \int_{\Sigma}
d^{2}z \eta_{\mu\nu} L^{\mu} \bar L^{\nu} + \int_{M_{3}}
d^3y\e^{IJ} L^{\mu} L^{\a}_{J} (\gamma_{5}\gamma_{\m})_{\a\b}
L^{\b}_{J} \ , \ee where $\Sigma = \p M_{3}$ and we work in the
conformal gauge on the worldsheet. The second term is
 \be S_{\kappa}= \int_\Sigma d^2z\,(\delta^{ij} + i \e^{ij}) d_{\a
i} \bar L^{\a}_{j} + (\delta^{ij} - i \e^{ij}) \bar d_{\a i}
L^{\a}_{j} + q_{RR}\bar d_{a i} \gamma^{5 \a\b} d_{\b j}
\delta^{ij}.
 \ee
It contains the kinetic term for the fermions (recall that the
first line does not give a good kinetic term for the $\theta$'s
because of the $\kappa$-symmetry) and the coupling with the RR
field. The RR 4-form of the $AdS_{4}$ background produces a
bispinor of the form $\delta_{ij}\gamma^{\mu\nu\rho\sigma}_{\a\b}
F_{\mu\nu\rho\sigma}$ which can be written as $\delta_{ij}
\gamma^{5}_{\a\b} q_{RR}$, where $q_{RR}=\e^{\mu\nu\rho\sigma}
F_{\mu\nu\rho\sigma}$ is the constant flux.

The third term in the action
 \bea
S_{gh}=&\int_\Sigma d^2z\,(\delta^{ij} + i \e^{ij}) w_{\a i}
\bar\p \l^{\a}_{j} + (\delta^{ij} - i \e^{ij}) \bar w_{\a i} \p
\bar \l^{\a}_{j} \nn\\
&+ N_{\mu\nu} \bar L^{\mu\nu} + N_{ij} \bar L^{ij} + \bar
N_{\mu\nu} L^{\mu\nu} + \bar N_{ij} L^{ij}\nn\\&  + N_{\mu\nu}
\bar N_{\rho\sigma} (-4 \eta^{\mu[\rho} \eta^{\sigma]\nu}) +
N_{ij} \bar N_{kl} \eta^{i[k} \eta^{l]j},
 \eea
contains the free action for the pure spinor ghost fields and the
interaction with the Lorentz generators of $SO(1,3)$ and of
$SO(2)$ and in the last line the coupling with the Riemann tensor
is also described. The value of the Riemann tensor is easily given
by the fact that the background is coset manifold.

The BRST operator reads
 \be\label{gsC} Q = \int d\sigma \left(
\l^{\a}_{I} d_{\a}^{I} \right) \ ,
 \ee
where the eight pure spinor variables  $\l^{\a}_{I}$ satisfy the
pure spinor conditions (\ref{pscond4}) we used in the description
of the four-dimensional linear dilaton background, both in the
left and in the right moving sectors. The $d$'s are the ones
computed from the action (\ref{gsA}).

One needs to show that the BRST operator is conserved and
nilpotent on the four-dimensional pure spinor constraints and that
this sigma model on $AdS_4$ is a consistent string theory
background at all orders in the worldsheet perturbation theory,
along the lines of \cite{Berkovits:2004xu}. A complete analysis of
non-critical superstrings on $AdS_4$ will appear in a future
publication.

\section{Discussion and open problems}

In the paper we presented a pure spinor formalism to describe
non-critical superstrings. We explicitly constructed the pure
spinor description of the non-critical superstrings in a linear
dilaton background, which can be further used to study more
general backgrounds. We have shown that, by mapping the bosonic
and fermionic linear dilaton RNS variables to pure spinor
variables, we get a description of a patch of the pure spinor
space. A basic requirement of the map is a doubling of the
superspace. We achieved this by enlarging the linear dilaton
superspace structure to include superspace coordinates and their
conjugate momenta, which are not BRST invariant on the RNS side,
although they are conserved. Working in a doubled superspace in
the pure spinor variables required an appropriate projection to
the physical superspace.

We consider the doubled superspace as a feature, since it allows
us to study pure spinor superstrings in backgrounds with twice as
many supersymmetries as the linear dilaton ones. As a concrete
example we presented the action for the type IIA non-critical
superstring on $AdS_4$ with RR four form flux, described by the
supercoset
$$OSp(2|4)/(SO(1,3)\times SO(2)) \ ,
 $$
that has eight real supercharges, which is double the
supersymmetry of the non-critical superstring on the linear
dilaton background $\mathbb{R}^{1,1}\times \mathbb{R}_\vp\times
U(1)_x$. We leave for a future work the proof of the consistency
of this background, along the lines of \cite{Berkovits:2004xu},
and the study of its spectrum and its holographic interpretation.
We just point out that this formulation of the non-critical
superstring raises the possibility of addressing a new class of
non-critical holographic backgrounds, such as the ones proposed by
Polyakov \cite{Polyakov:2004br}, which up to now have not been
accessible to worldsheet tools.

It would also be interesting to consider also the non-critical
pure spinor superstring on the eight-dimensional linear dilaton
background
 $$
 \RR^{1,5}\times \RR_\vp\times U(1)_x.
 $$
The construction of the worldsheet pure spinor formulation can be
repeated along the lines of the lower dimensional cases and we
expect it to be straightforward.

The realization of the pure spinor variables $(\lambda^{\alpha},
w_\a)$ as a beta-gamma system living on a curved pure spinor space
has important consequences. This has been discussed in
\cite{Witten:2005px,Nekrasov:2005wg}. By using a field
redefinition from RNS to pure spinor variables, we confirmed this
ten-dimensional  analysis, by computing the modifications of the
stress tensor due to the holomorphic top form on the pure spinor
space, as well as saturation rules for tree level correlators. We
then extended this map to the non-critical pure spinor spaces and
analyzed the global obstructions to define the pure spinor system
on the worldsheet and on space-time, reflected by quantum
anomalies in the worldsheet and pure spinor space holomorphic
diffeomorphisms. The non-critical pure spinor spaces have a
singularity at $\lambda^{\alpha} = 0$. Removing the origin left a
non-anomalous theory. However, for non-critical superstrings in
two and four dimensions, this resulted in a disconnected pure
spinor space.

There are various other open issues that deserve further study. We
have not performed a complete analysis of the BRST cohomology in
the pure spinor formalism. In the ten-dimensional case, the
argument for the equivalence between the RNS and pure spinor
cohomologies used in an essential way a similarity transformation
\cite{Berkovits:2001us}. It would be of importance to find such
similarity transformations in the various non-critical dimensions
as well. A direct map of the RNS unintegrated vertex operators $V$
reveals the interesting structure
$$V \sim \lambda P_{\theta}(\theta,\del \theta, ...) P_{p}(p,\del p, ...) \ ,$$
with   the $P$'s being some polynomials  and RNS GSO projection
being implemented automatically. However, for most vertex
operators the direct map does not give the pure spinor vertex
operators in the simple form expected. Here we expect to see again
the importance of the similarity transformation.

In the two-dimensional superstrings, there is a special set of
operators in the BRST cohomology at spin zero and ghost number
zero, known as the ground ring \cite{Bouwknegt:1991am,Ita:2005ne}.
In the pure spinor formulation of two-dimensional type II
non-critical string, as discussed in section \ref{2dsection}, the
cohomology at ghost number zero and weight zero is not empty: it
contains several operators constructed as  explained in
(\ref{ground}). It would be interesting to explore this further.

Another issue is the removal of the pure spinor constraints
\cite{Grassi:2001ug}. It has been shown in in
\cite{Berkovits:2005hy} that removing the pure spinor constraint
might lead to an infinite tower of ghost-for-ghosts, and it seems
that, except a finite number of them, the rest of the
ghost-for-ghosts are the same in all dimensions (see
\cite{Connes:2002ya}). It would be interesting to see if the
removal of the pure spinor constraint in lower dimensions leads to
the same ghost-for-ghosts and what are the differences.

A crucial issue that requires further study is the projection from
the doubled superspace to the physical linear dilaton superspace.
We expect this to be of much importance also for the study of pure
spinor critical superstrings on backgrounds with less than maximal
supersymmetry. Indeed, a similar lack of understanding exists for
instance when trying to study pure spinor superstrings
compactified on Calabi-Yau manifolds.\footnote{In appendix
\ref{projecti} we present an example of a possible way to perform
a projection. However, this is not the linear dilaton background
case.} A possible way to gain insight into this problem would be
to consider the ten-dimensional pure spinor superstring on
$\RR^{1,5}\times \RR^{4}/\ZZ_2$ and understand how the twisted
states arise in the cohomology computation.

Finally, we note that we discussed in this paper only the
non-normalizable vertex operators.
It is clearly of importance to analyze the normalizable vertex
operators in the pure spinor formalism \footnote{For a recent analysis
of normalizable states in the linear dilaton background see
\cite{Aharony:2004xn}.}.

\acknowledgments We would like to thank O.~Aharony, N.~Berkovits,
A.~Giveon, N. Itzhaki, M.~Kroyter, D.~Kutasov, S.~Murthy,
N.~Nekrasov, J.~Sonnenschein and E.~Witten for discussions. LM
would like to thank in particular S.~Murthy for enjoyable
explanations. This work is supported in part by DIP (grant H.52)
and ISF (grant number 03200306). LM is supported by the EU
Contract MRTN-CT-2004-512194.


\appendix

\section{Notations}

In this appendix we summarize for the convenience of the reader
the notations that we use in the construction of the non-critical
pure spinor superstrings on $\RR^{1,d-1}\times \RR_\vp\times
U(1)_x$.

There are three relevant groups : The Lorentz $SO(1,d-1)$ group of
symmetries of $\RR^{1,d-1}$, the $SO(d+2)$ group of symmetries of
$\RR^{1,d+1}$ and $U({d+2\over2})$.

The target space coordinates are denoted by
 \be
 \begin{array}{ccc}
 x^m,\quad&m=0,\ldots,d-1,x,\vp,&\quad(d+2)-\textrm{dimensional\,\, vector}\\
 x^\mu,\quad&\mu=0,\ldots,d-1,&\quad\textrm{flat}\,\,d-\textrm{dimensional\,\,
 vector}
 \end{array}
 \ee
We denote by $\Gamma^m$, $m=1,\ldots,d+2$, the $(d+2)$-dimensional
Dirac matrices.

In order to describe the doubling of superspace we introduce  an
index $I=+,\pd$ that keeps track of the physical and unphysical
fermionic coordinates
 \be
 I=\left\{\begin{array}{cc}
 +&\textrm{physical\,\,superspace},\\
 \pd&\textrm{unphysical\,\,superspace}.\end{array}\right\}
 \ee

Consider next the different cases. \vskip 0.2cm {$\RR_\vp\times
U(1)_x$} \vskip 0.2cm

The two-dimensional superstring has no spacetime Lorentz symmetry,
that is $d=0$. The only index is $I$.

\vskip 0.2cm {$\RR^{1,1}\times\RR_\vp\times U(1)_x$} \vskip 0.2cm

The four-dimensional superstring has an $SO(1,1)$ Lorentz
symmetry, that is $d=2$. The fermionic coordinates we that we use
 in the pure
spinor formulation are four Majorana-Weyl spinors $\t^{Ii}$. The
two physical fermionic coordinates, $I=+$, $i=1,2$, have the same
chirality. The two unphysical supercoordinates, $I=\pd$, $i=1,2$,
have the same chirality, but opposite to that of the physical
ones. Therefore, in this case the index $I=+,\pd$ takes care of
the spinor chirality as well.

In the text we passed from this $SO(1,1)$ notation to the $SO(4)$
notation. In the $SO(4)$ notation we reshuffle the fermionic
coordinates into a Dirac spinor $\t^A$, $A=1,\ldots,4$, that
splits into a pair of a Weyl and an anti-Weyl spinors: $\t^\a$ in the
$(\underline{2},0)\in SO(4)$ and $\t^\ad$ in the
$(0,\underline{2})\in SO(4)$.\footnote{Since a Weyl spinor is
complex, to recover the correct degrees of freedom this has to be
understood in the closed string picture.} The mapping of the
indices goes as follows
 \be
 \begin{array}{cc}
 (I,i)=(+,1)\rightarrow \a=1,\quad &(I,i)=(+,2)\rightarrow \dot\a=\dot1 \ ,\\
 (I,i)=(\dot+,1)\rightarrow
\a=2,\quad & (I,i)=(\dot+,2)\rightarrow \ad=\dot2 \ .
 \end{array}
 \ee
Finally, in the map from the RNS we used the $U(2)$ notations,
where an $SO(4)$ Dirac spinor $(\l^\a,\l^\ad)$ decomposes into
$(\l^+,\l^a,\l_{ab})$ of $U(2)$, where $a=1,2$ and $\l_{ab}$ is
the antisymmetric representation with only one component. The
relation between this and the Weyl notation is
 \be
 \l^\a=\l^a,\qquad \l^\ad=(\l^+,\l_{ab}) \ .
 \ee

The four-dimensional pure spinor constraint can be cast in the
following different ways
 \be
 \begin{array}{cc}
 SO(4):& \l^A\Gamma^m_{AB}\l^B\quad\Rightarrow\quad \l^\a\l^\ad=0,\\
 U(2) :& \l^+ \l^a=0,\quad \l_{ab}\l^a=0 \ ,\\
 SO(1,1):& \tau_{ij}\l^{Ii}\l^{Jj}=0.
 \end{array}
 \ee

 \vskip 0.2cm {$\RR^{1,3}\times\RR_\vp\times U(1)_x$}
\vskip 0.2cm

The six-dimensional superstring has an $SO(1,3)$ Lorentz symmetry,
that is $d=4$. The fermionic coordinates that we use in the pure
spinor formulation are two copies of four dimensional Dirac
spinors $(\l^{I\a},\l^{I\ad})$, for $I=+,\pd$. The two physical
fermionic coordinates, $I=+$, are a Weyl $\l^{+\a}$ and an
anti-Weyl $\l^{+\ad}$ spinor. The same applies for the two
unphysical supercoordinates, $\l^{\pd\a}$ and $\l^{\pd\ad}$.

In the text we pass from this $SO(1,3)$ notation to the space
$SO(6)=SU(4)$ notation. In the $SU(4)$ notation we reshuffle the
fermionic coordinates into two Weyl spinors $\l^{A}_i$ in the
$\underline{4}\in SU(4)$, for $A=1,\ldots,4$ and $i=1,2$. Here the
index $i$ simply enumerates different $SU(4)$ Weyl spinors with
the same chirality. The mapping of the indices goes as follows
 \be
 \l^{A}_1=(\l^{+\a},\l^{\pd\ad}) \ ,\qquad
 \l^{A}_2=(\l^{\pd\a},\l^{+\ad}).
 \ee
The reason for this is the match of their R-charge, as shown in
(\ref{dcharges}).

Finally, in the map from the RNS we use  the $U(3)$ notations.
First, in the pure spinor formulation we use  two Weyl spinors
$\l^A_i$ of $SU(4)$ that decompose into $U(3)$ representations
according to $\l^A_i=(\l^+_i,\l^a_i)$, for $i=1,2$, where $\l^+$
is a $U(3)$ singlet and $\l^a$ is a $U(3)$ vector. We can fit the
$SO(1,3)$ spinors into these $U(3)$ representations as follows
 \be
 \begin{array}{cccccc}
 (\l^+_1,\l^a_1)&\in U(3) : \qquad         &  \l^+_1      & \l^1_1 & \l^2_1
 &\l^3_1\\
 (\l^{+\ad},\l^{\pd\a})&\in SO(1,3): \qquad&  \l^{+\dot1} &
 \l^{\pd1}&\l^{\pd2} &\l^{+\dot2}
 \end{array}
 \ee
 \be
\begin{array}{cccccc}
 (\l^+_2,\l^a_2)&\in U(3) :\qquad          &  \l^+_2      & \l^1_2 &
 \l^2_2
 &\l^3_2\\
 (\l^{+\a},\l^{\pd\ad})&\in SO(1,3): \qquad&  \l^{\pd\dot1} &
 -\l^{+1}&-\l^{+2} &\l^{\pd\dot2}
 \end{array}
 \ee
There is a little abuse of notation here: the $+$ denotes the
singlet representations of $U(3)$, while on the $SO(1,3)$ it
denotes the $I=+$ physical superspace.

The six-dimensional pure spinor constraint can be cast in the
following ways
 \be
 \begin{array}{cc}
 U(3) :& \e^{ij}\l_i^+\l^a_j=0 \ ,\qquad \e_{abc}\e^{ij}\l^a_i\l^b_j=0 \ , \\
 SO(1,3):& \d_{IJ}\l^{I\a}\l^{J\ad}=0,\\
 & \e_{IJ}\e_{\a\b}\l^{I\a}\l^{J\b}=0,\\
 &
 \e_{IJ}\e_{\ad\dot\b}\l^{I\ad}\l^{J\dot\b}=0 \ .
 \end{array}
 \ee


\section{Non-critical RNS superstrings}

In this appendix we collect the RNS notations in $d=2n$ flat
dimensions (corresponding to a $2n+2$ dimensional background).

\subsection{The matter system}

The matter stress energy tensor of the system reads \bea T_{m} =&
\sum^{2n}_{\mu=1} \Big( - \half (\p x^{\mu})^{2} - \half
\psi^{\mu} \p \psi^{\mu} \Big) - \half (\p x)^{2} - \half \psi^x\p
\psi^{x} +\nn\\ & - \half  (\p \vp )^{2} + {Q(n)\over 2} \p^{2}
\vp - \half \psi_l \p \psi_l  \ , \eea

The OPE's conventions that we will be using are \be x^{i}(z)
x^{j}(0) \sim - \eta^{ij} \log z \,, ~~~~ \vp(z)  \vp(0) \sim -
\log z \ , \ee

\be \psi^{i}(z) \psi^{j}(0) \sim \frac{\eta^{ij}}{z} \ ,\quad
\psi_{l}(z) \psi_{l}(0) \sim \frac{1}{z} \ . \ee
\be T(z) e^{r x}(0) \sim \Big( { -r^{2}/2 \over z^{2}} + {\partial
\over z} \Big) e^{r x}(0)\ , \quad T(z) e^{s \vp}(0) \sim \Big( {-
s(s - Q(n))/2 \over z^{2} } + {\partial \over z} \Big) e^{s
\vp}(0)\,. \ee \label{eq:TmatterRNS}

We define $\Psi = \psi_l + i \psi$ and $\Psi^{I} = \psi^I + i
\psi^{I+n}$ (with $I = 1, \dots, n$). These are bosonized in the
usual way by introducing the bosonic fields $H$, $H^I$:
\begin{eqnarray}
  \Psi & = & \sqrt{2} e^{i H} \ , \quad \Psi^I = \sqrt{2} e^{i H^I} \
  , \nonumber \\
  \Psi \Psi^\dagger & = & 2 i \p H \ , \quad \Psi \Psi^{I\dagger} = 2 i \p H^I
  \ ,
\end{eqnarray}
where $\dagger$ denotes Hermitian conjugation in field space
without interchanging left- and right-movers: $\Psi^{\dagger} =
\psi - i \psi_{l}$, $\Psi^{\dagger I} = \psi^I - i \psi^{I+n}$. We
have
\begin{equation}
H^I(z) H^J(0) \sim - \delta^{I J} \log z \ , \quad H(z) H(0) \sim
- \log z \ .
\end{equation}
We define the spin fields $\Sigma^{\pm} = e^{{\pm } {i \over 2}
H}$ and $\Sigma^{a} = e^{\pm {i \over 2}H^{1}  \dots \pm {i \over
2} H^{n}}$, where the index $a$ runs over the independent spinor
representation of $SO(2n)$.

\subsection{The ghost system}

We have a fermionic $(b,c)$ ghost system of weights $(2,-1)$ and a
bosonic $(\beta,\gamma)$ ghost system of weights
$(\frac{3}{2},-\frac{1}{2})$. The OPE's and stress-energy tensor
are
\begin{eqnarray}
c(z) b(0) & \sim & \frac{1}{z} \ , \quad
\gamma(z) \beta(0) \sim \frac{1}{z} \ , \\
T_\mathrm{ghost} & = & - 2 b \p c - \p b c - {3\over 2} \b \p \g -
{1\over 2} \p \b \g\,,
\end{eqnarray}

Let us bosonize the ghost systems.  We define \be c =
e^{\chi}\,,~~~~~ b = e^{-\chi}\,, \ee with \be T_{b,c}  =
T_{\chi}= \half (\p \chi)^{2} + {3\over 2} \p^{2} \chi \ , \ee

\be \chi(z) \chi(0) \sim \log z \ , \ee

\be
 T_{\chi}(z) e^{a \chi}(0) \sim
 \Big(
 {a(a - 3)/2 \over z^{2}} +  {\p \over z}
 \Big) e^{a\chi}(0) \,.
\ee

For the superghosts, we have \be \gamma = e^{\phi} \eta, \beta =
\p \xi  e^{-\phi} \ee

\be T_{\b,\g} = T_{\phi} + T_{\eta,\xi} = -\half (\p \phi)^{2} -
\p^{2} \phi  - \eta \p\xi\,, \ee \be \phi(z) \phi(0) \sim {- \log
z}\,, ~~~~ \eta(z) \xi(0) \sim {1\over z}\,, ~~~ \ee

\be T_{\phi}(z) e^{b \phi}(0) \sim
 \Big(
 {-b(b + 2)/2 \over z^{2}} +  {\p \over z}
 \Big) e^{b\phi}(0) \,.
\ee

We further bosonize the fermions into \be \eta =
e^{\kappa}\,,~~~~~ \xi=e^{- \kappa}\,, \ee and \be T_{\eta,\xi}  =
T_{\kappa} = \half (\p \kappa)^{2} - \half \p^{2} \kappa \ , \ee

\be \kappa(z) \kappa(0) \sim \log z\,, \ee

\be T_{\kappa}(z) e^{c \kappa}(0) \sim
 \Big(
 {c(c +1)/2 \over z^{2}} +  {\p \over z}
 \Big) e^{c\kappa}(0) \,.
\ee

\subsection{Supersymmetry}

\subsubsection{The $N=2$ superconformal algebra}

In addition to (\ref{eq:TmatterRNS}) the $N=2$ superconformal
algebra includes the supercurrents
\begin{eqnarray}
  G^+ & = & \frac{i}{\sqrt{2}} \sum_{I=1}^n \Psi^{I\dagger} \p (x^I +
  i x^{I + n}) + \frac{i}{\sqrt{2}} \Psi^\dagger \p (\varphi + i x) -
  \frac{i Q}{\sqrt{2}} \p \Psi^\dagger \ , \\
  G^- & = & \frac{i}{\sqrt{2}} \sum_{I=1}^n \Psi^I \p (x^I - i x^{I +
  n}) + \frac{i}{\sqrt{2}} \Psi \p (\varphi - i x) - \frac{i
  Q}{\sqrt{2}} \p \Psi \ ,
\end{eqnarray}
and the $U(1)$ current
\begin{equation}
  J = \frac{1}{2} \sum_{I=1}^n \Psi^{I\dagger} \Psi^I + \frac{1}{2}
  \Psi^\dagger \Psi + i Q \p x \ .
\end{equation}

In terms of the fermion bosonization these currents take the form
\begin{eqnarray}
  G^+ & = & i \sum_{I=1}^n e^{-i H^I} \p (x^I + i x^{I+n}) + i e^{-i
  H} \p (\varphi + i x + i Q H) \ , \\
  G^- & = & i \sum_{I=1}^n e^{i H^I} \p (x^I - i x^{I+n}) + i e^{i H}
  \p (\varphi - i x - i Q H) \ , \\
  J & = & - i \sum_{I=1}^n \p H^I - i \p H + i Q \p x \ .
\end{eqnarray}

\subsubsection{The twisted $N=2$ algebra}

Out of the matter and  ghost superconformal generators we can
construct a twisted $\hat{c} =2$ $N=2$ superconformal algebra
whose generators are
 \bea G^{\prime +} = \gamma G_{m} + c
\Big(T_{m} - {3\over 2} \beta \p \gamma - {1\over 2} \gamma \p\b -
b\p c\Big) - \g^{2} b + \p^{2}c +
\p(c \xi \eta) \,, \nonumber\\
G^{\prime -} = b\,, ~~~~~~~ J^{\prime} = c b + \eta \xi\,, ~~~~~~
T^{\prime} =  T_{matter} + T_{ghost}\,,
 \eea
where $G_m=G_++G_-$ is the sum of the two matter supercurrents. The
dimension one current $G^{\prime +}$ is the BRST current of the
RNS superstring and $J^{\prime}$ is the ghost current.


\section{The RNS spectra}

In this appendix we show the details of the computation of the
spectrum in the RNS formalism on linear dilaton backgrounds. We
are interested just in operators that do not break space-time
supersymmetry. These are the primary operators of the worldsheet
$N=2$ superconformal algebra.

\subsection{d=2}
\label{spectra4d}

Let us discuss the spectrum on $\RR^{1,1}\times \RR_{\vp}\times
U(1)_x$. First we consider the holomorphic sector and then we
match holomorphic and antiholomorphic sectors to get the closed
superstring, for the type IIB case.

\subsubsection{Holomorphic sector}

\vskip 0.3cm {\it NS sector} \vskip 0.3cm

The tachyon vertex operator in the $-1$ picture is
 \be
 \label{tacop}
 T=e^{-\phi+ik_\mu x^\mu+ipx+\beta\varphi},
 \ee
and the GSO projection requires $Qp\in2\ZZ+1$. The condition
$\Delta(T)=1$ reads
 \be
 k_\mu^2-(\beta-{Q\over2})^2+p^2+{Q^2\over4}-1=0 \ ,
 \ee
 and the lowest lying state has
 $p=\pm{1\over Q}$. The
 operator
 \be
 T_\pm=e^{-\phi+{1\over Q}(\varphi\pm ix)},
 \ee
is a worldsheet  (anti)chiral primary $\Delta_{matter}(T_\pm) =
\pm\frac{q}{2} = \frac{1}{2}$ annihilated by $G^\pm$, with
space-time R-charge $R=\pm\frac{2}{3}$. It turns out that $T_+$
and $T_-$ are not mutually local. However, we are interested in
the mutual locality only when matching holomorphic and
antiholomorphic sectors, so we will discuss locality only below.

The other NS operators are analogous to the ``vectors" in the ten
dimensional superstring
 \bea
 \label{gravi4}
 J^\pm=e^{-\phi\pm H+ik_\mu x^\mu+ipx+\beta\varphi},\nn\\
 J^\mu=e^{-\phi\pm H_1+ik_\mu x^\mu+ipx+\beta\varphi},
 \eea
where $\mu$ is an $SO(1,1)$ Lorentz vector index. The GSO
projection requires $Qp\in2\ZZ$, and $\Delta=1$ gives
$k_\mu^2+p^2-(\beta-Q/2)^2+Q^2/4=0$. The lowest lying states with
$p=\beta=0$ are $J^\mu=e^{-\phi\pm H_1}$, in particular they are
worldsheet $N=2$ primaries (they have only single poles with
$G^\pm$) and are not charged under $U(1)_x$.

\vskip 0.3cm {\it R sector} \vskip 0.3cm

The operators in the Ramond sector are
 \be
 R=e^{-{\phi\over2}+{i\over2}\epsilon H+{i\over2}\epsilon_1 H_1+
 ik_\mu x^\mu+ipx+\beta\varphi},
 \ee
 where $\epsilon,\epsilon_1=\pm1$. The GSO projection requires
$Qp\in 2\ZZ+\half$ for $\e=\e_1=\pm1$ and $Qp\in 2\ZZ-\half$ for
$\e=-\e_1=\pm1$. The $\Delta(R)=1$ condition fixes
$k_\mu^2+p^2-(\beta-Q/2)^2=0$. Then we need to impose the Dirac
equation (\ref{diracco}), which fixes separately for each
direction the sign of the momentum according to the sign of the
corresponding spin component. In the $(x^1,x^2)$--plane we find
$ik_1=\e_1 k_2$, while in the $(x,\varphi)$--plane we have
$\beta=Q/2+\e p$. The last condition, together with the Seiberg
bound $\beta\leq Q/2$, imposes a restriction on the allowed
momenta in the $x$ direction. Introducing the notation
$R_{\e,\e_1}$ we find
 \bea
 \label{R4}
R_{++}=&e^{-\half\phi+{i\over2}(H+H_1)+ik_\mu x^\mu-i
px+({Q\over2}-p)\varphi},\,\,p={1\over Q}(2n+{3\over2})\geq0,\nn\\
R_{--}=&e^{-\half\phi-{i\over2}(H+H_1)+ik_\mu x^\mu+i
px+({Q\over2}-p)\varphi},\,\,p={1\over Q}(2n+{1\over2})\geq0,\nn\\
R_{+-}=&e^{-\half\phi+{i\over2}(H-H_1)+ik_\mu x^\mu-i
px+({Q\over2}-p)\varphi},\,\,p={1\over Q}(2n+{1\over2})\geq0,\nn\\
R_{-+}=&e^{-\half\phi-{i\over2}(H-H_1)+ik_\mu x^\mu+i
px+({Q\over2}-p)\varphi},\,\,p={1\over Q}(2n+{3\over2})\geq0,
 \eea
and we note that the $R_{--}$ operator is non local with respect
to (\ref{tacop}). We are interested in the lowest lying components
of the R vertex operators at zero momentum $k_\mu=0$ in the
transverse $\RR^{1,1}$ directions, which are listed in the main
text in (\ref{lowR4}).

\subsubsection{Closed superstring}

We match left and right vertex operators in IIB for concreteness,
the antiholomorphic sector being a copy of the holomorphic one we
just described. In type IIB we have two dimensional $\N=(4,0)$
spacelike SUSY in the flat noncompact directions. Because of the
requirement of mutual locality of the vertex operators, the
spectrum is not just the left right product of the sectors. Let us
denote each vertex operators in the left sector by $(\alpha,F)$,
where $\alpha$ is the space-time fermion index, $0$ in the $NS$
and $1$ in the R and $F$ is the worldsheet spinor index, given by
the sum of the picture plus the eigenvalues of the Lorentz Cartan
generators.\footnote{E.g. the vertex $e^{-\half
\phi+{i\over2}H-{i\over2}H_1}$ has $F=-1/2+1/2-1/2=-1/2$. $F$ is
defined only modulo $2$.} In addition we have the momentum $n$ and
winding $w$ in the compact $x$ direction. A closed string vertex
operator is denoted by $(\alpha,F,\bar \alpha,\bar F)$ and
$(n,w)$. Following \cite{Polchinski:1998rr,Itzhaki:2005zr}, the
mutual locality condition reads
 \be
 \label{local4}
F_2\alpha_1-F_1\alpha_2-\bar F_1\bar \alpha_2+\bar
F_2\bar\alpha_1+
\half(\alpha_1\alpha_2-\bar\alpha_1\bar\alpha_2)+2(n_1w_2+n_2w_1)\in2\ZZ.
 \ee
The total $U(1)$ R charge of a closed string vertex operator is
the sum of the left and right charges. We are again interested in
the short space-time $\N=(4,0)$ supermultiplets.

\medskip {\it NS--NS sector}
\vskip 0.3cm

The tachyon, denoted by $(\alpha,F,\bar \alpha,\bar
F)=(0,-1,0,-1)$, has momenta $p_{L,R}={1\over
Q}(n\pm{3w\over2})\in {1\over Q}(2\ZZ+1)$ and the $\Delta =1$
condition gives
 \be k_\mu^2-(\beta-{Q\over2})^2+({n\over
Q})^2+({Qw\over2})^2+{Q^2\over4}-1=0 \ .
 \ee
The lowest lying tachyons are (\ref{taclow4}).

The other NS--NS operators with  $(0,0,0,0)$ are the left right
product of the states in (\ref{gravi4})
 \bea
 \label{graviti}
G_{\pm\pm}=&e^{-\phi-\bar\phi\pm iH\pm i\bar H+ik_\mu
x^\mu+ip_Lx+ip_R\bar{x}+\beta\varphi},\nn\\
 G_{\mu\pm}=&e^{-\phi-\bar\phi\pm
iH_1\pm i\bar H+ik_\mu x^\mu+ip_Lx+ip_R\bar{x}+\beta\varphi},\nn\\
 G_{\pm \mu}=&e^{-\phi-\bar\phi\pm
iH\pm i\bar H_1+ik_\mu x^\mu+ip_Lx+ip_R\bar{x}+\beta\varphi},\nn\\
 G_{\mu\nu}=&e^{-\phi-\bar\phi\pm
iH_1\pm i\bar H_1+ik_\mu x^\mu+ip_Lx+ip_R\bar{x}+\beta\varphi},
 \eea
 where $\mu,\nu=1,2$ are $SO(1,1)_L$ Lorentz vector indices,
 the momenta are $p_{L,R}\in{1\over Q}2\ZZ$ and we introduced the notation $J\bar J\equiv G$.
The mass shell condition is
 $k_\mu^2+({n\over Q})^2+({Qw\over2})^2-(\beta-Q/2)^2+Q^2/4=0$.
The mutual locality condition between the NSNS states is given by
$n_1w_2+n_2w_1\in \ZZ$. We are interested in the lowest lying
states among (\ref{gravi4}) that belong to a short space-time
multiplet. At zero momentum in the transverse directions we find
(\ref{gr}).

The windings and momenta can be non integer. Indeed, they may not
have an interpretation in terms of actual windings and momenta in
the $x$ direction, but are just a useful notation for simplifying
the mutual locality computation.

\vskip 0.3cm
\medskip {\it R--R, R--NS and NS--R sectors}
\vskip 0.3cm

The matching of the lowest lying operators in these sectors has
been shown in the main text.

\subsection{$d=4$}

Here we discuss some details of the $SO(1,3)$ spectrum of the six
dimensional superstring. First let us collect the second set of
physical supercharges, which are nonlocal with respect to the ones
in (\ref{susy6})
\begin{eqnarray}
   q_{-1} & =  e^{-\half\phi + i(H^1 - H^2 + H - Q x)/2} \ ,\quad
   q_{-2} & =  e^{-\half\phi + i(-H^1 + H^2 + H - Q x)/2} \ , \nn\\
   q_{-\dot1} & =  e^{-\half\phi + i(H^1 + H^2 - H + Q x)/2} \
   ,\quad
   q_{-\dot2} & =  e^{-\half\phi + i(-H^1 - H^2 - H + Q x)/2}  \ \nn.\\
\end{eqnarray}
The choice of $q_+$ in the holomorphic sector and $q_-$ in the
antiholomorphic defines the type IIA superstring. However, we will
stick to the type IIB case.

\subsubsection{Holomorphic sector}

The first NS state is the tachyon whose vertex operator is
\begin{equation} \label{eq:6dtachyon1}
  T_p = e^{-\phi + i k_\mu x^\mu + i p x + \beta \varphi} \ .
\end{equation}
From requiring it to be of weight $\Delta(T)=1$ we obtain the
condition
\begin{equation}
  k_\mu k^\mu + p^2 - \beta (\beta -Q) = 1 \ ,
\end{equation}
and mutual locality with the chosen set of supercharges requires
$p Q \in 2 \mathbb{Z} + 1$.  It is a primary field and for it to
be a chiral primary of the $N=2$ SCA it must satisfy
\begin{equation}
  k_1 = - i k_3 \ , \quad k_2 = - i k_4 \ , \quad  p = \beta \ ,
\end{equation}
while the same relations with opposite signs will give an
anti-chiral primary.

Restricting to $k_\mu = 0$ states with the lowest R-charge we have
$\beta = \frac{1}{Q}$ and $p = \pm \frac{1}{Q}$. $\beta$ saturates
the Seiberg bound $\beta \le \frac{Q}{2}$ \cite{Seiberg:1990eb},
the vertex operator (\ref{eq:6dtachyon1}) is normalizable and does
not represent a microscopic state and we have to replace it with
the non-normalizable state\footnote{We can understand the
appearance of the factor $\vp$ in (\ref{laplace}) by the fact that
this is the non-normalizable solution of the Laplace equation in
six dimensions.}
\begin{equation}
\label{laplace}
  T_p = \varphi e^{-\phi + i k_\mu x^\mu + i p x + \beta \varphi} \ .
\end{equation}
Note that the lowest lying mode of this vertex operator has
$\beta=p$, so that the condition $\Delta(T)=1$ gives $k_\mu^2=0$.
This is different from the $d=2$ case: there, with respect to the
flat $SO(1,1)$ momenta, the tachyon was off-shell with a
continuous mass above a gap. Here, the tachyon is again off-shell,
but now the lowest value for the mass is zero. The tachyon being
massless is a specific feature of the $d=4$ case. The two lowest
lying such tachyons are given in (\ref{tachy6}).

At the next level there are the NS vectors with one NS oscillator
excitation given by
\begin{equation}
 J_\mu = e^{-\phi + i \epsilon H^I + i k_\mu x^\mu + i p x + \beta \varphi}
\end{equation}
with the $\Delta(J)=1$ condition
\begin{equation}
  k_\mu k^\mu + p^2 - \beta (\beta - Q) = 0 \ .
\end{equation}
Mutual locality with the supercharges requires that $p Q \in 2
\mathbb{Z}$. In order for the vertex operator to represent a
physical state the momentum must satisfy $k_I = - i \epsilon k_{I
+ 2}$ which is nothing more than the transversality condition. The
zero momentum states are given in (\ref{vector6}). The vector
polarized in the $\varphi x$-plane is
\begin{equation}
J_\pm = e^{-\phi + i \epsilon H + i k_\mu x^\mu + i p x + \beta
\varphi}
\end{equation}
with the same conditions for having weight one and being mutually
local with the supercharges and $\beta = \epsilon p + Q$. From
these requirements and the bound $\beta \le \frac{Q}{2}$ it
follows that such states always have momentum in
$\mathbb{R}^{1,3}$ so they do not appear in the zero momentum
cohomology.

We now turn to the Ramond sector. The R ground state is of the
form
\begin{equation}
  R = e^{-\phi / 2 + i (\epsilon_1 H^1 + \epsilon_2 H^2 + \epsilon H)
  / 2 + i p x + \beta \varphi} \ ,
\end{equation}
where $k_\mu k^\mu + p^2 = \left( \beta - \frac{Q}{2} \right)^2$.
The $F=1$ states are mutually local with the supercharges for $p Q
\in 2 \mathbb{Z} + 1$ while for the $F=0$ states the $x$ momentum
is $p Q \in 2 \mathbb{Z}$. From the Dirac equation one obtains
\begin{equation} \label{eq:6dDiracEq}
  k_1 = - i \epsilon_1 k_3 \ , \quad k_2 = - i \epsilon_2 k_4 \ ,
  \quad \beta = \epsilon p + \frac{Q}{2} \ .
\end{equation}
The zero momentum R states with $F=1$ are given in (\ref{ram1})
and the ones with $F=0$ are in (\ref{ram0}).

\subsubsection{Closed superstring}

The closed superstring operators are obtained as the product of
holomorphic and antiholomorphic vertex operators subject to level
matching and mutual locality conditions. Using the same convention
$(\alpha, F, \bar  \alpha, \bar F)$ as in the previous
sections \cite{Itzhaki:2005zr}, the mutual locality condition
reads now
 \be
 \label{local6}
F_1\alpha_2-F_2\alpha_1-\bar F_1\bar \alpha_2+\bar
F_2\bar\alpha_1+
\alpha_1\alpha_2-\bar\alpha_1\bar\alpha_2+2(n_1w_2+n_2w_1)\in2\ZZ.
 \ee
The level matching $L_0 = \tilde L_0$ requires
\begin{equation}
  N - \tilde N = \frac{\tilde p^2 - p^2}{2} \ ,
\end{equation}
and since we only consider the lowest R-charge states  the range
of $x$-momenta is $p = 0, \pm \frac{1}{Q}$. The zero momentum
supermultiplets are quoted in the main text.


\section{Pure spinor cohomology}

\subsection{$d=0$ gauge supermultiplet}

In this appendix we show how the supermultiplet which the tachyon
sits in is obtained from the pure spinor cohomology as in
(\ref{tacchy}). In order to simplify the notations we define the
complex coordinates $Z=\varphi + i x$ and $\bar Z = \varphi - i x$
such that $Z(z) \bar Z(0) \sim -2 \log z$. We will also use a
parametrization of the $d_I$ in which
\begin{equation}
  d_+ = p_+ - \theta^{\dot +} \partial \bar Z + Q \partial \theta^{\dot
  +} \ , \quad
  d_{\dot +} = p_{\dot +} \ .
\end{equation}
In order to prevent confusion with the background charge, we will
denote the BRST operator by $Q_B=\oint \l^Id_I$.

The tachyon appears from what would naively be the first level of
massive states, i.e.\ the weight one ghost number one vertex
operator
\begin{equation}
  \mathcal{U}^{(1)} = \del \lambda^I A_I + \lambda^I \del \theta^J
  B_{IJ} + \lambda^I d_J C_I^J + \lambda^I \Pi^{\bar Z} H_I +
  \lambda^+ J^+_+ F_{++}^+ + \lambda^{\dot +} J_{\dot +}^{\dot +}
  F_{\dot + \dot +}^{\dot +} \ ,
\end{equation}
where $J_I^J=w_I \lambda^J$, but where the wavefunctions in the
various superfields are all equal to $e^{-Z/Q}$. The above vertex
operator then has weight zero. We note that due to the gauge
invariance $\delta w_+ = \Lambda_{+ \dot +} \lambda^{\dot +} +
\Lambda'_{+ \dot +}
\partial \lambda^{\dot+}$, $\delta w_{\dot +} = \Lambda_{\dot + +}
\lambda^+ + \Lambda'_{\dot + +} \partial \lambda^+$ and the pure
spinor constraint the last two terms are the only allowed such
terms. The gauge transformation of this operator are given by
$\delta \mathcal{U}^{(1)} = Q_B {\Lambda}^{(0)}$, where
$\Lambda^{(0)}$ is the ghost number zero operator
\begin{equation}
  \Lambda^{(0)} = \partial \theta^I \Omega_I + d_I \Lambda^I +
  \Pi^{\bar Z} \Gamma_{\bar Z} + J_+^+ \Phi_+^+ + J_{\dot +}^{\dot +}
  \Phi_{\dot +}^{\dot +} \ .
\end{equation}

The equations of motion obtained by requiring $\mathcal{U}^{(1)}$
to be $Q_B$-closed when the pure spinor constraint and the Ward
identities are taken into account are
\begin{eqnarray}
  && D_+ A_+ + B_{++} - Q C_+^{\dot +} + 2 \partial_Z C_+^{\dot
  +} = 0\ ,\\
  && D_{\dot +} A_{\dot +} + B_{\dot + \dot +} + Q C_{\dot +}^+ = 0 \ ,\\
  && D_+ B_{+I} = D_{\dot +} B_{\dot + I} = 0 \ ,\\
  && D_+ H_+ - C_+^{\dot +} = 0 \ , \quad
  D_{\dot +} H_{\dot +} - C_{\dot +}^+ = 0 \ ,\\
  && D_+ C_+^+ - F_{++}^+ = 0 \ , \quad
  D_{\dot +} C_{\dot +}^{\dot +} - F_{\dot + \dot +}^{\dot +} = 0 \ ,
  \quad
  D_+ C_+^{\dot +} = D_{\dot +} C_{\dot +}^+ = 0 \ ,\\
  && D_+ F_{++}^+ = D_{\dot +} F_{\dot + \dot +}^{\dot +} = 0 \ .
\end{eqnarray}
These equations of motion are left invariant under the gauge
transformations
\begin{eqnarray}
  \delta A_I & = & \Omega_I - \epsilon_{IJ} Q \Lambda^J + 2 \delta_I^+
  \p_Z\Lambda^{\dot +} \ ,\\
  \delta B_{IJ} & = & - D_I \Omega_J  \ ,\\
  \delta C_I^J & = & - D_I \Lambda^J - \delta_I^+ \delta_+^J \Phi_+^+
  - \delta_I^{\dot +} \delta_{\dot +}^J \Phi_{\dot +}^{\dot +} \ ,\\
  \delta H_I & = & - \sigma^1_{IJ} \Lambda^J + D_I \Gamma_{\bar Z} \
  ,\\
  \delta F_{++}^+ & = & D_+ \Phi_+^+ \ , \quad
  \delta F_{\dot + \dot +}^{\dot +} = D_{\dot +} \Phi_{\dot +}^{\dot +} \ .
\end{eqnarray}

The gauge transformations can be used to choose a gauge in which
$A_I = 0$, $C_+^+ = C_{\dot +}^{\dot +} = 0$ and $H_I = 0$. In
such a gauge the equations of motion yield that
\begin{equation}
  B_{++} = B_{\dot + \dot +} = 0 \ , \quad
  C_+^{\dot +} = C_{\dot +}^+ = 0 \ , \quad
  F_{++}^+ = F_{\dot + \dot +}^{\dot +} = 0 \ .
\end{equation}
and the only remaining fields are $B_{+ \dot +}$ and $B_{\dot +
+}$ satisfying
\begin{equation}
  D_+ B_{+ \dot +} = 0 = D_{\dot +} B_{\dot + +}  \ ,
\end{equation}
each with two degrees of freedom. Then we need to project onto the
physical supercoordinates $\t^+$, so we keep only $B_{\dot+ +}$ in
the cohomology.

The closed string spectrum is a product of the above superfields
on the left and right:
$\l^\pd\bar\l^\pd\p\t^+\bar\p\bar\t^+B_{\dot ++ } \bar B_{\dot ++
}$ leading to two bosonic and two fermionic degrees of freedom,
thus matching the count obtained in the RNS cohomolgy.

\subsection{$d=2$ gauge supermultiplet}

In this appendix we show how the four-dimensional tachyon and its
supermultiplet are obtained. In order to simplify the computation
we use $SO(4)$ spinor notations and the complex coordinates
$Z=\varphi+i x$, $W=x^1+i x^2$ and their complex conjugates $\bar
Z$ and $\bar W$. The details about how to pass from the $SO(4)$ to
the $SO(1,1)$ notations are in appendix A. The superderivatives in
these conventions are cast in the form
\begin{eqnarray}
  d_\alpha & = & p_\alpha - \frac{1}{2} \delta_{\alpha \dot \alpha}
  \theta^{\dot \alpha} \partial W - \frac{1}{2} \sigma^1_{\alpha \dot
  \alpha} \theta^{\dot \alpha} \partial \bar Z + \frac{Q}{2}
  \epsilon_{\alpha \dot \alpha} \partial \theta^{\dot \alpha} \ ,\\
  d_{\dot \alpha} & = & p_{\dot \alpha} - \frac{1}{2} \delta_{\alpha
  \dot \alpha} \theta^\alpha \partial W - \frac{1}{2} \sigma^1_{\alpha
  \dot \alpha} \theta^\alpha \partial \bar Z - \frac{Q}{2}
  \epsilon_{\alpha \dot \alpha} \partial \theta^\alpha
\end{eqnarray}
and their non-singular OPE's read
\begin{equation}
  d_\alpha (z) d_{\dot \alpha} (0) \sim - \frac{Q \epsilon_{\alpha
  \dot \alpha}}{z^2} - \frac{1}{z} \delta_{\alpha \dot \alpha}
  \partial W(0) - \frac{1}{z} \sigma^1_{\alpha \dot \alpha} \partial \bar
  Z(0) \ .
\end{equation}
The BRST operator is $Q_B = \oint (\lambda^\alpha d_\alpha +
\lambda^{\dot \alpha} d_{\dot \alpha})$.

The ghost number one and weight zero vertex operator is obtained
with the procedure explained in the previous section. We take the
operator one would write for the first massive level in the
critical case \cite{Grassi:2005jz}
\begin{eqnarray}
\label{vertes}
  \mathcal{U}^{(1)} & = & \partial \lambda^A A_A + \lambda^A\partial
  \theta^B B_{AB}+ \lambda^A d_BC_A^B + \lambda^A \Pi^W H_{W\, A} +\nonumber\\
  && {} + \lambda^A \Pi^{\bar W} H_{\bar W \,A} +
  \lambda^A \Pi^{\bar Z} K_{\bar Z A} +  \lambda^\alpha J_\beta^\gamma F_{\alpha \gamma}^\beta +
  \lambda^{\dot \alpha} J_{\dot \beta}^{\dot \gamma} F_{\dot \alpha
  \dot\gamma}^{\dot \beta} \ ,
\end{eqnarray}
where $A=(\a,\ad)$ and $B=(\b,\dot\b)$ are $SO(4)$ Dirac indices
and $J_\alpha^\beta = :w_\alpha \lambda^\beta:$ and $J_{\dot
\alpha}^{\dot \beta} = :w_{\dot \alpha} \lambda^{\dot \beta}:$ are
the worldsheet currents invariant under the gauge transformations
in $w_\alpha$ and $w_{\dot \alpha}$. Then, the wavefunctions of
the various superfields appearing in the vertex operator all
contain the weight $-1$ vertex operator $e^{-Z/Q}$, in order to
have the weight zero vertex operator needed for the massless
tachyon. The gauge transformation of this vertex operator is
$\delta \mathcal{U}^{(1)} = Q_B \Lambda^{(0)}$, where $\Lambda$ is
the ghost number zero operator
\begin{eqnarray}
  \Lambda^{(0)} & = & \partial \theta^A \Omega_A+ d_A\Lambda^A
  + \Pi^W \Gamma_W + \Pi^{\bar
  W} \Gamma_{\bar W} + \Pi^{\bar Z} \Gamma_{\bar Z}  + J_\alpha^\beta \Phi_\beta^\alpha + J_{\dot \alpha}^{\dot
  \beta} \Phi_{\dot \beta}^{\dot \alpha} \ .\nn\\
\end{eqnarray}

Using the pure spinor constraint and the Ward identities, $Q_B
\mathcal{U}^{(1)} =0$ implies the equations of motion
\begin{eqnarray}
  && D_\beta A_\alpha + B_{\beta \alpha} - \epsilon_{\alpha \dot \beta} Q
  C_\beta^{\dot \beta} + \delta_{\alpha \dot \beta} \partial_{\bar
  W}
  C_\beta^{\dot \beta} + \sigma^1_{\alpha \dot
  \beta} \partial_Z C_\beta^{\dot \beta} +
  \delta_{\alpha \dot \alpha} \theta^{\dot \alpha} H_{W \beta} = 0
  \ ,\\
  && D_{\dot \beta} A_{\dot \alpha} + B_{\dot \beta \dot \alpha} +
  \epsilon_{\beta \dot \alpha} Q C_{\dot \beta}^\beta + \delta_{\beta
  \dot \alpha} \partial_{\bar W} C_{\dot \beta}^\beta +
  \sigma^1_{\beta \dot \alpha} \partial_Z C_{\dot
  \beta}^\beta + \delta_{\alpha \dot \alpha}
  \theta^\alpha H_{\bar W \dot \beta} = 0 \ ,\\
  && D_{(\beta} C_{\alpha)}^\gamma + F_{(\alpha \beta)}^\gamma = 0 \ ,
  \;\;
  D_{(\beta} C_{\alpha)}^{\dot \beta} = 0 \ , \;\;
  D_{(\dot \beta} C_{\dot \alpha)}^\beta = 0 \ , \;\;
  D_{(\dot \beta} C_{\dot \alpha)}^{\dot \gamma} + F_{(\dot \alpha \dot
  \beta)}^{\dot \gamma} = 0 \ ,\\
  && D_{(\alpha} B_{\beta) \gamma} = 0 \ , \quad
  D_{(\alpha} B_{\beta) \dot \alpha} - H_{\bar W (\alpha}
  \delta_{\beta) \dot \alpha} = 0 \ ,\\
  && D_{(\dot \alpha} B_{\dot \beta) \dot \gamma} = 0 \ , \quad
  D_{(\dot \alpha} B_{\dot \beta) \alpha} - \delta_{\alpha (\dot
  \alpha} H_{\bar W \dot \beta)} = 0 \ ,\\
  && D_{(\alpha} H_{W \beta)} - C_{(\alpha}^{\dot \beta} \delta_{\beta)
  \dot \beta} = 0 \ , \quad
  D_{(\dot \alpha} H_{W \dot \beta)} - \delta_{\beta (\dot \beta}
  C_{\dot \alpha)}^\beta = 0 \ ,\\
  && D_{(\alpha} H_{\bar W \beta)} = 0 \ , \quad
  D_{(\dot \alpha} H_{\bar W \dot \beta)} = 0 \ ,\label{hhhw}\\
  && D_{(\alpha} K_{\beta)} - C_{(\alpha}^{\dot \beta} \sigma^1_{\beta)
  \dot \beta} = 0 \ , \quad
  D_{(\dot \alpha} K_{\dot \beta)} - \sigma^1_{\beta (\dot \beta}
  C_{\dot \alpha)}^\beta = 0 \ ,\\
  && D_{(\alpha} F_{\beta \gamma)}^\delta = 0 \ , \quad
  D_{(\dot \alpha} F_{\dot \beta \dot \gamma)}^{\dot \delta} = 0 \ ,
\end{eqnarray}
which posses the gauge symmetry
\begin{eqnarray}
  \delta A_\alpha & = & \Omega_\alpha - \epsilon_{\alpha \dot \alpha}
  Q \Lambda^{\dot \alpha} + \delta_{\alpha \dot \alpha}
  \partial_{\bar W}
  \Lambda^{\dot \alpha} + \sigma^1_{\alpha \dot
  \alpha} \partial_Z \Lambda^{\dot \alpha} +
  \delta_{\alpha \dot \alpha} \theta^{\dot \alpha} \Gamma_{\bar W}
  \ , \label{eq:4d-eom1}\\
  \delta A_{\dot \alpha} & = & \Omega_{\dot \alpha} + \epsilon_{\alpha
  \dot \alpha} Q \Lambda^\alpha + \delta_{\alpha \dot \alpha}
  \partial_{\bar W} \Lambda^\alpha + \sigma^1_{\alpha
  \dot \alpha} \partial_Z \Lambda^\alpha+
  \delta_{\alpha \dot \alpha} \theta^\alpha \Gamma_{\bar W} \ ,
  \label{eq:4d-eom2}\\
  \delta B_{\alpha \beta} & = & - D_\alpha \Omega_\beta \ , \quad
  \delta B_{\alpha \dot \beta} = - D_\alpha \Omega_{\dot \beta} +
  \delta_{\alpha \dot \beta} \Gamma_{\bar W} \ ,\\
  \delta B_{\dot \alpha \beta} & = & - D_{\dot \alpha} \Omega_\beta +
  \delta_{\beta \dot \alpha} \Gamma_{\bar W} \ , \quad
  \delta B_{\dot \alpha \dot \beta} = - D_{\dot \alpha} \Omega_{\dot
  \beta} \ ,\\
  \delta C_\alpha^\beta & = & - D_\alpha \Lambda^\beta -
  \Phi_\alpha^\beta \ , \quad
  \delta C_\alpha^{\dot \beta} = - D_\alpha \Lambda^{\dot \beta} \ ,\\
  \delta C_{\dot \alpha}^\beta & = & - D_\alpha \Lambda^\beta \ ,
  \quad
  \delta C_{\dot \alpha}^{\delta \beta} = - D_{\dot \alpha}
  \Lambda^{\dot \beta} - \Phi_{\dot \alpha}^{\dot \beta} \ ,\\
  \delta H_{W \alpha} & = & - \delta_{\alpha \dot \alpha}
  \Lambda^{\dot \alpha} + D_\alpha \Gamma_W \ , \quad
  \delta H_{W \dot \alpha} = - \delta_{\alpha \dot \alpha}
  \Lambda^\alpha + D_{\dot \alpha} \Gamma_Z \ ,\\
  \delta H_{\bar W \alpha} & = & D_\alpha \Gamma_{\bar W} \ , \quad
  \delta H_{\bar W \dot \alpha} = D_{\dot \alpha} \Gamma_{\bar W} \
  ,\\
  \delta K_\alpha & = & - \sigma^1_{\alpha \dot \alpha} \Lambda^{\dot
  \alpha} + D_\alpha \Gamma_{\bar Z} \ , \quad
  \delta K_{\dot \alpha} = - \sigma^1_{\alpha \dot \alpha}
  \Lambda^\alpha + D_{\dot \alpha} \Gamma_{\bar Z} \ ,\\
  \delta F_{\alpha \gamma}^\beta & = & D_\alpha \Phi_\gamma^\beta \ ,
  \quad
  \delta F_{\dot \alpha \dot \gamma}^{\dot \beta} = D_{\dot \alpha}
  \Phi_{\dot \gamma}^{\dot \beta} \ .
\end{eqnarray}

By using the gauge parameters $\Omega_\alpha$, $\Omega_{\dot
\alpha}$, $\Gamma_{\bar W}$, $\Phi_\alpha^\beta$, $\Phi_{\dot
\alpha}^{\dot
  \beta}$ $\Lambda^\alpha$ and $\Lambda^{\dot \alpha}$ we can set
\begin{equation}
  A_\alpha = A_{\dot \alpha} =C_\alpha^\beta = C_{\dot
  \alpha}^{\dot \beta} = H_{W \alpha} = H_{W \dot \alpha} =
  H_{\bar W\a}=H_{\bar W\ad}=0,
\end{equation}
where $H_{\bar W}$ are set to zero after using (\ref{hhhw}). From
the equations of motion in this gauge we get that
\begin{equation}
  F_{(\alpha \beta)}^\gamma = F_{(\dot \alpha \dot \beta)}^{\dot
  \gamma} = 0 \ , \quad C_1^{\dot 1} = C_2^{\dot 2} = C_1^{\dot 2}
  + C_2^{\dot 1} = C_{\dot 1}^1 = C_{\dot 2}^2 = C_{\dot 1}^2 + C_{\dot
  2}^1 = 0 \ .
\end{equation}
Using the equations of motion now we can solve for
 \be
 B_{\a\dot\b}=D_\a T_{\dot\b},\qquad B_{\ad\b}=D_{\ad}T_\b \ .
 \ee
After taking into account all the gauge invariances and the
equations of motion, of the whole (\ref{vertes}) we are left with
the following field content. Out of the four $K_{\bar Z\,A}$, we
keep three. We expect that these fields correspond to some higher
states in the superstring cohomology. The fields sitting in the
$B$'s, that we identify with the supermultiplet the tachyon sits
in, boiled down to $T_\a,T_\ad$. If we focus on these last fields,
the part of the vertex operator which contains the tachyon reads
${\cal U}^{(1)}_T=\l^\a\p\t^\ad D_\a T_\ad+\l^\ad\p\t^\a D_\ad
T_\a$. We have to project the vertex operator to the physical
supercoordinates, that in our $SO(4)$ notation are
$(\t^1,\t^{\dot1})$. Then the vertex operator reduces to
 \be
{\cal U}^{(1)}_T=\l^\a\p\t^{\dot1} D_\a
T_{\dot1}(\t^1,\t^{\dot1})+\l^\ad\p\t^1 D_\ad T_1(\t^1,\t^{\dot1})
\ ,\label{tachyve}
 \ee
where the two physical superfields $T_1$ and $T_{\dot1}$ contain
$2\oplus2$ degrees of freedom each. At this point, the structure
of the pure spinor space crucially comes into play. As we
discussed, the four dimensional pure spinor space $\l^\a\l^\ad=0$
is the union of two disconnected patches and our would-be vertex
operator (\ref{tachyve}) is symmetric under the exchange of the
two patches $\l^\a\leftrightarrow\l^\ad$. The RNS formalism is
mapped onto just one of the two patches, let us choose the
$\l^\ad\ne 0$ patch. If we want to compare the RNS cohomology with
the pure spinor result, we are forced to do that on one of the two
disconnected patches, so the tachyon supermultiplet on the patch
$\l^\ad\ne 0$ has the form
 \be
 {\cal U}^{(1)}=\l^\ad \p\t^{1}D_\ad T_1(\t^1,\t^{\dot1}) \ ,
 \ee
containing $2\oplus2$ degrees of freedom. In the closed string,
the vertex operator is just the product of the holomorphic and
antiholomorphic vertex operators, which contains the tachyon
supermultiplet.

\section{The six-dimensional map}

In this appendix we present some details about the map from the
RNS to the pure spinor variables in the six dimensional case of
section \ref{map6d} in the main text. In the following discussion
it will be convenient to break the $SO(6)$ group into $U(3)$ and
classify the different supercharges and pure spinor components
using their representation in $SU(3)_{U(1)}$, in terms of which
$q_{+ \dot 1}$ is in the representation $1_\frac{3}{2}$, while
$q_{+ 1}$, $q_{+ 2}$ and $q_{+ \dot 2}$ form a $3_{-\frac{1}{2}}$
representation.

We will work in the patch in which the $1_\frac{3}{2}$ component
of the pure spinor is non-zero and raise the $1_\frac{3}{2}$
supercharge $q_{+ \dot 1}$ to picture $\frac{1}{2}$ and obtain
\begin{equation}
  q_{+ \dot 1} = b \eta e^{3 \phi /2 + i (H^1 + H^2 + H - Q x) / 2} +
  \dots \ ,
\end{equation}
where $\dots$ are terms with lower exponentials of the field
$\phi$. The $3_{-\frac{1}{2}}$ is left in the $-\frac{1}{2}$
picture. We further define the fermionic momenta
\begin{eqnarray}
  p_{+ \dot 1} & =  b \eta e^{3 \phi / 2+ i (H^1 + H^2 + H - Q x) / 2}
  \ , \quad
  p_{+ 1} & =  e^{-\phi / 2 + i (H^1 - H^2 - H + Q x) / 2} \ ,\nn\\
  p_{+ 2} & =  e^{-\phi / 2 + i (-H^1 + H^2 - H + Q x) / 2} \
  ,\quad
  p_{\dot + \dot 2} & =  e^{-\phi / 2 + i (-H^1 - H^2 + H + Q x) / 2}
  \ .\nn
\end{eqnarray}
The last momentum is not a physical supercharge. It is taken
instead of $q_{+ \dot 2}$, which is singular with $q_{+1}$ and
$q_{+2}$ because of the supersymmetry algebra. The non-physical
$p_{\dot + \dot 2}$ is non-singular with all the others. We also
define the coordinates conjugate to these momenta
\begin{eqnarray}
  \theta^{+ \dot 1} & =  c \xi e^{-3 \phi / 2 - i (H^1 + H^2 + H - Q
  x) / 2} \ ,\quad
  \theta^{+ 1} & =  e^{\phi / 2 - i (H^1 - H^2 - H + Q x) / 2} \ ,\nn\\
  \theta^{+ 2} & =  e^{\phi / 2 - i (-H^1 + H^2 - H + Q x) / 2} \
  ,\quad
  \theta^{\dot + \dot 2} & =  e^{\phi / 2- i (-H^1 - H^2 + H + Q x) /
  2} \ .\nn
\end{eqnarray}
 We now map
 \begin{equation}
   \eta = e^{\tilde \phi + \tilde \kappa} p_{+ \dot 1} \ , \quad b =
   e^{(\tilde \phi - \tilde \kappa) / 2} p_{+ \dot 1} \ ,
 \end{equation}
from which we obtain
\begin{eqnarray}
  \tilde \phi & = & -\frac{i}{4} (3 H + 3 H^1 + 3 H^2 - 3 Q x - 4 i
  \kappa - 9 i \phi + 2 i \chi) \ ,\\
  \tilde \kappa & = & \frac{i}{4} (H + H^1 + H^2 - Q x - 4 i \kappa -3
  i \phi - 2 i \chi) \ ,
\end{eqnarray}
which are non-singular with the fermionic momenta and satisfy the
OPE's
\begin{equation}
  \tilde \phi (z) \tilde \phi (0) \sim - \log z \ ,
  \quad \tilde \kappa (z) \tilde \kappa (0) \sim \log z
\end{equation}
and all the other OPE's are non-singular.

The coordinate $x$ is singular with the new variables. This is
solved by performing the shift
\begin{equation}
  x' = \frac{1}{\sqrt{2}} (i \phi - H^1 - H^2 - H) \ .
\end{equation}
(A definition with the opposite signs is also possible.) It is
curious that the new $x'$ is independent of the original $x$.

Hence, the RNS bosonic fields $x$, $\beta$ and $\gamma$ are mapped
into the pure spinor bosons $x'$, $\tilde \phi$ and $\tilde
\kappa$, while the bosonic coordinates $\varphi$ and $x^\mu$
($\mu=1,\dots,4$ ) are mapped into themselves. The eight fermionic
RNS variables $\psi_l$, $\psi$, $\psi^\mu$, $b$ and $c$
($\mu=1,\dots,4)$ are mapped to the eight fermionic coordinates
and momenta.
One can turn the RNS energy-momentum tensor into pure spinor
fields on this patch and obtain
\begin{eqnarray}
  T' & = & -\frac{1}{2} \del x^\mu \p x^\nu\eta_{\mu\nu}- \frac{1}{2} (\del
  \varphi)^2 - \frac{1}{2} (\del x')^2 + \frac{Q}{2} \del^2 (\varphi -
  i x') - \nonumber\\
  && {} - p_{+ \dot 1} \del \theta^{+ \dot 1} - p_{+ 1} \del \theta^{+
  1} - p_{+ 2} \del \theta^{+ 2} - p_{\dot + \dot 2} \del \theta^{\dot
  + \dot 2} - \nonumber\\
  && {} - \frac{1}{2} (\del \tilde \phi)^2 + \frac{1}{2} (\del
  \tilde \kappa)^2 + \del^2 \tilde \phi + \del^2 \tilde \kappa \
  . \label{eq:RNS6d-energy-momentum}
\end{eqnarray}
whose total central charge still vanishes. Next, one uses the
standard bosonization of a $\beta \gamma$-system in order to
relate $\tilde \phi$ and $\tilde \kappa$ to the pure spinor
variables
\begin{equation}
  \lambda^{+ \dot 1} = e^{\tilde \phi + \tilde \kappa} \ , \quad
  w_{+ \dot 1} = \del \tilde \kappa e^{-\tilde \phi - \tilde \kappa}
\end{equation}
and write the $(\tilde\phi,\tilde\kappa)$ part of the
energy-momentum tensor as
\begin{equation}
  T_\lambda = w_{+ \dot 1} \del \lambda^{+ \dot 1} - \frac{1}{2}
  \del^2 \log \Omega \ ,
\end{equation}
where $\Omega$ is the top dimensional form on the pure spinor
space \cite{Nekrasov:2005wg,Witten:2005px}. By comparison with
(\ref{eq:RNS6d-energy-momentum}) the top form is
\begin{equation}
  \Omega = e^{-3 (\tilde \phi + \tilde \kappa)} = (\lambda^{+ \dot 1})^{-3}
  \ .
\end{equation}

The above expressions are not covariant. They can be covariantized
by adding the missing $3_{-\frac{1}{2}}$ and $1_\frac{3}{2}$
components of the weight $(1,0)$ $bc$-systems $(p_{2 a},
\theta_2^a)$ and $(p_{2 +}, \theta_2^+)$ and the weight $(1,0)$
pure spinor $\beta \gamma$-systems $(w_{1 a}, \lambda_1^a)$ and
$(w_{2 +}, \lambda_2^+)$ (we switched back to the notation
introduced in the beginning of this section) allowing us to write
covariantly the energy-momentum tensor (\ref{6dstress}). The
addition of the quartet does not modify the central charge due to
the cancellation between the $bc$-systems and the $\beta
\gamma$-systems. The BRST operator should be modified so that
these additional fields do not modify the cohomology.


\section{A projection}
\label{projecti}

 Consider the four dimensional non-critical superstring we constructed in section \ref{map4d}.
In this appendix we propose a way to project out half of the
doubled superspace. However, we do not know what target space this
string theory describes.

Consider the
 pure spinor BRST operator we considered in the main text
 \be\label{qb}
 Q_B=\oint\l^{Ii} d_{Ii} \ ,
 \ee
and add the following contribution
 \be
 Q'=\e^{I1}Q_{I1} \ ,
 \label{qnew}
 \ee
where $Q_{I1}=\oint q_{I1}$ are two of the doubled supercharges.
These are defined to anticommute with the $d$'s in (\ref{ds4}),
and are given by
 \be
 q_{Ij}=p_{Ij}+\half\tau_{ij}\left(\d_{IJ}\t^{Ji}\p(x_1+ix_2)+\tau_{IJ}\t^{Ji}\p(\vp-ix)-Q\e_{IJ}
 \p\t^{Ji}\right),
 \ee
such that the OPE's of the $q_{Ii}$ among themselves is the same
as (\ref{ddope4}) but with opposite signs.

Since the supercharges and the superderivatives anticommute, the
new BRST charge ${\cal Q}=Q_B+Q'$ is nilpotent on the pure spinor
constraint (\ref{ps4}). Now let us look at the cohomology of this
theory. Instead of computing the cohomology of (\ref{qb}) and then
removing the unphysical $\t^{\dot+i}$ from the spectrum, as we did
in the main text, we now take ${\cal Q}$ as the full BRST operator
of the theory, without the need of a further projection. The
cohomology computation for the ghost number one and weight zero
vertex operators (\ref{ghostfir}) can be computed in two steps.
First, we compute the cohomology of the old $Q_B$, by which we
obtain the ``off-shell" four dimensional vector supermultiplet
(\ref{cohove}) with $4\oplus4$ states, that we can pack in a real
four dimensional superfield $V$. Secondly, the action of the new
term $Q'$ is just to remove from the spectrum the dependence on
half of the supercoordinates $\t^{I1}$. However, this is a
different projection with respect to the one that realizes the
linear dilaton background!

Requiring that the four dimensional vector superfield $V$ belongs
to the cohomology of ${\cal Q}= Q_B+Q'$ restricts it to
 \be
 \label{spectrum4}
V|_{\t^{I1}=0}=\phi+\t^{I2}\psi_{I2}+\t^2 F,
 \ee
giving a total of $2\oplus2$ degrees of freedom as before, but
with different space-time charges.

The interesting observation is that the BRST charge ${\cal Q}=
Q_B+Q'$ can be derived by a GS-like action through the usual
Oda--Tonin trick \cite{Oda:2001zm}.

Consider now the four dimensional pure spinor action in the linear
dilaton background (\ref{fouraction}). It is invariant with
respect to the pure spinor operator (\ref{qb}) but not with
respect to the total BRST operator ${\cal Q}=Q_B+Q'$. In fact, the
variation of the action with respect to $Q'$ in (\ref{qnew}) is
$[Q',S]=\e^{I1}\int d^2z\bar\p d_{I1}$. By applying a simple
descent method, we find that to restore the invariance of the
action under the total BRST symmetry we need to add the term
 \be
 \label{snew}
 S_{new}=\e^{I1}\int d^2z\bar\p w_{I1},
 \ee
 whose variation is $[Q_B,S_{new}]=-[Q',S]$. The total action
$S+S_{new}$ is now invariant with respect to the total BRST
operator. This analysis seems to produce a consistent projection.
What is the resulting string theory is not clear.

One may go the other way to get the pure spinor BRST invariant
action from the GS action, where we start with a GS--like action
$S_{GS}$ and consider the BRST fermion $\Psi=\int d^2z
\,w_{I1}\bar\p\t^{I1}$, following the procedure of
\cite{Oda:2001zm,Matone:2002ft}. Then the total pure spinor action
is given by
 $$
S_{ps}=S_{GS}+\{{\cal Q},\Psi\} \ .
 $$
In our case ${\cal Q}=Q_B+Q'$.


\bibliography{PhD}
\bibliographystyle{JHEP}

\end{document}